\documentclass{book}
\usepackage{epsfig}
\usepackage{mathrsfs}
\usepackage{amssymb}
\usepackage[T1]{fontenc}
\usepackage{textcomp} 
\newcommand{\bq}{\begin{equation}}
\newcommand{\eq}{\end{equation}}
\newcommand{\isperdef}{ \: \raisebox{-0,25 mm}{$\stackrel{\scriptscriptstyle \mathrm{def}}{=}$} \: }

\newcounter{gnol}

\newtheorem{prop}{Proposition}
\newtheorem{defin}{Definition}
\newtheorem{cor}{Corollary}[prop]
\newtheorem{corzel}{Corollary}[gnol]
\newtheorem{inter}{Interpretation}
\newtheorem{post}{Postulate}
\newtheorem{ques}{Question}
\newtheorem{ans}{Answer}

\newtheorem{lemma}[prop]{Lemma}

\begin{document}

\pagestyle{empty}

\begin{titlepage}
$ $\\[2.8 cm]
\begin{center}
{ \Huge \Huge Quantum Measurement}
\\[1,2 mm]{\huge A Coherent Description}
\\[0,8 cm]
{\Large Bas Janssens}
\\[2,2 cm]
{ \Huge $$\left( 
	\begin{array}{ccccc}
	{}*&*&0&0&0 \\
	{}*&*&0&0&0 \\
	0&0&*&*&* \\
	0&0&*&*&* \\
	0&0&*&*&* \\
	\end{array}
\right)$$ }
\end{center}
\end{titlepage}

\cleardoublepage

\title{{\Huge Quantum Measurement}\\{\huge A Coherent Description}}
\author{
	Master's thesis in mathematical physics\\
	Radboud University Nijmegen\\
	\\[-0,2 cm]	
	Author: Bas Janssens \\ 
	Supervisor: Hans Maassen}
\date{December 2004}
\maketitle

\cleardoublepage

\begin{quote}
${}$\\[5 cm]
\begin{center}
\emph{`Is it your opinion, Winston, that the past has real existence?'}
\\[0.3 cm]
George Orwell\\*
Nineteen Eighty-Four
\end{center}
\end{quote}

\cleardoublepage

\frontmatter 

\pagestyle{plain}

\chapter{Prologue}

Quantum mechanics is weird. 
I had never realized this until the spring of 2003, when dr.~Maassen clearly and 
carefully explained to me, in a two-hour lecture, why quantum mechanics cannot be a simple hidden variables theory.

This was something of a shock to me.
But as the shock subsided, I realized that weirdness and lack of objective determinism 
were the least of my problems. There is the horrible threat of inconsistency!

Quantum mechanics is intrinsically probabilistic. 
Observation however involves a single outcome.
In order to handle this consistently, quantum systems \emph{must} 
exhibit a so-called collapse of the wave function.
There are loads of specific theoretical models which show collapse on a measured system.
But is this really \emph{necessary}?
Or would it be possible to perform a measurement without collapse?

Let's kill the tension right away. The answers are yes and no respectively.
If you bear with me for \pageref{dolsnok} short and exciting pages, I'll prove to you 
that transfer of information out of a system always causes 
collapse on that system.
Along the way, we will gain quantitative insight into the balance between 
information gain and state disturbance.

Well then. Now that I've skillfully tricked you into reading the rest of this thesis, I am left with 
the pleasant task of thanking people.
It goes without saying that I am grateful to my friends and family for such diverse matters 
as giving birth or money (which only goes for a fairly restricted class of family members)
and tolerating or even supporting my (rather obnoxious) enthusiasm.
(The latter goes for anyone having had even remote contact with me over the past year.)

But here and now, I would like to express my gratitude to those who made a direct contribution to this thesis:
Prof.~Ronald Kleiss for agreeing to be my official first and unofficial second supervisor. 
Prof.~Klaas Landsman for a careful reading of this text, and for many useful suggestions. 
M$\breve{\mathrm{a}}$d$\breve{\mathrm{a}}$lin Gu\k{t}$\breve{\mathrm{a}}$ for suggesting a simple proof of
lemma~(\ref{cs}) in the case of completely positive maps, putting me on track for proposition~(\ref{appred}).
Janneke Blokland for useful advice on the editing.
And most of all, I would like to thank dr.~Hans Maassen.\newpage

The two-hour lecture I told you about was part of his course in \emph{`Quantum Probability, Quantum Information Theory
and Quantum Computing'} which lies at the very heart of this thesis. Indeed, insiders will recognize chapter~\ref{heis}
as a mere extension of the lecture notes \cite{maa}.
I am thankful for the excellent guidance and for the countless conversations we had, shaping my perception of 
quantum mechanics into its present state.
They were sometimes slightly confusing, but always pleasant and fertile.
I have learnt much from Hans over the past year, and I would be proud if his style may be seen, reflected in my writing. 
\\*\\*
Bas Janssens\\*
December 2004

\chapter{Definitions and Conventions}
\begin{tabular}{p{2 cm}|  p{9,3 cm}}
$\mathbb{P}$&
\rule[-2 mm]{0 mm}{5 mm}Generically denotes a classical probability distribution.  \\
\hline
$\mathbb{E}_{\mathbb{P}}(\mathfrak{a})$ & 
\rule[-1,5 mm]{0 mm}{5,5 mm}$\mathbb{E}_{\mathbb{P}}(\mathfrak{a}) = 
\int_{\Omega} \mathfrak{a}(\omega) \mathbb{P}(d\omega)$,
the expectation of random variable $\mathfrak{a}$ under $\mathbb{P}$.\\
$\mathbf{Var}_{\mathbb{P}} (\mathfrak{a})$ &
\rule[-1,5 mm]{0 mm}{5 mm}$\mathbf{Var}_{\mathbb{P}} (\mathfrak{a}) = 
\mathbb{E}_{\mathbb{P}}(\mathfrak{a}^2) - 
\mathbb{E}_{\mathbb{P}}(\mathfrak{a})^2$,
the variance under $\mathbb{P}$ of random variable $\mathfrak{a}$.\\
$\mathbf{Cov}_{\mathbb{P}} (\mathfrak{a},\mathfrak{b})$ &
\rule[-1,5 mm]{0 mm}{5 mm}$\mathbf{Cov}_{\mathbb{P}} (\mathfrak{a},\mathfrak{b}) =
\mathbb{E}_{\mathbb{P}}(\mathfrak{a}\mathfrak{b}) - 
\mathbb{E}_{\mathbb{P}}(\mathfrak{a})\mathbb{E}_{\mathbb{P}}(\mathfrak{b})$,
the covariance under $\mathbb{P}$ of $\mathfrak{a}$ and $\mathfrak{b}$.\\
$\mathscr{C}(V)$&
\rule[-1,5 mm]{0 mm}{5 mm}If $V \subset \mathbb{C}$, then $\mathscr{C}(V)$ is the space of continuous 
functions \mbox{on $V$.}\\
\end{tabular}\\*[0.5 cm]
\begin{tabular}{p{2 cm}|p{9,3 cm}}
$\mathscr{A,B,C,D}$&
\rule[-2 mm]{0 mm}{5 mm}Script letters denote C$^*\!$-algebras.\\
\hline
$\alpha$&
\rule[-1,5 mm]{0 mm}{5,5 mm}Automorphisms are usually denoted by $\alpha$. 
By an automorphism of a C$^*\!$-algebra, we \emph{always} mean a C$^*\!$-automorphism, i.e. we assume 
$\alpha(A)^{\dagger} = \alpha(A^{\dagger})$. \\
$\mathscr{C}(A)$&
\rule[-1,5 mm]{0 mm}{5,5 mm}If $A \in \mathscr{A}$, then $\mathscr{C}(A) \subset \mathscr{A}$ is
the C$^*\!$-sub-algebra generated by $A$ and $\mathbb{I}$.\\
$\mathscr{S(A)}$&
\rule[-1,5 mm]{0 mm}{5 mm}Denotes the convex \emph{state space} of normalized positive linear functionals 
$\mathscr{A} \to \mathbb{C}$.\\
$\overline{z}$&
\rule[-1,5 mm]{0 mm}{5 mm}The complex conjugate of a complex number $z \in \mathbb{C}$.\\
$A^{\dagger}$&
\rule[-1,5 mm]{0 mm}{5 mm}The Hermitean conjugate of $A \in \mathscr{A}$.\\ 
$\Re A$ & \rule[-1,5 mm]{0 mm}{5 mm}$\Re A = \frac{1}{2} (A + A^{\dagger})$: 
the real part of $A \in \mathscr{A}$.\\
$\Im A$ & \rule[-1,5 mm]{0 mm}{5 mm}$\Im A = \frac{1}{2i} (A - A^{\dagger})$: 
the imaginary part of $A \in \mathscr{A}$.\\
$\mathbf{M}^*$&
\rule[-1,5 mm]{0 mm}{5 mm}If $\mathbf{M}$ is a positive linear mapping $\mathscr{B} \to \mathscr{A}$, 
then its dual $\mathbf{M}^*$ is a
$\mathscr{S(A)} \to \mathscr{S(B)}$ mapping
defined by $\mathbf{M}^*(\rho) = \rho \circ \mathbf{M}$.\\
$\mathbf{Var}_{\rho} (A)$ &
\rule[-1,5 mm]{0 mm}{5 mm}$\mathbf{Var}_{\rho} (A) = \rho(A^{\dagger}A) - \overline{\rho(A)}\rho(A) $,
the variance under $\rho$ of \mbox{$A \in \mathscr{A}$}.\\ 
$\mathbf{Cov}_{\rho} (A)$ &
\rule[-1,5 mm]{0 mm}{5 mm}$\mathbf{Cov}_{\rho} (A,B) = \rho(A^{\dagger}B) - \overline{\rho(A)}\rho(B) $,
the covariance under $\rho$ of $A,B \in \mathscr{A}$.\\
$\mathbf{Spec}(A)$&
\rule[-1,5 mm]{0 mm}{5 mm}The spectrum of $A \in \mathscr{A}$.\\
$Y'$&
\rule[-1,5 mm]{0 mm}{5 mm}$Y' = \{\, A \in \mathscr{A} \, | \, [A,Y] = 0 \,\}$,
the relative commutant of $Y$.\\
\end{tabular}\\*[0.5 cm]
\begin{tabular}{p{2 cm}|p{9,3 cm}}
$M_n$&
\rule[-2 mm]{0 mm}{5 mm}The algebra of $n \times n$-matrices acting on $\mathbb{C}^n$.  \\
\hline
$\psi_{+}$&
\rule[-2,5 mm]{0 mm}{9 mm}
$\psi_+ = \left( 	\begin{array}{c}
 			1	\\
			0	\\
			\end{array}
		\right)$ in $\mathbb{C}_2$\\
$\psi_{-}$&
\rule[-2,5 mm]{0 mm}{9 mm}
$\psi_- = \left( 	\begin{array}{c}
 			0	\\
			1	\\
			\end{array}
		\right)$ in $\mathbb{C}_2$\\		
$\sigma_{x}$&
\rule[-2,5 mm]{0 mm}{9 mm}%
$\sigma_x = 	\left( 	\begin{array}{cc}
 			0	& 1	\\
			1 	& 0	\\
			\end{array}
		\right)$ in $M_2$.\\
$\sigma_y$ &		 
\rule[-2,5 mm]{0 mm}{9 mm}%
$\sigma_y = \left( \begin{array}{cc}
 			0	& -i	\\
			i 	& 0	\\
	\end{array}		
\right)$ in $M_2$.\\
$\sigma_z$ &
\rule[-2,5 mm]{0 mm}{9 mm}%
$\sigma_z = \left( \begin{array}{cc} \label{spiderman}
 			1	& 0	\\
			0	& -1	\\
	\end{array}
\right)$ in $M_2$.\\
$\mathbf{P}_{+}$&
\rule[-2,5 mm]{0 mm}{9 mm}%
$\mathbf{P}_+ = \left( \begin{array}{cc}
 			1	& 0	\\
			0 	& 0	\\
	\end{array}
\right)$ in $M_2$.\\
$\mathbf{P}_{-}$&
\rule[-2,5 mm]{0 mm}{9 mm}%
$\mathbf{P}_- = \left( \begin{array}{cc}
 			0	& 0	\\
			0 	& 1	\\
	\end{array}
\right)$ in $M_2$.\\
\end{tabular}

\newpage

\tableofcontents

\chapter{Introduction}
In the prologue, I already made a brief sketch of the subject of this thesis.
Allow me to add a few details.

This is a Master's thesis in mathematical physics, written in the period 
September 2003 -- October 2004 at the Radboud University Nijmegen, under 
supervision of dr.~Hans Maassen. Its aim is twofold:
\begin{itemize}

\item[-] First of all, I intend to prove general theorems, showing that state collapse on a measured 
system is a necessary consequence of transporting information out of that system.
  
\item[-] Afterwards, we shall investigate the balance between information gain and state 
disturbance in a more quantitative way. 

\end{itemize}
The first point is in contrast with authors like Joos, Zeh and Zurek who, if I understand correctly,
endeavour to find  specific models of decoherence on a system, independent of the information transfer.

Another group of authors (Hepp, Lieb, Sewell, Rieckers)  
transports information to a central pointer in an infinite system. 
Although centrality of the pointer enables them to model a global collapse on all observables, 
it inhibits them from using automorphic time evolution. 

I on the other hand will use finite (not necessarily finite dimensional) systems, transporting 
information to non-central pointers. 
I do not restrict attention to automorphic time evolution, but it is allowed as 
a special case in each proposition in this thesis.    
We will see that collapse of the wave function then automatically occurs on the examined system. But we will 
also show that 
an approximate collapse occurs on a much wider range of observables, including the observables 
of the measurement apparatus.

This brings us to the similarities with this second group of authors. 
Most of chapter~\ref{macro} is based on a most original idea, due to Hepp, that 
collapse has to do with the difference in size between the pointer (macroscopic) and the  
observable on which collapse is supposed to occur (microscopic). In fact, the whole point of 
using finite systems was originally just to get a quantitative estimate of how the idealization of 
infinite systems is reached in the realistic case of a large but finite system. 
Exactly how large must the system be? What observables defy collapse?
You will find answers in chapter~\ref{macro}.\newpage    

But before entering the bulk of this thesis, I would like to caution the reader about two points which 
might seem essential at first sight, but are in fact merely a matter of personal preference of the author:
\begin{itemize}

\item[-] In the postulates of quantum mechanics, systems are modelled by C$^*\!$-algebras.
	 This is not essential: I might just as well have chosen von Neumann algebras.
	 If you are not familiar with operator algebra techniques altogether,
	 you may take in mind $\mathscr{B(H)}$, the algebra of all bounded operators on the Hilbert space $\mathscr{H}$. 
	 This example will serve you well throughout the text.

\item[-] I also wield a rather unorthodox interpretation of quantum mechanics. 
	 \mbox{I do} this simply because it is in my eyes the simplest possible interpretation.
	 Do not be distracted: the issue of interpreting quantum mechanics is quite separate from 
	 the issue of state collapse after information transfer. If you do not like my interpretation of quantum mechanics, 
	 simply take your own favourite interpretation and apply it to the mathematics in this thesis.
	 The result will probably be satisfactory.  

\end{itemize}

Finally, a short note on source material.
This thesis contains a grand total of 26 lemmas and propositions, plus another 10 corollaries.
Of course not all of these are new. There are three possibilities. 

Sometimes, I prove theorems already proven by others before. In that case of course, 
I refer to this person explicitly.
I have also formulated a number of results which have been widely known for a long time.   
In that case I explicitly mention that it is a `standard result'.
This leaves a total of 19 lemmas and propositions plus 7 
corollaries that are neither attributed to one particular person nor
explicitly labelled `standard result'. 
These are of my own invention.  
The reader will understand however that there exist no guarantees that no one else has invented them before.
If so, I have not been able to track this down.

Now, without more ado, we finally move from the disclaimer to the actual physics. 
Enjoy the ride\ldots

\mainmatter

\chapter{Quantum Measurement} \label{schrod}
In order to investigate quantum measurement, we dwell on the foundations of quantum mechanics 
for a short while.  

\section{Postulates of Quantum Mechanics}
Regardless of their interpretation, we will postulate the existence of the three 
mathematical protagonists of quantum theory: an algebra, a state and a one-parameter group of automorphisms. 
\begin{post} \label{post1}
A quantum mechanical system will be modelled mathematically by a unital C\/$^*$-algebra 
$\mathscr{A}$, the algebra of observables.
\end{post}
Quite often, $\mathscr{A} = \mathscr{B}(\mathscr{H})$, the algebra
of all bounded linear operators on some Hilbert space $\mathscr{H}$. 
If you're not familiar with C$^*\!$-algebras, this is a good example to keep in mind.
In general, any C$^*\!$-algebra $\mathscr{A}$ has a faithful representation on 
some Hilbert space $\mathscr{H}$, see \cite[p.~281]{ka1}.     

\begin{post} \label{post2}
A physical state of this system will be modelled mathematically by a (normalized) positive linear functional 
$\rho$ on $\mathscr{A}$.  
\end{post}
The set of all possible states on $\mathscr{A}$ makes up state space, $\mathscr{S}(\mathscr{A})$.
A unit vector $|\psi \rangle  \: \in \: \mathscr{H}$, for example, induces a state $\rho$ on 
$\mathscr{B}(\mathscr{H})$ by 
$ \rho(A) 
\isperdef
\langle \psi|A|\psi\rangle $.
Respecting conventional abuse of language instead of common sense, we will not always distinguish between vector states and vectors. 

But these are not the only states allowed for the system. If for each
positive \mbox{integer $i$,} we have a normalized vector $|\psi_i \rangle \in \mathscr{H}$ and a number $p_i \in [0,1]$ such that 
$\sum_{i = 1}^{\infty} p_i = 1 $, we may form the state 
$ \rho(A) 
\isperdef  
\sum_{i = 1}^{\infty} p_i \langle  \psi_i |A|\psi_i \rangle $ 
on $\mathscr{B}(\mathscr{H})$.
This is a positive linear functional on $\mathscr{A}$ as well, and thus 
perfectly acceptable as a physical state.
This particular state happens to be normal, i.e. continuous in the weak
operator topology. But we also accept non-normal functionals as states,
according to the postulate above. 

\begin{post} \label{post3}
Time evolution in an isolated system is modelled mathematically by a one-parameter group of automorphisms of $\mathscr{A}$: 
$t \mapsto \alpha_t$. That is, $\alpha_{t + s} = \alpha_t \circ \alpha_s
$ for all $t,s \in \mathbb{R} $. 
\end{post}
Let $\alpha^* : \mathscr{S}(\mathscr{A}) \to \mathscr{S}(\mathscr{A})$ denote 
the dual action of $\alpha$ on state space: $\alpha^* (\rho) \isperdef \rho \circ \alpha$.
Then a state $\rho$ on time $t_0$ will evolve to a state $\alpha_{(t_1 -
t_0)}^{*}\rho$ on time $t_1$.

A unitary $U \in \mathscr{A}$ induces an automorphism $\alpha$ of $\mathscr{A}$ by 
$\alpha (A) = U^{\dagger} A U$.
Usually the one-parameter group of automorphisms describing time-evolution is induced by a one-parameter 
group of unitaries $t \mapsto U_t \in \mathscr{A}$.
So after a time $t$ the observable $A$ will evolve to $\alpha_t (A) = U_t ^{\dagger}AU_t$.
Translating to the Schr\"odinger picture, a state $ \rho $ will evolve to $\rho_t$: 
$ \rho_t (A) = \alpha_{t}^{*}\rho(A) = \rho(U_t ^{\dagger} A U_t)$.
If $ \rho $ is the vector state induced by $|\psi \rangle $, then it evolves to 
$\rho_t (A) = \langle  \psi | U_t ^{\dagger} A U_t |\psi \rangle $. 
In other words, $\rho_t$ is the vector state induced by $U_t |\psi \rangle $.

\subsubsection{Induced Probability Measures}

States on a C$^*\!$-algebra have the pleasant property of inducing probability measures.
This is clear from the following standard result:

\begin{prop}[Induced Probability Measure] \label{kloek}
Let $\mathscr{A}$ be a C\/$^*$-algebra.
Let $X \in \mathscr{A}$ be Hermitean. Then each state 
$\rho \in \mathscr{S}(\mathscr{A})$ defines uniquely a probability measure
$\mathbb{P}_{\rho, X}$ on the Borel $\sigma$-algebra of $ \mathbf{Spec}(X)$
such that $\rho(f(X)) = \int f(x) \mathbb{P}_{\rho, X}(dx)$ for all $f \in \mathscr{C}(\mathbf{Spec}(X))$.
\end{prop}
\textbf{Proof}:
\begin{quote}
By the Gel'fand transform (see e.g. \cite[p.~271]{ka1}), we have an
injective C$^*\!$-homomorphism 
$f \mapsto f(X)$ of $ \mathscr{C}(\mathbf{Spec}(X)) $, the continuous functions on 
the spectrum of $X$, into $\mathscr{A}$. We compose this with the state $\rho$ :
$\mathscr{A} \to \mathbb{C}$ to yield a functional 
$\mathbb{E}$ on $ \mathscr{C}(\mathbf{Spec}(X)) $.
In short, $\mathbb{E}(f) \isperdef \rho(f(X))$.
$\mathbb{E}$ is positive: if $f \geq 0$, then $f(X) \geq 0$ in the operator ordering,
hence $\mathbb{E}(f) = \rho(f(X)) \geq 0$ since $\rho$ is a positive functional.

By the Riesz representation theorem (see \cite[p.~209]{coh}), $\mathbb{E}$ defines a unique
Borel measure $\mathbb{P}_{\rho, X}$
on $\mathbf{Spec}(X)$ satisfying $\mathbb{E}(f) = \int f(x)\mathbb{P}_{\rho, X} (dx)$. That this is a probability measure
indeed can be seen from $\mathbb{P}_{\rho, X}(\mathbf{Spec}(X)) = \rho(\mathbb{I}) = 1$. 
\begin{flushright}
\emph{q.e.d.}
\end{flushright}
\end{quote}
For example, consider the physically relevant case of a normal (i.e. weakly continuous) state $\rho$ 
on a von Neumann algebra $\mathscr{A}$.
Then $X$ defines\footnote
{V is a Borel subset of $\mathbf{Spec}(X)$. In this thesis, all subsets of spectra
will be assumed Borel measurable.} 
a projection valued measure $ V \mapsto \mathbf{P}(V)$. 
In this setting, $\mathbb{P}_{\rho, X}$ is simply defined by 
$\mathbb{P}_{\rho, X}(V) \isperdef \rho(\mathbf{P}(V))$.  
In case $\rho$ is a vector state $|\psi\rangle$ and $X$ has discrete spectrum with 
non-degenerate eigenspaces, $X = \sum_{i} x_i |\psi_i\rangle \langle \psi_i|$, 
this amounts to $\mathbb{P}_{\rho, X}( \{ x_i \} ) = |\langle\psi_i| \psi \rangle|^2 $.

\newpage \section{Interpretation of the Postulates}

Up until now, we only have the existence of mathematical objects. 
In order to link mathematics to physical experiment, we seek an interpretation of the 
postulates above.

\subsection{An Inconsistent Interpretation} 

Proposition~(\ref{kloek}) cries out for an interpretation of postulates (\ref{post1}) and (\ref{post2}). 
The first that comes to mind would be:
\begin{inter}\label{oelmogor}
${}$\\*
\\[-7 mm]
\begin{itemize}
\item[-]A quantum mechanical system will be represented by a unital C\/$^*$-algebra $\mathscr{A}$,
	the algebra of observables.
\item[-]Each random variable $\mathfrak{a}$ is represented by a Hermitean $A \in \mathscr{A}$. 
\item[-]The random variable $\mathfrak{a}$ objectively takes values in $\mathbf{Spec}(A)$. If the system is in state $\rho \in \mathscr{S(A)}$, 
then the probability that $\mathfrak{a}$ takes value in $V$ is $\mathbb{P}_{\rho, A}(V)$.
\end{itemize}
\end{inter}

This interpretation allows us to interpret the state 
$\rho(A) = \sum_{i = 1}^{\infty} p_i \langle \psi_i |A|\psi_i \rangle$ as a system in state $|\psi_i \rangle $ 
with probability\footnote{ This interpretation of $\rho$ is slightly less straightforward than it seems at first sight, since 
the decomposition of $\rho$ into pure states may not be unique.} $p_i$.

Unfortunately, the interpretation above is inconsistent\label{niretsch}, at least for $\mathscr{A} = M_2 \otimes M_2$. 
For each Borel subset $V \subset \mathbf{Spec}(A)$, 
$\mathbb{P}_{\rho, X}(V)$ gives the probability that $\mathfrak{a}$ takes
value in $V $.
Therefore $\mathfrak{a}$ is a random variable on the probability space 
$(\mathbf{Spec}(A), \mathbb{P}_{\rho, A}(V))$.
Similarly, $\mathfrak{b}$ is a random variable on the probability space 
$(\mathbf{Spec}(B), \mathbb{P}_{\rho, B}(V))$. 

But if $\mathfrak{a}$ and $\mathfrak{b}$ both take objective values, then there must exist \emph{some} 
probability distribution $\mathbb{P}$ on $\mathbf{Spec}(A) \times \mathbf{Spec}(B)$ such that 
$\mathbb{P} (V \times W)$ is the probability that $\mathfrak{a}$ lies in $V
$ \emph{and} $\mathfrak{b}$ in $W $.
So $\mathfrak{a}$ and $\mathfrak{b}$ must be random variables on \emph{the same} probability space  
$(\mathbf{Spec}(A) \times \mathbf{Spec}(B),\mathbb{P})$, having $\mathbb{P}_{\rho, A}$ and 
$\mathbb{P}_{\rho, B}$ as marginal probability distributions.
 
This means that each set of random variables has to satisfy Bell's inequalities. 
(See \cite[p.~116]{jau} for a very thorough and \cite[p.~673]{bj} for a very accessible version.)
But in $M_2 \otimes M_2$, Bell's inequalities are violated for certain
choices of $\rho$.
As a  result, interpretation~(\ref{oelmogor}) is inconsistent.

This is not exclusively the case for $\mathscr{A} = M_2 \otimes M_2$. Interpretation~(\ref{oelmogor}) 
is inconsistent for $\mathscr{A} = \mathscr{B(H)}$ with
$\mathit{dim} ( \mathscr{H} ) > 2$ (see \cite{koc}). 

\subsubsection{Induced Joint Probability Measures}

In order to pave the way for a consistent interpretation, we will extend proposition~(\ref{kloek}) with
the following standard result:
\newpage
\begin{prop}[Induced Joint Probability Measures] \label{simple}
Let $X$,$Y \in \mathscr{A}$ be Hermitean such that $[X,Y] = 0$. Then each state 
$\rho \in \mathscr{S}(\mathscr{A})$ defines uniquely a probability measure
$\mathbb{P}_{\rho, X , Y}$ on the Borel $\sigma$-algebra of $ \mathbf{Spec}(X) \times
\mathbf{Spec}(Y)$
such that $\rho(f(X)g(Y)) = \int f(x)g(y) \mathbb{P}_{\rho, X , Y}(dx,dy)$.
\mbox{In particular:}
$$
\mathbb{E}_{\mathbb{P}}(X) = \rho(X), \quad \quad \mathbb{E}_{\mathbb{P}}(Y) = \rho(Y)
$$
$$
\mathbf{cov}_{\mathbb{P}}(X, Y) = \rho(XY) - \rho(X)\rho(Y)
$$
\end{prop}
\textbf{Proof}
\begin{quote}
For continuous $f$ and $g$ on the spectra of $X$ and $Y$, we have 
once again $f(X)$ and $g(Y)$ by the Gel'fand transform (see \cite[p.~271]{ka1}). 
We define a functional 
$\mathbb{E}$ on $ \mathscr{C}(\mathbf{Spec}(X) \times \mathbf{Spec}(Y)) = 
\mathscr{C}(\mathbf{Spec}(X)) \otimes \mathscr{C}(\mathbf{Spec}(Y))$ by 
$\mathbb{E}(f \otimes g) \isperdef \rho(f(X)g(Y))$.
$\mathbb{E}$ is positive: if $f \otimes g \geq 0$, choose $f,g \geq 0$. Then,
because $[f(X),g(Y)] = 0$, $f(X)g(Y) = \sqrt{f(X)} g(Y) \sqrt{f(X)} \geq 0$ in the operator ordering.
Now since $\rho$ is a positive functional,
$\mathbb{E}(f \otimes g) = \rho(\sqrt{f(X)} g(Y) \sqrt{f(X)}) \geq 0$.

By the Riesz representation theorem (see \cite[p.~209]{coh}), $\mathbb{E}$ defines a unique
Borel measure $\mathbb{P}_{\rho, X , Y}$
on $\mathbf{Spec}(X) \times \mathbf{Spec}(Y)$ satisfying $\mathbb{E}(f
\otimes g) =$ \\*$\int f(x)g(y) \mathbb{P}_{\rho, X , Y}(dx,dy)$. Of course 
$\mathbb{P}_{\rho, X , Y}(\mathbf{Spec}(X) \times \mathbf{Spec}(Y)) = \rho(\mathbb{I}) = 1$. 
\begin{flushright}
\emph{q.e.d.}
\end{flushright}
\end{quote}
For example, let $\rho$ be a normal (i.e. weakly continuous) state 
on a von Neumann \mbox{algebra $\mathscr{A}$.}
Then $X$ and $Y$ define commuting projection valued measures 
$ V \mapsto \mathbf{P}(V)$ and $ W \mapsto \mathbf{Q}(W)$. 
In this setting, $\mathbb{P}_{\rho, X, Y}$ is simply defined by the formula
$\mathbb{P}_{\rho, X, Y}(V \times W) \isperdef$\\*$ 
\rho(\mathbf{P}(V) \mathbf{Q}(W))$.  

\subsection{A Traditional Interpretation} 

A standard interpretation of postulates (\ref{post1}) through (\ref{post3}) is the following:
\begin{inter}\label{inter}
${}$\\*
\\[-7 mm]
\begin{itemize}
\item[-]A quantum mechanical system will be represented by a unital C\/$^*$-algebra 
	$\mathscr{A}$, the algebra of observables\footnote {In this case,
	we will denote both system
	and algebra by $\mathscr{A}$.}.
\item[-]At any fixed time, there is one state $\rho \in \mathscr{S(A)}$,
representing all knowledge concerning $\mathscr{A}$.
\item[-]Each observable is represented by a Hermitean element $A$ of
$\mathscr{A}$\footnote{Again, both the observable and the Hermitean element
will be referred to by $A$.}. 
\item[-]There is an action called `measurement'.
	Observables only take on objective values if they are measured. 
	Joint measurement of commuting observables $A$ and $B$ yields 
	values of $A$ in $V \subset \mathbf{Spec}(A)$ and of $B$ in 
	$W \subset \mathbf{Spec}(B)$ 
	with probability \mbox{$\mathbb{P}_{\rho, A, B}(V \times W)$.}
\item[-]Time evolution on an undisturbed system $\mathscr{A}$ is represented by a 
	one-parameter group of automorphisms of $\mathscr{A}$: $t \mapsto \alpha_t$. 
	A state $\rho$ at time $t_0$ will evolve to the state 
	$\alpha_{(t_1 - t_0)}^{*}\rho$ at time $t_1$.
\end{itemize}
\end{inter}
Bell's inequalities do not apply here because it is not possible to perform a simultaneous measurement 
on non-commuting observables. One problem solved.

In order to interpret a state $\rho$ on a system $\mathscr{A}$, an outside observer is introduced, performing 
this abstract `measurement of $A$'. This has the effect 
of forcing $A$ to take on an objective value.
Neither the observer, nor the measurement are described within the framework of quantum mechanics. 
But they do have a physically observable effect on the system.

\subsubsection{State Reduction}
We will demonstrate this with a simple example. 
Let $\mathscr{A}$ be $M_2$, the algebra of $2 \times 2$-matrices acting on $\mathscr{H} = \mathbb{C}^2$.
This describes an electron, having only spin-properties.
Let the observable $A$ be $\sigma_z \in M_2$, the spin in the $z$-direction\footnote{ 
For notation on spin systems, see page~\pageref{spiderman}. 
}.
\mbox{$\sigma_z$ has} spectrum $\mathbf{Spec}(\sigma_z) = \{ 1 , -1 \}$.
Suppose that $\sigma_z$ is measured.   
According to interpretation~(\ref{inter}), either $\sigma_z = 1$ or $\sigma_z = -1$.
(With probabilities $\mathbb{P}_{\rho, \sigma_z}(\{ 1 \} )$ and $\mathbb{P}_{\rho, \sigma_z}( \{ -1 \} )$ 
respectively.)

Suppose that the measurement is \emph{repeatable}. This means that a second measurement of 
$\sigma_z$, performed immediately after the first, would yield the same result.
Then after measurement, knowledge of the system has increased:  
if the measurement has revealed $\sigma_z = 1$, then we know that any future measurement 
of $\sigma_z$ will yield $\sigma_z = 1$ again.
According to interpretation (\ref{inter}), 
we must now describe the system by a different mathematical state than before, one 
yielding $\sigma_z = 1$ with certainty.
The \emph{only} state on $\mathscr{A}$ which does this is the vector \mbox{state $|\psi_{+}\rangle$}. 
This change of state forced by measurement is called \emph{state reduction}.

\subsubsection{Classical State Reduction}

In classical probability theory measurement is also possible, and the reduction it produces 
is called 'conditioning'. 
A classical spin-system is described by a probability distribution $\mathbb{P}$ on a classical 
probability space $\Omega = \{ +1 , -1 \}$. Repeatable measurement can be performed on 
the random variable $\sigma_z : \Omega \to \mathbb{R}$ defined by $ \sigma_z(\omega) = \omega$. 
If $\sigma_z = 1$, the observer will update $\mathbb{P}$ to
the conditioned probability distribution 
$\mathbb{P}( \: \bullet \: | [\sigma_z = 1])$ defined by
\bq
\mathbb{P}( \: V \: | [\sigma_z = 1] ) \isperdef 
\frac
		{  \mathbb{P}(  \: V \:  \cap [\sigma_z = 1])  }
		{  \mathbb{P}( [\sigma_z = 1] )  }
= \delta_{+1}(V)		
\eq
where $\delta_{+1}$ is the point measure on $\omega = +1$.
Another observer, unaware of the measurement outcome, will describe 
the system by a distribution
$\mathbb{P}' = \sum_{x = \pm 1} \mathbb{P}([\sigma_z = x]) 
\mathbb{P}( \: \bullet \: | [\sigma_z = x])$.
This equals the original distribution $\mathbb{P}$.
So in a classical probability space, state reduction is \emph{subjective}. 
It can be attributed entirely to the increase of knowledge of the first observer.

\subsubsection{Quantum State Reduction}\label{oegsnok}

In quantum mechanics the situation is radically different: 
objective collapse after measurement can be verified experimentally. 
Suppose $M_2$ is in vector state $ \alpha |\psi_{+}\rangle +\beta |\psi_{-}\rangle $.
Repeatable $\sigma_z$-measurement is then performed on $M_2$.
$\sigma_z = 1$ will occur with probability $|\alpha|^2$ and $\sigma_z = -1$ with probability $|\beta|^2$.
So after measurement, the system is in state $\psi_+$ with probability $|\alpha|^2$ and in state $\psi_-$ 
with probability $|\beta|^2$.
The situation is now objectively different from the one before: 
whether we know the value of $\sigma_z$ or not, we may perform 
$\sigma_x$-measurement. In  both $\psi_+$ and $\psi_-$, the probability of finding $\sigma_x = 1$ equals $1/2$.
But before measurement, in the state $ \alpha |\psi_{+}\rangle  + \beta |\psi_{-}\rangle $, this probability 
would have been $1/2 + \Re(\overline{\alpha} \beta)$. 

So experimental verification of collapse can be achieved as follows: 
start with a system in state $ 1/\sqrt{2} |\psi_{+}\rangle  + 1/\sqrt{2} |\psi_{-}\rangle $. 
The first observer measures $\sigma_z$, the second $\sigma_x$. 
This is repeated a number of times. As soon as the second observer measures 
$\sigma_x = -1$, the point is made. Collapse is verified objectively. 

In summary, repeatable measurement of $\sigma_z$ always causes the state of $M_2$ to jump:  
\begin{itemize}  
\item[-]If the observer learns that $\sigma_z = + 1$, the state jumps from
	$ \alpha |\psi_{+}\rangle  + \beta |\psi_{-}\rangle $ to $\psi_{+}$. 
	We will call this change `state \emph{reduction}' 
\item[-]If the observer is ignorant of the outcome, the state jumps from 
	the vector state $ \alpha |\psi_{+}\rangle  + \beta |\psi_{-}\rangle $ to the 
	mixed state $|\alpha|^2 |\psi_+\rangle\langle\psi_+| + |\beta|^2
	|\psi_-\rangle\langle\psi_-|$. 
	We will call this change `state \emph{collapse}'
\end{itemize}
In the literature, each jump is commonly referred to as both collapse and reduction. 
In order to avoid confusion, we shall keep these notions separate. 

\subsubsection{Comments}

There is a sharp and physically observable schism between the situation before and after
measurement.
Therefore, it is important to know if measurement takes place and if so, exactly when\footnote
{J.Bell puts it like this (see \cite{bel}): ``... so long as the wave packet reduction is an essential component, and so long as 
we do not know exactly when and how it takes over from the Schr\"odinger equation, we do not have an exact 
and unambiguous formulation of our most fundamental physical theory.''}. 
In practice, there is hardly any doubt as to when it takes place. And if
you feel comfortable with interpretation (\ref{inter}), you may read 
the rest of this thesis as an attempt to explain why, in practice, the exact point in time where the actual 
reduction takes place is not of much importance.

But personally, I feel rather uncomfortable with the need for outside observers, not described within quantum theory, 
exerting influence on a system that \emph{is} described by quantum theory.
I would like my physical theory to be a universe in itself.
It should describe all the observables that can be measured.
But also all observers, and the act of measurement itself. 

\subsection{My Favourite Interpretation} 

First of all then,
we want to describe all interference with a system $\mathscr{A}$ within the framework of quantum mechanics.
This does not mean that time evolution on $\mathscr{A}$ is always automorphic. 
But it does mean that there is always a system $\mathscr{D} \supseteq \mathscr{A}$ such that 
time evolution is automorphic on $\mathscr{D}$. 
Think of $\mathscr{D}$ as the entire universe, if you have to.

Secondly, observables never take on objective values \emph{at all}.
Physics is not about objective events.
Physics intends to predict the observations made by observers. So a physical theory should have:
\begin{itemize}
\item[-]A list of all observers $\mathscr{C}$.
\item[-]For each $\mathscr{C}$, a list of observables that are directly observed by $\mathscr{C}$.
\end{itemize}
Then it should predict the probabilities of the observations made by each separate
observer without having to make any 
reference to other observers or objective reality. 
\begin{inter} \label{mijnes}
${}$\\*
\\[-7 mm]
\begin{itemize} 
\item[-]There is one largest universal system. It is represented by a
\mbox{C\/$^*\!$-algebra $\mathscr{D}$}.

\item[-]
	Each observable is represented by a Hermitean element $A$ of $\mathscr{D}$.
	Each observer is represented by an abelian C\/$^*\!$-subalgebra $\mathscr{C}\subset\mathscr{D}$.
	$\mathscr{C}$ directly observes all Hermitean $A$ in $\mathscr{C}$. 
	$\mathscr{C}$ cannot directly observe $A$ if $A \not\in \mathscr{C}$.

\item[-]At any fixed time, there is one $\rho \in \mathscr{S}(\mathscr{D})$ representing the physical state of $\mathscr{D}$.
	Each direct observation of any $A \in \mathscr{C}$ made by $\mathscr{C}$ has a value in $\mathbf{Spec}(A)$, 
	the spectrum \mbox{of $A$.}  
	If $A,B \in \mathscr{C}$, then the probability that $\mathscr{C}$ observes a value of 
	$A$ in $V \subset \mathbf{Spec}(A)$ 
	and a value of $B$ in $W \subset \mathbf{Spec}(B)$ is given by 
	$\mathbb{P}_{\rho, A, B}(V \times W)$. 

\item[-]Even while observation takes place, time evolution is represented by a one-parameter group of 
	automorphisms of $\mathscr{D}$: 
	$t \mapsto \alpha_t$. A state $\rho$ at time $t_0$ will 
	evolve to the state $\alpha_{(t_1 - t_0)}^{*}\rho$ at time $t_1$.
\end{itemize}
\end{inter}
All $A \in \mathscr{C}$ are observed by $\mathscr{C}$ and all probabilities of 
finding joint values are given by the theory.
This means that the observables in $\mathscr{C}$ may be considered random variables on some 
classical probability space.  
If $A \in \mathscr{D}$ and $B \in \mathscr{D}$ do not commute, then they cannot be directly 
observed by the same observer.
Both are random variables, but not on the same probability space.
Therefore Bell's inequalities do not apply.

Each subsystem of $\mathscr{D}$ is of course represented by some subalgebra $\mathscr{A} \subset \mathscr{D}$.
If this subsystem happens to be invariant under the time evolution of $\mathscr{D}$, then 
we can regard $\mathscr{A}$ as an isolated system with time evolution $\alpha_t |_{\mathscr{A}}$.

\subsubsection{Direct and Indirect Observation}

Notice that one single observer $\mathscr{C}$ cannot directly observe all
$A \in \mathscr{D}$ 
if $\mathscr{D}$ is not abelian.
Suppose for example that the observer is an eye. 
This eye observes directly the voltage on each of its neurons.
Indirectly, it can also observe say a painting on the other side of the room:
rays of light carry information from the painting to the retina and the eye observes voltages in the retina directly.
There is a radical difference between direct and indirect observation.

Direct observation is the most primitive form. 
It is needed to link mathematics to experience. 
It is restricted to observables $A$ in the observer $\mathscr{C}$, and it does 
not result in any objective collapse. 

Indirect observation of observables outside $\mathscr{C}$ is possible. 
However, this requires some pre-formed image of the outside world: the eye must trust 
photons to travel in straight lines.  
We will call this indirect observation \emph{measurement}, and we will come to it later on.    

\subsubsection{Reduction and Collapse}

Given a state $\rho$ on $\mathscr{A}$, we now formally define its \emph{reduced} 
state $\rho_{Y}$ on $\mathscr{A}$:
\begin{defin}[Reduced State]\label{condit}
Let $\mathscr{A}$ be a C\/$^*!$-algebra.
Let $Y \in \mathscr{A}$, let $\rho \in \mathscr{S(A)}$ and let $\rho(Y^{\dagger}Y) \neq 0$.
Then we define\footnote
{We will see that if $\rho(Y^{\dagger}Y) = 0$, there will never be any need for a reduced state.
From now on, when mentioning reduced states, I will tacitly assume their existence. Even in theorems. } 
the state $\rho_{Y}$ on $\mathscr{A}$ by
$$
\rho_{Y}(A) \isperdef \frac{\rho(Y^{\dagger}AY)}{\rho(Y^{\dagger}Y)}.
$$
\end{defin}
Suppose that a countable decomposition $\{\, V_i \,| \, i \in I \,\}$ of the spectrum of $Y$ is given.
This means that $V_i$ are Borel subsets of $\mathbf{Spec}(Y)$ such that $V_i \cap V_j = 0$
for $i \neq j$, and $\bigcup_{i \in I} V_i = \mathbf{Spec}(Y)$. 
Then we also have a \emph{collapsed} state $\mathbf{C}^{*}\rho$:
\begin{defin}[Collapsed State]\label{koliek} 
Let $\mathscr{A}$ be a von Neumann algebra. Let $Y \in \mathscr{A}$ be Hermitean,
and let $\{\, V_i \,| \, i \in I \,\}$ be a countable decomposition of its spectrum.
Then if $\rho \in \mathscr{S(A)}$, its collapsed state $\mathbf{C}^* (\rho)$ is 
defined by
$$
\mathbf{C}^* (\rho) (A) \isperdef \rho \big( \sum_{I} \mathbf{P}(V_i) A \mathbf{P}(V_i) \big)
$$
where $V \mapsto \mathbf{P}(V)$ is the projection valued measure of $Y$.  
\end{defin}
Note that $\mathbf{C}^* (\rho) (A) = \sum_{I} \mathbb{P}_{\rho, Y}(V_i) \rho_{\mathbf{P}(V_i)}(A)$.
The collapsed state is the sum of reduced states, weighed over the probability distribution. 

\subsubsection{Conditioning}

Suppose that $\rho$ is a normal state on a von Neumann algebra $\mathscr{A}$.
Suppose $\mathscr{C}$ directly observes both $A$ and $B$ so that $[A , B] = 0$.
Then $A$ has projection valued measure $V \mapsto \mathbf{P}(V)$ and $B$ has commuting 
projection valued measure $W \mapsto \mathbf{Q}(W)$.
One can then calculate the probability distribution of  $B \in \mathscr{C}$
provided that $\mathscr{C}$ observes a value of $A$ in $V \subset \mathbf{Spec}(A)$:
\begin{eqnarray*}
\mathbb{P}_{\rho, A, B}( \, [B \, \mathrm{in} \, W] \, | \, [A \, \mathrm{in} \, V] \,) & \isperdef &
\frac{\mathbb{P}_{\rho, A, B}([B \, \mathrm{in} \, W \,\, \mathrm{and} \,\, A \, \mathrm{in} \, V])}
{\mathbb{P}_{\rho, A, B}([A \, \mathrm{in} \, V])}\\ 
& = &  \frac{\rho \big( \mathbf{P}(V) \mathbf{Q}(W) \big) }{\rho \big( \mathbf{P}(V) \big) } \\
&=& \frac{\rho \big( \mathbf{P}(V) \mathbf{Q}(W) \mathbf{P}(V) \big) }{\rho \big( \mathbf{P}(V) \big) }\\
& = & \rho_{\mathbf{P}(V)} \big( \mathbf{Q}(W) \big) \\
& = & \mathbb{P}_{\rho_{\mathbf{P}(V)}, B}([B \, \mathrm{in} \, W]).
\end{eqnarray*}
The reduced state $\rho_{\mathbf{P}(V)}$ induces the conditional probability 
measure on any $B \in A'$, i.e. any $B$ such that $[B , A] = 0$. 
We thus have an interpretation of $\rho_\mathbf{P}(V)$\label{condex} 
considered as a state on $A'$.

\subsubsection{A Benevolent Word of Caution to the Reader}
Since this is my thesis, I will proceed with my favourite interpretation.
But keep in mind that this is merely a way of interpreting the mathematics to come.
As such, theorems are universal and do not hinge on any interpretation.

\newpage \section{Measurement}

We have stated that an observer $\mathscr{C}$ can observe $X \in \mathscr{D}$ indirectly, even if 
$X \not\in \mathscr{C}$. This is accomplished by transferring information from $X$ to some
$Y \in \mathscr{C}$, the so-called `pointer observable', and then observing $Y$ directly.
If $\mathscr{D}$ is in state $\rho$, time evolution $\alpha_t$ must be such that 
$\mathbb{P}_{\alpha_{t}^*(\rho), Y} = \mathbb{P}_{\rho, X}$: 
the probability distribution that $\mathscr{C}$ finds when 
observing $Y$ at time $t$ exactly equals the one any $\tilde{\mathscr{C}} \ni X$  
would find when observing $X$ at time $0$.  
 
Since we only need the automorphism $\alpha_t$ at the fixed time $t$ when measurement is completed,
we will drop the suffix $t$ from now on.

\subsection{Automorphic Measurement}

If $\alpha$ is such that $\alpha(Y) = X$, then 
$\mathbb{P}_{\alpha^*(\rho), Y} = \mathbb{P}_{\rho, X}$ for all $\rho \in \mathscr{S(D)}$.
It is immediately clear that the averages are the same, $\alpha^*(\rho)(Y) = \rho(\alpha(Y)) = \rho(X)$.
But since $\alpha$ is an automorphism, $\alpha(f(A)) = f(\alpha(A))$ for all $A \in \mathscr{D}$ and 
$f \in \mathscr{C}(\mathbf{Spec}(A))$.
Therefore the expectation values $\mathbb{E}$ used in the proof of proposition~(\ref{kloek})
to construct the probability measures are automatically identical: $\alpha^{*}\rho(f(Y)) = \rho(f(X))$.
Then the probability distributions themselves must be identical.  

\subsubsection{Example}

Let $\mathscr{D} = M_2 \otimes M_2$. Think of $M_2 \otimes \mathbb{I}$ as an electron with spin.
$X = \sigma_z \otimes \mathbb{I}$ will be measured.
Think of $\mathbb{I} \otimes M_2$ as a small computer memory, capable of storing one bit of information.
$\mathscr{C}$ is the commutative algebra generated by $Y = \mathbb{I} \otimes \sigma_z$: the actual memory.
$\alpha$ is the automorphism defined by $\alpha(A \otimes B) = B \otimes A$.
Then $\alpha^*(\rho)(Y) = \rho(\alpha(Y)) = \rho(X)$: the information that was in the electron 
prior to measurement has now arrived inside $\mathscr{C}$. 

Of course this is not repeatable: the states of the electron 
and the computer memory are interchanged, so that a second measurement of $X$ will 
in general yield a different result.
 
\subsection{More General Measurement}

For automorphic measurement, we require that 
$\mathbb{P}_{\alpha^*(\sigma), Y} = \mathbb{P}_{\sigma, X}$ holds for all
$\sigma$ in $\mathscr{S(D)}$.
But this may only be necessary for a restricted class of states $\sigma$ in $\mathscr{S(D)}$.
An experiment often consists of two parts:
a system $\mathscr{A}$ to be examined in an unknown state $\rho$ and a measurement apparatus 
$\mathscr{B}$ in a known default-state\footnote{This default-state $\tau$ certainly need not be pure.}  $\tau$.
$\mathscr{A}$ contains the observable $X$ that is to be measured. 
$\mathscr{B}$ contains some `pointer-observable' $Y$.
Automorphic time evolution on $\mathscr{D = A \otimes B}$ may now take place
in such a way that \mbox{$\mathbb{P}_{\alpha^*(\rho \otimes \tau), \mathbb{I} \otimes Y} =
\mathbb{P}_{\rho \otimes \tau, X \otimes \mathbb{I}} =  \mathbb{P}_{\rho,
X}$.}
Then $Y$ is observed, directly or indirectly.
The set of states $\sigma$ in $\mathscr{S(D)}$ for which 
$\mathbb{P}_{\alpha^*(\sigma), \mathbb{I} \otimes Y} = \mathbb{P}_{\sigma, X \otimes \mathbb{I}}$
must hold is in this case 
$\{ \, \sigma = \rho \otimes \tau \, | \, \rho \in \mathscr{S(A)} \, \}$.  

We define
$\mathbf{M}^*$ : $\mathscr{S(A)} \to \mathscr{S(A \otimes B)}$ by 
$\mathbf{M}^*(\rho) \isperdef \alpha^{*}(\rho \otimes \tau)$.
Because it is affine, $\mathbf{M}^*$ can be extended to a linear mapping $\mathscr{A}^* \to (\mathscr{A\otimes B})^*$
on all continuous linear functionals on $\mathscr{A}$.
It is therefore the dual of a linear map $\mathbf{M}$ : $\mathscr{A \otimes B} \to \mathscr{A}$.
(Hence the notation $\mathbf{M}^*$.)
Because $\mathbf{M}^*$ respects normalization, $\mathbf{M}$ is unital: $\mathbf{M}(\mathbb{I}) = \mathbb{I}$.
And because $\mathbf{M}^*$ maps states to states, $\mathbf{M}$ is positive: $B \geq 0 \Rightarrow \mathbf{M}(B) \geq 0$.
$\mathbf{M}$ is even completely positive. (See chapter~\ref{heis}.)

In summary, an affine map $\mathbf{M}^*$ : $\mathscr{S(A) \to S(A \otimes
B)}$ of the form $\mathbf{M}^* (\rho)  \isperdef \alpha^{*}(\rho \otimes \tau)$ 
is by definition a \emph{perfect} measurement iff
$\mathbb{P}_{\mathbf{M}^*(\rho), Y} = \mathbb{P}_{\rho, X}$ for all $\rho \in
\mathscr{S(A)}$. 

\subsubsection{Example}
This example is basically due to Hepp (see \cite{hep}):
let $\mathscr{D} = M_2 \otimes M_2$. $M_2 \otimes \mathbb{I}$ is as an electron with spin
$X = \sigma_z \otimes \mathbb{I}$. 
It is in an unknown state
$\rho \in \mathscr{S}(M_2)$. 
$\mathbb{I} \otimes M_2$ is again a computer memory, capable of storing one bit of information.
It is in the default-state $\tau = \psi_+$.
The actual memory $\mathscr{C}$ is generated by $Y = \mathbb{I} \otimes \sigma_z$.

In the notation of page~\pageref{spiderman}, $\alpha$ is induced by the unitary operator 
$\mathbf{P}_+ \otimes \mathbb{I} + \mathbf{P}_- \otimes \sigma_x$.
It is the controlled not-gate.
One easily checks that $\alpha(Y) $ equals $ \sigma_z \otimes \sigma_z$ and not $X$.
However, for all $\rho$ in $\mathscr{S}(M_2)$ we have 
	$$\mathbf{M}^* (\rho)(Y) =
	\alpha^*(\rho \otimes \tau)(Y) = 
	\rho \otimes \psi_+ (\sigma_z \otimes \sigma_z) = 
	\rho \otimes \psi_+ (\sigma_z \otimes \mathbb{I}) =
	\rho(X).$$ 
$\mathscr{C}$ can now observe $Y$ directly, finding the same probability distribution that direct observation of 
$X$ would have delivered. 

\subsubsection{Unbiased Measurement}\label{klantenklop}

We will also consider  \emph{unbiased} measurement: the average is transferred from $X$
to $Y$, but not necessarily the entire probability distribution. Automorphic measurement is automatically perfect.

Now $\mathbf{M}^* (\rho)(Y) = \rho(X) \quad \forall \rho \in \mathscr{S(A)} \Leftrightarrow \mathbf{M}(Y) = X$.
This characterizes unbiased measurement.
In contrast to automorphic measurement, this does not imply
$\mathbf{M}\big(f(Y)\big) = f(X)$ for all continuous $f$.
This means that in general $\mathbb{P}_{\mathbf{M}^* (\rho) , Y} \neq \mathbb{P}_{\rho , X}$, 
although the averages do coincide.
For perfect measurement, equality holds: $\mathbb{P}_{\mathbf{M}^* (\rho) , Y} = \mathbb{P}_{\rho , X}$ 
or equivalently $\mathbf{M}(f(Y)) = f(X) \quad \forall f \in \mathscr{C}(\mathbf{Spec}(X))$.
Yet even for perfect measurement, it may 
well be that $\alpha(Y) \neq X$. In the example above, 
$\mathbf{M}^*$ is an unbiased measurement of $X$: $\mathbf{M}(Y) = X$. It is even perfect: $\mathbf{M}(f(Y)) = f(X)$ for functions
on $\mathbf{Spec}(X)$.
But it is not automorphic: 
$\alpha(\mathbb{I} \otimes \sigma_z) = \sigma_z \otimes \sigma_z$, so $\alpha (Y) \neq X$.

In summary: all automorphic measurements are perfect. All perfect
measurements are unbiased. And both statements cannot be reversed.  

\newpage \section{State Reduction}

Before plunging into the generalities of state reduction, let us look at an example.

\subsection{An Example of Reduction}\label{polpokje}

Let $\mathscr{D} = M_2 \otimes M_2 \otimes M_2$. 
Again, $M_2 \otimes \mathbb{I}\otimes \mathbb{I}$ is an electron in unknown state $\rho \in \mathscr{S}(M_2)$.
It has spin $X = \sigma_z \otimes \mathbb{I} \otimes \mathbb{I}$ to be measured.
$\mathbb{I} \otimes M_2 \otimes M_2$ is a computer memory in default state $\tau = \psi_+ \otimes \psi_+$.
It is capable of storing two bits of information.
$\mathscr{C}$ is the commutative algebra generated by $Y_1 = \mathbb{I} \otimes \sigma_z \otimes \mathbb{I}$
and $Y_2 = \mathbb{I} \otimes \mathbb{I} \otimes \sigma_z$.

First, in exactly the same way as above, $X$ is measured and the information is stored on $Y_1$:
$\alpha_{t_1}$ is induced by the unitary operator
$\mathbf{P}_+ \otimes \mathbb{I} \otimes \mathbb{I} + \mathbf{P}_- \otimes \sigma_x \otimes \mathbb{I}$. 
Then another, similar measurement of $X$ is performed using $Y_2$ as pointer:
$\alpha_{t_2 - t_1}$ is induced\footnote
{In realistic models, $\alpha_{t_1}$ and $\alpha_{t_2 - t_1}$ belong to the same dynamical semi-group,
so that $\alpha_{t_1} \circ \alpha_{t_2 - t_1} = \alpha_{t_2 - t_1} \circ \alpha_{t_1}$.
This is satisfied in this example.} 
by 
\mbox{$\mathbf{P}_+ \otimes \mathbb{I} \otimes \mathbb{I} + \mathbf{P}_-
\otimes \mathbb{I} \otimes \sigma_x$.}

Finally, $\mathscr{C}$ directly observes $Y_1$ and $Y_2$ in state $ \mathbf{M}^* (\rho) = 
\alpha_{t_2}^* (\rho \otimes \tau)$. The results will be distributed according to the probability 
distribution $\mathbb{P}_{\mathbf{M}^* (\rho), Y_1, Y_2}$:
$$
\begin{array}{lrclr}
\mathbf{M}^* (\rho) (\mathbb{I} \otimes \mathbf{P}_+ \otimes \mathbf{P}_+)  = & \quad \rho(\mathbf{P}_+) \!\!\!&,
& 0 \quad & =  \mathbf{M}^* (\rho) (\mathbb{I} \otimes \mathbf{P}_+ \otimes \mathbf{P}_-)\\
\mathbf{M}^* (\rho) (\mathbb{I} \otimes \mathbf{P}_- \otimes \mathbf{P}_+)  = & \quad 0 &,
&\!\!\! \rho(\mathbf{P}_-) \quad& = \mathbf{M}^* (\rho) (\mathbb{I} \otimes \mathbf{P}_- \otimes \mathbf{P}_-) \\
\end{array}
$$
In other words, $\mathscr{C}$ observes: \\*
$Y_1 = +1$ and $Y_2 = +1$ with probability $\mathbb{P}_{\rho, X}(\{ +1 \})$\\*
$Y_1 = -1$ and $Y_2 = -1$ with probability $\mathbb{P}_{\rho, X}(\{ -1 \})$\\*
$Y_1 = -1$ and $Y_2 = +1$ with probability 0\\*
$Y_1 = +1$ and $Y_2 = -1$ with probability 0.

$\mathscr{C}$ may interpret this correlation causally: 
the first measurement outcome influences the second.
Correlation can also be seen with the help of the reduced state: 
\begin{eqnarray*}
\lefteqn{(\mathbf{M}^* (\rho))_{\mathbb{I} \otimes \mathbf{P}_+ \otimes \mathbb{I}} (A)=} \\
&=& \frac
{	
\rho \otimes \tau \Big(\alpha_{t_2}
\big( (\mathbb{I} \otimes \mathbf{P}_+ \otimes \mathbb{I}) A (\mathbb{I} \otimes \mathbf{P}_+ \otimes \mathbb{I}) 
\big) \Big) }
{ \mathbf{M}^* (\rho)(\mathbb{I} \otimes \mathbf{P}_+ \otimes \mathbb{I})} \\
&=&
\frac{	
\rho \otimes \tau \Big(	\big( (\mathbf{P}_+ \otimes \mathbf{P}_+ + \mathbf{P}_- \otimes \mathbf{P}_-)\otimes \mathbb{I} \big)
			\alpha_{t_2}(A)
			\big((\mathbf{P}_+ \otimes \mathbf{P}_+ + \mathbf{P}_- \otimes \mathbf{P}_-)\otimes \mathbb{I} \big) \Big)
}
{	\rho \otimes \tau ( \mathbf{P}_+ \otimes \mathbf{P}_+ \otimes \mathbb{I} + \mathbf{P}_- \otimes \mathbf{P}_- \otimes \mathbb{I})
}\\
&=&
\frac{	
\rho \otimes \tau\Big(	\big((\mathbf{P}_+ \otimes \mathbb{I} )\otimes \mathbb{I} \big)
			\alpha_{t_2}(A)
			\big((\mathbf{P}_+ \otimes \mathbb{I} )\otimes \mathbb{I}\big)\Big)
}
{	\rho(\mathbf{P}_+)
}\\
&=&
\alpha_{t_2}^* (\rho_{\mathbf{P}_+} \otimes \tau)(A)\\
& = & \mathbf{M}^*(\rho_{\mathbf{P}_+})
\end{eqnarray*}
In the third step, we have made special use of $\tau = \psi_+ \otimes \psi_+$.
According to the discussion following definition~\ref{condit}, the above equation has the following
significance:
\begin{quote}
	\emph{Observations made by $\mathscr{C}$, conditioned on the first measurement outcome 
	\mbox{$Y_1 = +1$}, will be as if the electron had originally been in the reduced state $\rho_{\mathbf{P}_+}$.}
\end{quote}
Since $\rho_{\mathbf{P}_+} = \psi_+$ for all 
$\rho$ in $\mathscr{S}(M_2)$,
and because $\mathbf{M}^* (\psi_+) = \psi_+ \otimes \psi_+$, this explains that if, according to $\mathscr{C}$, 
the first measurement yields $+1$, so does the second.    

\subsubsection{Nota Bene}
One would be tempted to pose the following question:
\begin{quote}
\emph{Suppose that $\mathscr{C}$ observes $Y_1 = +1$ at time $t_1$.
	Does $\mathscr{C}$ then necessarily observe \mbox{$Y_1 = +1$} at time $t_2$?}
\end{quote}
This question is metaphysical in nature because it cannot be answered by experiment.
At time $t_2$, how do you know what you observed at time $t_1$? You must consult some 
memory.

Any experiment one could possibly devise involves a memory (possibly
external; a piece of paper or a hard-disk) storing information on $Y_1$.
Above, this memory is simply $Y_1$ itself. And just as above, the result of observing this memory at time $t_2$ 
will never yield discrepancies within the memory, independent of the observation made at time $t_1$.
 
\subsection{Reduction as a Consequence of Measurement}

Many examples of the kind above have been described (\cite{hep}, \cite[p.~292]{boh}, \cite[p.~678]{bj}). 
But in fact, reduction is not just
\emph{possible}, as has been known for a long time (see \cite{neu}). It is a \emph{necessary} consequence
of transferring information from $X$ to some pointer $Y$.

\subsubsection{The Origin of Reduction and Collapse}

Let $\mathbf{M} : \mathscr{A \otimes B} \to \mathscr{A}$ be such that 
$\mathbf{M}^* (\rho) = \alpha(\rho \otimes \tau)$ for some automorphism $\alpha$ 
and $\tau \in \mathscr{S(B)}$.
Suppose $\mathbf{M}$ is a perfect measurement of 
$X \in \mathscr{A}$ with\/\footnote
{This includes automorphic measurement if $\mathscr{B}$ happens to be 
$\mathbb{I}$ and $\mathbf{M}$ automorphic.} pointer $Y \in \mathscr{A \otimes B}$.
Suppose that $\mathscr{A}$ and $\mathscr{B}$ are von Neumann algebras, as in the previous example. 
Then $X$ and $Y$ have
projection valued measures $V \mapsto \mathbf{P}(V)$ and $W \mapsto \mathbf{Q}(W)$ respectively.
If $\rho$ is normal, then 
$\mathbb{P}_{\mathbf{M}^* (\rho) , Y} = \mathbb{P}_{\rho , X}$ 
implies $\mathbf{M}^* (\rho)(\mathbf{Q}(V)) = \rho(\mathbf{P}(V)) $ for all subsets $V$ 
of $\mathbf{Spec}(X)$:
the spectral projections of $X$ are measured, using the corresponding ones of $Y$ as pointers.
In this situation we can apply: 

\begin{prop}[Reduction] \label{redrum}
Let $\mathscr{A}$ and $\mathscr{B}$ be C\/$^*\!$-algebras, $\mathbf{P} \in \mathscr{A}$ and 
$\mathbf{Q} \in \mathscr{A \otimes B}$ projections.
For $\rho \in \mathscr{S(A)}$, let 
$\mathbf{M}^* (\rho) \isperdef \alpha^* (\rho \otimes \tau)$ for some
automorphism $\alpha$ of $\mathscr{A} \otimes \mathscr{B}$ and $\tau \in \mathscr{S(B)}$.
Suppose $\mathbf{M}^* (\rho)(\mathbf{Q}) = \rho(\mathbf{P})$ for all 
$\rho \in \mathscr{S(A)}$. Then for any $\rho \in \mathscr{S(A)}$:
$$
	(\mathbf{M}^* (\rho))_{\mathbf{Q}} = \mathbf{M}^* (\rho_{\mathbf{P}}).
$$
\end{prop}
\textbf{Proof}:
\begin{quote}
By moving to the GNS-representation, we may assume $\rho \otimes \tau$ to be a vector state $|\psi \rangle$.
Now by assumption,
$$
\alpha^*(\rho_{\mathbf{P}} \otimes \tau)(\mathbf{Q}) = 
\mathbf{M}^* (\rho_{\mathbf{P}})(\mathbf{Q}) = \rho_{\mathbf{P}}(\mathbf{P}) = 1.
$$
Since $(\rho_{\mathbf{P}} \otimes \tau) = (\rho \otimes \tau)_{\mathbf{P} \otimes \mathbb{I}}$ 
corresponds to the vector state 
$\frac{\mathbf{P} \otimes \mathbb{I} |\psi \rangle} {\| \mathbf{P} \otimes \mathbb{I} |\psi \rangle \|}$, 
\\*we have
\bq \label{sn243}
\alpha^*(\rho_{\mathbf{P}} \otimes \tau)(\mathbf{Q}) = 
\frac
{\langle \mathbf{P} \otimes \mathbb{I} \psi| \alpha(\mathbf{Q}) |\mathbf{P} \otimes \mathbb{I} \psi \rangle} 
{\| \mathbf{P} \otimes \mathbb{I} |\psi \rangle \|^2}=1. 
\eq
$\alpha(\mathbf{Q})$ is a projection. Therefore equation~(\ref{sn243}) implies
$$
\| \alpha(\mathbf{Q}) \mathbf{P}\otimes \mathbb{I} |\psi \rangle\|^2 =
\| \mathbf{P} \otimes \mathbb{I} |\psi \rangle \|^2
$$ 
which entails, again because $\alpha(\mathbf{Q})$ is a projection, that 
\bq \label{knoestwak}
 \alpha(\mathbf{Q}) \mathbf{P}\otimes \mathbb{I} |\psi \rangle =
 \mathbf{P} \otimes \mathbb{I} |\psi \rangle. 
\eq 
In a similar fashion, 
$\mathbf{M}^* (\rho_{\mathbb{I} - \mathbf{P}})(\mathbf{Q}) = 
\rho_{\mathbb{I} -\mathbf{P}}(\mathbf{P}) = 0$ leads to
\bq \label{worst}
\alpha(\mathbf{Q}) \big( ( \mathbb{I} - \mathbf{P})\otimes \mathbb{I} \big)
|\psi \rangle = 0.
\eq    
Equations \ref{knoestwak} and \ref{worst} imply
$$
\alpha(\mathbf{Q}) |\psi \rangle = 
\mathbf{P}\otimes \mathbb{I} |\psi \rangle.
$$  
Thus for all $D \in \mathscr{A \otimes B}$:
\begin{eqnarray*}
(\mathbf{M}^* (\rho))_{\mathbf{Q}}(D) &=& 
\frac
{\rho \otimes \tau \big( \alpha ( \mathbf{Q} D \mathbf{Q} ) \big) }
{\mathbf{M}^* (\rho)(\mathbf{Q})} \\
&=&
\frac
{\langle \alpha (\mathbf{Q}) \psi | \alpha (D) |\alpha (\mathbf{Q}) \psi \rangle}
{\rho(\mathbf{P}) }\\
&=&
\frac
{\langle(\mathbf{P} \otimes \mathbb{I}) \psi | \alpha(D) |(\mathbf{P} \otimes \mathbb{I}) \psi \rangle}
{\rho \otimes \tau (\mathbf{P} \otimes \mathbb{I}) } \\
&=&
(\rho \otimes \tau)_{\mathbf{P} \otimes \mathbb{I}}\big(\alpha(D)\big)=
\alpha^*(\rho_{\mathbf{P}} \otimes \tau)(D) \\
&=& 
\mathbf{M}^* (\rho_{\mathbf{P}})(D).
\end{eqnarray*}
\begin{flushright}
\emph{q.e.d.}
\end{flushright}
\end{quote}
\subsubsection{State Reduction}

This has two major consequences. The first is subjective:
\begin{quote}
\emph{Suppose that a perfect measurement of $X \in \mathscr{D}$ is performed with pointer 
	$Y \in \mathscr{C} \subset \mathscr{D}$. If $\mathscr{D}$ was in state $\rho \in \mathscr{S(A)}$
	before measurement,
	then all observations made by $\mathscr{C}$ after measurement, conditioned on the observation that
	the measurement outcome $Y$ is in the set $V$, will be as if the system had originally been in 
	the reduced state $\rho_{\mathbf{P}(V)}$ instead of $\rho$.}
\end{quote}
We now have an interpretation of the reduced state $\rho_{\mathbf{P}(V)}$ outside $Y'$.
Perhaps the following commutative diagram says more than a thousand words:\\[-0.3 cm]
\setlength{\unitlength}{1,2 cm}
\begin{picture}(5,3) \label{diagramke}
\put(0,1){\emph{reduction}}
\put(1.5,1.95){$\rho$}
\put(1.6,1.6){\vector(0,-1){1.2}}
\put(2.3,2){\vector(1,0){1.2}}
\put(1.3,0.1){$\rho_{\mathbf{P}(V)}$}
\put(4.3,1.6){\vector(0,-1){0.6}}
\put(2.3,0.1){\vector(1,0){1.2}}
\put(4,1.9){$\mathbf{M}^*(\rho)$}
\put(5.4,2){\vector(1,0){1.2}}
\put(3.7,0){\shortstack{
$(\mathbf{M}^*(\rho))_{\mathbf{Q}(V)}$\\ = \\$\mathbf{M}^*(\rho_{\mathbf{P}(V)})$}}
\put(5.4,0.1){\vector(1,0){1.2}}
\put(7.3,1.9){$\mathbb{P}_{\mathbf{M}^*(\rho) , Y}$}
\put(7.8,1.6){\vector(0,-1){0.6}}
\put(6.9,0){\shortstack{ $ \mathbb{P}_{\mathbf{M}^*(\rho) , Y}(\, \bullet \,| \, [Y \, \mathrm{in} \, V] \,) $ \\ =
\\ $ \mathbb{P}_{\mathbf{M}^*(\rho_{\mathbf{P}(V)}) , Y}$ }}
\put(8.4,1.3){\emph{conditioning}}
\end{picture}\\*[0.5 cm]
The map $\rho \to \rho_{\mathbf{P}(V)}$ is called state reduction.
The fact that we observe state reduction (left hand side of the diagram) results from 
harmless conditioning on the physically relevant probability distributions (right hand side). 

\subsubsection{State Collapse} \label{smoelgnoe}

The second consequence is objective. It is summarized in the diagram below:

\begin{center}
\setlength{\unitlength}{1,2 cm}
\begin{picture}(5,2) \label{diagramketje}
\put(0.2,0.8){\emph{collapse}}
\put(1.5,1.55){$\rho$}
\put(1.6,1.3){\vector(0,-1){0.8}}
\put(2.1,1.6){\vector(1,0){1.4}}
\put(1.3,0.1){$\mathbf{C}^* (\rho)$}
\put(2.1,0.2){\vector(1,0){1.4}}
\put(4,1.5){$\mathbf{M}^*(\rho)$}
\put(3.7,0.1){$\mathbf{M}^* \circ \mathbf{C}^* (\rho)$}
\put(3.9 , 0.8 ) {= on $Y'$}
\end{picture}
\end{center} 

\begin{cor}[Collapse]
Let $\mathscr{A}$ and $\mathscr{B}$ be von Neumann algebras, $X \in \mathscr{A}$, $Y \in \mathscr{A \otimes B}$
Hermitean. Let $\{\, V_i \,|\, i \in I \,\}$ be any countable decomposition of $\mathbf{Spec}(X)$.
Let $\mathbf{M}^*$ : $\mathscr{S(A)} \to \mathscr{S(A \otimes B)}$ be defined by 
$\mathbf{M}^* (\rho) \isperdef \alpha^* (\rho \otimes \tau)$ for some
automorphism $\alpha$ of $\mathscr{A \otimes B}$ and normal state $\tau \in \mathscr{S(B)}$.
Suppose $\mathbf{M}^*$ is a perfect measurement of $X$ with pointer $Y$, 
i.e. $\mathbb{P}_{\mathbf{M}^* (\rho), Y} = \mathbb{P}_{\rho, X}$ for all $\rho \in \mathscr{S(A)}$.
Then for all normal $\rho \in \mathscr{S(A)}$, and 
for all $D \in Y'$:
$$
\mathbf{M}^* (\rho)(D) = (\mathbf{M}^* \circ \mathbf{C}^*) (\rho) (D),
$$
where $\mathbf{C}^*$ is the collapse operation for $X$ and $\{\, V_i \,|\, i \in I \,\}$, as in definition~\ref{koliek}.  
\end{cor}
\textbf{Proof}:
\begin{quote}
Let $V \mapsto \mathbf{P}(V)$ and $V \mapsto \mathbf{Q}(V)$ be the spectral measures of 
$X$ and $Y$ respectively.
Since  
$\bigcup_{I} V_i = \mathbf{Spec}(X)$, we have \mbox{$\sum_{I} \mathbf{Q}(V_i) = \mathbb{I}$}.
Suppose $[D , Y] = 0$. Then also $[D , \mathbf{Q}(V_i)] = 0$ for all $i \in I$.
Therefore
\begin{eqnarray*}
\mathbf{M}^* (\rho) (D) &=&
\mathbf{M}^* (\rho) \Big( \big(\sum_{I} \mathbf{Q}(V_i) \big) D \Big) \\
&=& 
\mathbf{M}^* (\rho) \big( \sum_{I} \mathbf{Q}(V_i) D \mathbf{Q}(V_i) \big) \\
&=&
\sum_{I} \mathbb{P}_{\mathbf{M}^* (\rho) , Y}(V_i) \cdot (\mathbf{M}^* (\rho))_{\mathbf{Q}(V_i)} (D) \\
&=&
\sum_{I} \mathbb{P}_{ \rho , X}(V_i) \cdot \mathbf{M}^* ( \rho_{\mathbf{P}(V_i)})(D)\\
&=&
\sum_{I} \rho \big( \mathbf{P}(V_i) \mathbf{M}(D) \mathbf{P}(V_i) \big)\\
&=&
(\mathbf{M}^* \circ \mathbf{C}^*)(\rho)(D).
\end{eqnarray*}
\begin{flushright}
\emph{q.e.d.}
\end{flushright}
\end{quote}
Normally, a system $\mathscr{A}$ will be examined by an observer $\mathscr{C \subset B}$ outside 
$\mathscr{A}$. This means that $Y$ is of the form $\mathbb{I} \otimes \tilde{Y}$. 
Then all $A \in \mathscr{A \otimes B}$ of the form $\tilde{A} \otimes \mathbb{I}$ 
will commute with the pointer.
So \emph{regarded as a state on the examined system $\mathscr{A}\otimes \mathbb{I}$}, we have 
$$
\mathbf{M}^* (\rho) = \mathbf{M}^* \circ \mathbf{C}^* (\rho).
$$
In other words:
\begin{quote}
\emph{Suppose that a perfect measurement of $X \in \mathscr{A}$ is performed 
	by an observer $\mathscr{C \subset B}$ outside $\mathscr{A}$, using a pointer
	$Y \in \mathbb{I} \otimes \mathscr{B}$. 
	Then all measurements of any $A \in \mathscr{A}$ made by any second observer 
	$\tilde{\mathscr{C}}$  will be as if the 
	system had originally been in the collapsed state $\mathbf{C}^*
	(\rho)$
	instead of $\rho$.}
\end{quote}

For example, suppose that $\mathscr{A} = M_2$ is in vector state $\alpha |\psi_+\rangle +\beta |\psi_-\rangle$, 
and $\sigma_z$ is measured perfectly by an outside observer. 
Then all subsequent measurement of $\mathscr{A}$ will be as if $\mathscr{A}$ 
had originally been in the mixed state 
$|\alpha|^2 \cdot |\psi_+ \rangle \langle \psi_+| + |\beta|^2 \cdot |\psi_-\rangle \langle \psi_-|$.
     
\subsubsection{Summary}     
     
Reduction is subjective. It involves only one observer. 
Reduction occurs after both direct and indirect observation.

Collapse is objective. It involves at least two observers.    
Note the crucial role of $[A , Y] = 0$: if the first observer had been inside $\mathscr{A}$ 
instead of outside, none of the above would hold. Collapse only occurs with indirect observation. 
Both are not just possible, but necessary consequences of measurement. 
\subsection{Imperfect Reduction after Imperfect Measurement} 
 
Suppose an unbiased measurement is not perfect, but still rather good. Then one does not expect a perfect
reduction, but still a rather good one. Proposition~(\ref{redrum}) allows
such a generalized version. In contrast to generalized collapse (which
comes along quite naturally), generalized reduction is rather thorny and
uncomfortable. But in the end, if we work hard enough, we do obtain a hard
estimate of the reduction, even for biased measurement:
 
\begin{prop}[Generalized Reduction] \label{redrumdelta}
Let $\mathscr{A}$ and $\mathscr{B}$ be C\/$^*\!$-algebras, $\mathbf{P} \in \mathscr{A}$ and 
$\mathbf{Q} \in \mathscr{A \otimes B}$ projections.
For $\rho \in \mathscr{S(A)}$, let 
$\mathbf{M}^* (\rho) \isperdef \alpha^* (\rho \otimes \tau)$ for some
automorphism $\alpha$ of $\mathscr{A} \otimes \mathscr{B}$ and $\tau \in \mathscr{S(B)}$.
Suppose there is a $\Delta \geq 0$ such that $| \mathbf{M}^* (\rho)(\mathbf{Q}) - \rho(\mathbf{P}) | \leq \Delta$ for all 
$\rho \in \mathscr{S(A)}$. Then for any $\rho \in \mathscr{S(A)}$:
$$
\| (\mathbf{M}^* (\rho))_{\mathbf{Q}} - \mathbf{M}^* (\rho_{\mathbf{P}}) \| \leq
\frac
 	{\sqrt{\Delta}}
 	{\mathbf{M}^* (\rho) (\mathbf{Q})}
\left( 1 + 2\sqrt{\Delta} + \sqrt{1 + (1 + 2 \sqrt{\Delta})^2 } \right)
$$
\end{prop}
\textbf{Proof}:
\begin{quote} 
For notational convenience, define 
$
\mathbf{P}_0 = \mathbf{P}$, $\mathbf{P}_1 = \mathbb{I} - \mathbf{P} 
$.
Define $\epsilon_{i}^2 = |\rho_{\mathbf{P}_{i}} (\mathbf{P}) - \mathbf{M}^* ( \rho_{\mathbf{P}_{i}} )(\mathbf{Q})| = 
|\delta_{i,0} - \mathbf{M}^* ( \rho_{\mathbf{P}_{i}} ) (\mathbf{Q})|$. 
Here $\epsilon_{i}^2$ is the probability 
that a measurement of $\mathbf{P}$ in state $\rho_{\mathbf{P}_{i}}$ yields the wrong outcome.
The $\epsilon_{i}$ depend on $\rho$, but are always less than $\sqrt{\Delta}$.  
By constructing the GNS representation of $\rho \otimes \tau$, we may assume $\rho \otimes \tau$ to be a 
vector state $|\psi \rangle$.
Then the $\epsilon_{i}$ have geometrical significance: they regulate the length of the vector
\begin{eqnarray}
|\chi_{i}\rangle & \isperdef& 
\alpha(\mathbf{Q}) \mathbf{P}_i \otimes \mathbb{I} |\psi \rangle - 
\mathbf{P} \mathbf{P}_i \otimes \mathbb{I} |\psi \rangle  \nonumber\\
&=&
(\alpha(\mathbf{Q}) - \delta_{i,0} ) \mathbf{P}_i \otimes \mathbb{I} |\psi \rangle \label{finifini}
\end{eqnarray} 
by
\begin{eqnarray}
\|\chi_{i}\|^2 & \isperdef&
\| \alpha(\mathbf{Q}) \mathbf{P}_i \otimes \mathbb{I} |\psi \rangle - 
\delta_{i,0} \mathbf{P}_i \otimes \mathbb{I} |\psi \rangle \|^2 \nonumber\\
&=&
\langle  \mathbf{P}_i \otimes \mathbb{I} \psi| 
(\alpha(\mathbf{Q}) - \delta_{i,0})^{\dagger} 
(\alpha(\mathbf{Q}) - \delta_{i,0})
 | \mathbf{P}_i \otimes \mathbb{I}\psi \rangle \nonumber\\
 &=&
\langle  \mathbf{P}_i \otimes \mathbb{I} \psi| 
(1 - 2 \delta_{i,0})\alpha(\mathbf{Q})  + \delta_{i,0}
 | \mathbf{P}_i \otimes \mathbb{I} \psi \rangle \nonumber\\
 &=&
(1 - 2 \delta_{i,0})\rho \otimes \tau 
((\mathbf{P}_i \otimes \mathbb{I})  \alpha(\mathbf{Q}) (\mathbf{P}_i \otimes \mathbb{I})) +
\delta_{i,0} \rho \otimes \tau (\mathbf{P}_i \otimes \mathbb{I}) \nonumber\\
&=&
\rho (\mathbf{P}_i)( (1 - 2 \delta_{i,0}) \alpha^* (\rho \otimes \tau_{\mathbf{P}_{i} \otimes \mathbb{I}})(\mathbf{Q}) +
\delta_{i,0} ) \nonumber\\
&=&
\rho (\mathbf{P}_i) (| \alpha^* 
(\rho \otimes \tau_{\mathbf{P}_{i} \otimes \mathbb{I}})(\mathbf{Q}) - \delta_{i,0}|)\nonumber\\
&=&
\rho(\mathbf{P}_i) (| \mathbf{M}^* (\rho_{\mathbf{P}_{i}})(\mathbf{Q}) - \delta_{i,0} |)\nonumber\\
&=& \
\epsilon_{i}^2 \rho(\mathbf{P}_i) \label{ep}.
\end{eqnarray}
With this we will estimate
\begin{eqnarray}
\lefteqn{\mathbf{M}^* (\rho) (\mathbf{Q} X \mathbf{Q}) - 
	\mathbf{M}^* (\rho_{\mathbf{P}})(X) \cdot \mathbf{M}^* (\rho) (\mathbf{Q}) =} \label{enul}\\
&=&
\alpha^* (\rho \otimes \tau) (\mathbf{Q} X \mathbf{Q}) - \alpha^* (\rho_{\mathbf{P}} \otimes \tau) (X) 
\cdot \alpha^* (\rho \otimes \tau) (\mathbf{Q}) \nonumber\\
&=&
\alpha^* ( \rho \otimes \tau ) 
(\mathbf{Q} (X - \alpha^* (\rho_{\mathbf{P}}\otimes \tau) (X) )  \mathbf{Q}) \nonumber\\
&=&
\rho \otimes \tau ( \alpha( \mathbf{Q} ) \alpha(X - \alpha^* (\rho_{\mathbf{P}}\otimes \tau) (X) )  
\alpha(\mathbf{Q}) ) \nonumber\\
&=&
\sum_{k,l} \rho \otimes \tau( \mathbf{P}_k \otimes \mathbb{I} \alpha( \mathbf{Q} ) 
\alpha(X - \alpha^* (\rho_{\mathbf{P}}\otimes \tau) (X) )  
\alpha(\mathbf{Q}) \mathbf{P}_l \otimes \mathbb{I} ) \nonumber\\
&=&
\sum_{k,l} \langle  \alpha(\mathbf{Q}) \mathbf{P}_k \otimes \mathbb{I} \psi| 
\alpha(X - \alpha^* ( \rho\otimes\tau_{\mathbf{P}\otimes \mathbb{I}} ) (X))
| \alpha( \mathbf{Q} )\mathbf{P}_l\otimes \mathbb{I}\psi \rangle \nonumber\\
&=&
\sum_{k,l} \langle  \delta_{k,0} \mathbf{P}_k \otimes \mathbb{I} \psi + \chi_{k}| 
\alpha(X - \alpha^* ( \rho\otimes \tau_{\mathbf{P} \otimes \mathbb{I}} ) (X))
| \delta_{l,0} \mathbf{P}_l \otimes \mathbb{I} \psi + \chi_{l} \rangle. 
\nonumber
\end{eqnarray}
The last step goes by definition of $\chi_{k}$: from equation~(\ref{finifini}), we see that 
$\alpha(\mathbf{Q}) \mathbf{P}_k \otimes \mathbb{I} \psi 
= \delta_{k,0} \mathbf{P}_k \otimes \mathbb{I} \psi + \chi_{k}$.
We will examine the smallness of each term separately.
\begin{eqnarray}
\lefteqn{\langle  \mathbf{P} \otimes \mathbb{I} \psi | 
\alpha(X - \alpha^* ( \rho\otimes \tau_{\mathbf{P} \otimes \mathbb{I}} ) (X))
| \mathbf{P} \otimes \mathbb{I} \psi \rangle =}\nonumber\\
&=&
\rho \otimes \tau(\mathbf{P} \otimes \mathbb{I} \alpha(X) \mathbf{P} \otimes \mathbb{I}) - 
\rho \otimes \tau(\mathbf{P} \otimes \mathbb{I}) 
\alpha^* (\rho \otimes \tau_{\mathbf{P} \otimes \mathbb{I}})(X)\nonumber\\
&=&
0. \label{pardoes}
\end{eqnarray}
That's one down. We will estimate the cross-terms with the Cauchy-Schwarz inequality. 
For that, we need the length of both $\| \chi_k \|$ and the vector 
\begin{eqnarray}
\lefteqn{\| \alpha(X - \alpha^* ( \rho \otimes \tau_{\mathbf{P} \otimes \mathbb{I}} ) (X)) 
\mathbf{P} \otimes \mathbb{I} \psi \|^2 =} \nonumber\\
&=&
\langle  \mathbf{P} \otimes \mathbb{I} \psi | 
\alpha(X - \alpha^* ( \rho \otimes \tau_{\mathbf{P} \otimes \mathbb{I}} ) (X))^{\dagger} \nonumber\\
& &
\alpha(X - \alpha^* ( \rho \otimes \tau_{\mathbf{P} \otimes \mathbb{I}} ) (X))
| \mathbf{P} \otimes \mathbb{I} \psi \rangle \nonumber\\
&=&
\langle  \mathbf{P} \otimes \mathbb{I} \psi | 
\alpha(X)^{\dagger}\alpha(X) 
-
2\Re 
\big(
\alpha(X)^{\dagger} \alpha^* ( \rho \otimes \tau_{\mathbf{P} \otimes \mathbb{I}} ) (X) 
\big)
+\nonumber\\
& & +
| \alpha^* ( \rho \otimes \tau_{\mathbf{P} \otimes \mathbb{I}} ) (X) |^2
| \mathbf{P} \otimes \mathbb{I} \psi \rangle \nonumber\\
&=&
\langle  \mathbf{P} \otimes \mathbb{I} \psi | 
\alpha(X)^{\dagger}\alpha(X) 
| \mathbf{P} \otimes \mathbb{I} \psi \rangle
-
\frac
{
|\langle  \mathbf{P} \otimes \mathbb{I} \psi |
\alpha(X)
| \mathbf{P} \otimes \mathbb{I} \psi \rangle|^2
} 
{ \| | \mathbf{P} \otimes \mathbb{I} \psi \rangle \| }
\nonumber\\
&=&
\langle  \mathbf{P} \otimes \mathbb{I} \psi  | 
\alpha(X)^{\dagger} 
(\mathbb{I} - \frac 	{| \mathbf{P} \otimes \mathbb{I} \psi \rangle\langle  \mathbf{P} \otimes \mathbb{I} \psi |} 
			{\langle  \mathbf{P} \otimes \mathbb{I} \psi | \mathbf{P} \otimes \mathbb{I} \psi \rangle})
\alpha(X)
| \mathbf{P} \otimes \mathbb{I} \psi \rangle\nonumber\\
&\leq&
\| \mathbf{P} \otimes \mathbb{I} \psi \|^2 \| X \|^2.
\end{eqnarray}
Now that we have the length of both vectors, we see by Cauchy-Schwarz:
\bq \label{etwee}
\| \langle  \chi_{k}| 
\alpha(X - \alpha^* ( \rho \otimes \tau_{\mathbf{P} \otimes \mathbb{I}} ) (X)) 
\mathbf{P} \otimes \mathbb{I} \psi \rangle \|
\leq
\| X \| \cdot \| \mathbf{P} \otimes \mathbb{I} \psi \| \epsilon_{k} \|
\mathbf{P}_k \otimes \mathbb{I} \psi \|.  
\eq
And similarly 
\bq \label{edrie}
\| \langle  \mathbf{P} \otimes \mathbb{I} \psi| 
\alpha(X - \alpha^* ( \rho \otimes \tau_{\mathbf{P} \otimes \mathbb{I}} ) (X))
| \chi_{l}  \rangle \|
\leq
\| X \| \cdot \| \mathbf{P} \otimes \mathbb{I} \psi \| \epsilon_{l} \|
\mathbf{P}_l \otimes \mathbb{I} \psi \|.
\eq
Finally, from $\| \alpha(X) - \alpha^* ( \rho \otimes \tau_{\mathbf{P} \otimes \mathbb{I}} (X)\| \leq 2\|X\|$ we see that
\bq \label{evier}
\|\langle  \chi_{k}| 
\alpha(X - \alpha^* ( \rho \otimes \tau_{\mathbf{P} \otimes \mathbb{I}} ) (X))
|  \chi_{l} \rangle\|
\leq
2 \|X \| \epsilon_{k} \| \mathbf{P}_k \otimes \mathbb{I} \psi \| \epsilon_{l} \|
\mathbf{P}_l \otimes \mathbb{I} \psi \|. 
\eq
Putting inequalities (\ref{pardoes}), (\ref{etwee}), (\ref{edrie}) and (\ref{evier}) 
into (\ref{enul}), we finally obtain 
\begin{eqnarray*}
\lefteqn{\| \mathbf{M}^* (\rho) (\mathbf{Q} X \mathbf{Q}) 
- \mathbf{M}^* (\rho_{\mathbf{P}})(X) \cdot \mathbf{M}^* (\rho) (\mathbf{Q}) \| \leq}\\
&\leq&
	2 \| X \| \| \mathbf{P}_0 \otimes \mathbb{I} \psi \| 
		( \epsilon_0 \| \mathbf{P}_0 \otimes \mathbb{I} \psi \| + 
		\epsilon_1 \| \mathbf{P}_1 \otimes \mathbb{I} \psi \| ) +\\
&&	
2 \| X \| 	( \epsilon_0 \| \mathbf{P}_0 \otimes \mathbb{I} \psi \| + 
		\epsilon_1 \| \mathbf{P}_1 \otimes \mathbb{I} \psi \| )^2\\
&=&
2 \| X \| ( \epsilon_0 \| \mathbf{P}_0 \otimes \mathbb{I} \psi \| + 
	\epsilon_1 \| \mathbf{P}_1 \otimes \mathbb{I} \psi \| ) \times \\
	& &
( \| \mathbf{P}_0 \otimes \mathbb{I} \psi \| + 
\epsilon_0 \| \mathbf{P}_0 \otimes \mathbb{I} \psi \| + \epsilon_1 \| \mathbf{P}_1 \otimes \mathbb{I} \psi \|).
\end{eqnarray*}
To estimate this last expression, note that $\mathbf{P}_0$ and $\mathbf{P}_1$ are complementary projections, and 
$|\psi \rangle$ is of norm one. Therefore, there exists an angle $\theta$ such that 
$\cos \theta = \| \mathbf{P}_0 \otimes \mathbb{I} \psi \|$ and $\sin \theta = \| \mathbf{P}_1 \otimes \mathbb{I} \psi \|$.
Since both $\epsilon_0 , \epsilon_1 \leq \sqrt{\Delta}$, we have 
\bq \label{oeloel}
 \left\| \frac{\mathbf{M}^* (\rho) (\mathbf{Q} X \mathbf{Q})}{\mathbf{M}^* (\rho) (\mathbf{Q})} - 
\mathbf{M}^* (\rho_{\mathbf{P}})(X) \right\| 
 \leq 
 \frac
 	{\| X \| \sqrt{\Delta}}
 	{\mathbf{M}^* (\rho) (\mathbf{Q})}
  f(\theta)	
\eq
with $f(\theta) = 2 (\cos\theta + \sin\theta) (\cos\theta + \sqrt{\Delta}(\cos{\theta} + \sin{\theta}) )$.
With standard a\-na\-ly\-sis and goniometry, one can verify that $f$ takes maximal value 
$1 + 2\sqrt{\Delta} + \sqrt{1 + (1 + 2 \sqrt{\Delta})^2 }$, proving the proposition. 
\begin{flushright}
$\emph{q.e.d.}$
\end{flushright}
\end{quote}

Proposition~(\ref{redrum}) is contained in the above as the special case $\Delta = 0$.
Note that the bound disappears if the probability of observing 
measurement outcome $+1$ becomes less than $1/2 \sqrt{\Delta}$. 
This means that excellent measurement ($\Delta \sim 0$) without reduction (upon finding $+1$) is not
excluded, provided that the probability of outcome $+1$ remains small. Of course there is conservation of misery:
the probability of outcome $0$ is large, and upon finding $0$ there is very good reduction.

Nonetheless, this principle can be used nicely in so-called `knowingly reversible measurement' (see \cite{ari}).
This is a non-perfect measurement, leaving the state fixed with a certain probability. 
The observer obtains not only a measurement outcome, but also the information whether or not 
the state is conserved successfully.

\newpage \section{State Collapse} 

On page~\pageref{smoelgnoe}, we have obtained collapse from reduction in order to 
show the link between the two. 
But there is an easier way of proving the necessity of collapse, more suitable 
for generalization.

\subsection{Perfect Collapse after Perfect Measurement}

The setting is slightly different: an observer $\mathscr{C} \subset \mathscr{B}$ attempts 
to decide whether a system $\mathscr{A}$ is in state $\psi_1$ or in $\psi_2$.
In order to do that, a measurement $\mathbf{M}^* $ : $\mathscr{S(A)} \to \mathscr{S(A \otimes B)}$
is performed (of the form $\mathbf{M}^* (\rho) = \alpha^* (\rho \otimes \tau)$) in such a way 
that observation of a certain pointer-observable $Y \in \mathscr{C}$ yields with certainty $y_1$ in state 
$\mathbf{M}^* (\psi_1)$ and $y_2$ in state $\mathbf{M}^* (\psi_2)$.
 
\begin{lemma} \label{snarwak}
Let $|\phi_i\rangle $ $(i = 1,2)$ be vector states on some algebra $\mathscr{D}$.
Let $Y \in \mathscr{D}$ be Hermitean such that 
$$
\mathbf{Var}_{\phi_i}(Y) = 0 
\quad \mathrm{and} \quad
\langle\phi_i |Y|\phi_i\rangle = y_i \quad (i = 1,2)$$ 
with $y_1 \neq y_2$.
Then for all $A \in \mathscr{D}$ such that $[A,Y] = 0$:
$$
	\langle  \phi_1| A |\phi_2 \rangle  = 0.
$$
\end{lemma}
\textbf{Proof}:
\begin{quote}
$|\phi_1 \rangle $ and $|\phi_2 \rangle $ must be eigenvectors of $Y$ with eigenvalues 
$y_1$ \mbox{and $y_2$}. Since $[A,Y] = 0$, $A$ respects the eigenspaces of $Y$. 
We therefore have $|\phi_1 \rangle \perp |A \phi_2 \rangle$:
$$
(y_1 - y_2)\langle \phi_1 | A |\phi_2 \rangle = \langle y_1 \phi_1 | A |\phi_2 \rangle
- \langle \phi_1 | A |y_2 \phi_2 \rangle = \langle \phi_1 | [Y,A] |\phi_2
\rangle = 0.
$$
\begin{flushright}
$\emph{q.e.d.}$
\end{flushright}
\end{quote}
This standard result can be used in the following way:
\begin{prop}[Collapse] \label{snarklop}
Let $\mathbf{M}^* $ : $\mathscr{S(A) \to S(A \otimes B)}$ be of the form 
$\mathbf{M}^* (\rho) = \alpha^{*}(\rho \otimes \tau)$ for some 
automorphism $\alpha$ of $\mathscr{A \otimes B}$ and $\tau \in \mathscr{S(B)}$.
Let $\psi_i$, $(i = 1,2)$ be vector states on $\mathscr{A}$, let $|\alpha|^2 + |\beta|^2 = 1$
and let $Y \in \mathscr{A \otimes B}$ be Hermitean such that 
$$
\mathbf{Var}_{\mathbf{M}^*(\psi_i)}(Y) = 0 
\quad \mathrm{and} \quad
\mathbf{M}^*(\psi_i) (Y) = y_i \quad (i = 1,2)
$$ 
with $y_1 \neq y_2$.
Then for all $A \in \mathscr{A \otimes B}$ such that $[A,Y] = 0$:
$$
	\mathbf{M}^* ( |\alpha \psi_1 + \beta \psi_2\rangle 
	 \langle \alpha \psi_1 + \beta \psi_2| )(A) =
	\mathbf{M}^*(|\alpha|^2|\psi_1\rangle \langle \psi_1| + |\beta|^2
	|\psi_2\rangle \langle \psi_2|)(A).	
$$
\end{prop}
\textbf{Proof:}
\begin{quote}
By the GNS-representation, we assume $\tau$ to be a vector state $|\tau\rangle$.
Thus
\begin{eqnarray}
\lefteqn{| \mathbf{M}^*( |\alpha \psi_1 + \beta \psi_2\rangle 
	 \langle \alpha \psi_1 + \beta \psi_2| )(A) -
\mathbf{M}^*(|\alpha|^2|\psi_1\rangle \langle \psi_1| + |\beta|^2 |\psi_2\rangle \langle \psi_2|)(A)|} \nonumber\\
&=&
| \langle (\alpha \psi_1 + \beta \psi_2) \otimes \tau| \alpha(A)| (\alpha \psi_1 + \beta \psi_2) \otimes \tau\rangle -\nonumber\\
& &
	\left(|\alpha|^2\langle \psi_1 \otimes \tau| \alpha(A)| \psi_1 \otimes \tau \rangle  +
	|\beta|^2 \langle \psi_2 \otimes \tau| \alpha(A)| \psi_2 \otimes \tau\rangle  \right) | \nonumber\\
&\leq&
	2|\alpha||\beta||\langle \psi_1 \otimes \tau| \alpha(A) |\psi_2 \otimes \tau \rangle | \nonumber\\
&\leq& 
	|\langle \psi_1 \otimes \tau| \alpha(A) |\psi_2 \otimes \tau \rangle |. \label{snarhond}
\end{eqnarray}
The last step uses that $2|\alpha||\beta| \leq 1$ since $|\alpha|^2 + |\beta|^2 = 1$.

Finally, we come to lemma~(\ref{snarwak}), here with the vectors 
$|\psi_i \otimes \tau \rangle $, \mbox{$(i = 1,2)$} and with the observable $\alpha(Y)$:
\bq
\psi_i \otimes \tau (\alpha(Y)) = \mathbf{M}^* (\psi_i) (Y) = y_i
\eq
and 
\begin{eqnarray}
\mathbf{Var}_{\psi_i \otimes \tau}( \alpha(Y) ) &=& 
\psi_i \otimes \tau(\alpha(Y)^2) - \psi_i \otimes \tau(\alpha(Y))^2 \nonumber\\
&=&
\psi_i \otimes \tau(\alpha(Y^2)) - \psi_i \otimes \tau(\alpha(Y))^2 \nonumber\\
&=&
\mathbf{Var}_{\mathbf{M}^* (\psi_i)}(Y) \nonumber\\
&=& 0. 
\end{eqnarray}
Thus $\langle \psi_1 \otimes \tau | \alpha(A) | \psi_2 \otimes \tau \rangle  = 0$.
\begin{flushright}
$\emph{q.e.d.}$
\end{flushright}
\end{quote}
If $Y$ is of the form $\mathbb{I} \otimes \tilde{Y}$, then
$[A \otimes \mathbb{I} , \mathbb{I} \otimes \tilde{Y}] = 0$ for any $A \in \mathscr{A}$.
This means that after the measurement is performed on $\mathscr{A}$ in vector state 
$|\alpha \psi_1 + \beta \psi_2\rangle $, further measurement of any $A \in \mathscr{A}$ 
by other observers will be as if the state had originally been 
$|\alpha|^2|\psi_1\rangle \langle \psi_1| + |\beta|^2 |\psi_2\rangle \langle \psi_2|$.

Of course the same holds for other observables commuting with 
$Y$, such as observables in $\mathscr{C}$ for example, or\footnote
{
If the measuring device happens to be classical ($\mathscr{B}$ is commutative), then a complete and 
rigorous collapse has been achieved. This was proposed by Jauch (see \cite[p.~174]{jau}). 
Although this is an extremely useful remark (see e.g. \cite{hep}), I do not hold this to be a 
\emph{fundamental} solution to the problem of measurement for the following reasons:
\begin{itemize}
\item[-]Measurement apparatuses consist of particles. Particles have momentum and position. 
These do not commute, so Abelian $\mathscr{B}$ can only be an idealization.
\item[-]Automorphic time evolution always conserves purity of states (see \cite[lemma 2]{hep}).
	Even on Abelian algebras.  
\end{itemize}
In chapter~\ref{macro}, we will examine this idealization more closely.
} 
in $\mathscr{A} \otimes \mathscr{C}$.
All of this has an immediate generalization for the case of a less perfect measurement, 
and for observables not quite commuting with $Y$.

\subsection{Imperfect Collapse after Imperfect Measurement}

\begin{lemma} \label{vectorredux}

Let $\phi_i$, $(i = 1,2)$ be vector states on some algebra $\mathscr{D}$.
Let $Y$ be a Hermitean element of $\mathscr{A}$ such that $\phi_1 (Y) \neq \phi_2 (Y) $.
Let 
$$
\phi_{i}(Y) = y_i \quad \mathrm{and} \quad \mathbf{Var}_{\phi_i} (Y) = \sigma_{i}^2 \quad (i = 1,2)
$$
be the expectation and variance of $Y$ in the state $\phi_i$, $(i = 1,2)$. 
If $A$ is a Hermitean observable such that $\|[A,Y]\| \leq \delta \| A \| $, then 
\begin{displaymath}
|\langle \phi_1|A|\phi_2\rangle | \leq \frac{\delta + \sigma_1 + \sigma_2}{|y_1 - y_2|} \|A\|.
\end{displaymath} 
\end{lemma}
\setcounter{gnol}{\value{prop}}
\textbf{Proof}:
\begin{quote}
For $i = 1,2$,
$$
\sigma_{i}^2 = \langle \phi_i| Y^{\dagger}Y |\phi_i\rangle  - \langle \phi_i|
Y^{\dagger} |\phi_i\rangle \langle \phi_i| Y |\phi_i\rangle. 
$$
So
$$ 
\sigma_{i} ^2 = \langle Y\phi_{i}| \mathbb{I} - \mathbf{P}_{ // } | Y \phi_{i}\rangle     
$$     
where $\mathbf{P}_{ // }$ denotes the projection onto the one-dimensional vector-space spanned by $|\phi\rangle $.
So $\mathbb{I} - \mathbf{P}_{ // }$ is the projection orthogonal to $|\phi_i\rangle $. We denote it by $\mathbf{P}_{\perp}$.
From $\mathbf{P}_{\perp}^2 = \mathbf{P}_{\perp}^{\dagger} = \mathbf{P}_{\perp}$ we see that
$$
\sigma_{i} = \|\mathbf{P}_{\perp} Y \phi_{i} \|.
$$ 
Decomposing $Y|\phi_{i}\rangle$ into components parallel and  perpendicular to $|\phi_{i}\rangle$ we find
$$
Y|\phi_{i}\rangle  = \mathbf{P}_{//} Y |\phi_{i}\rangle  + \mathbf{P}_{\perp} Y
|\phi_{i}\rangle. 
$$   
Denoting $ \mathbf{P}_{\perp} Y|\phi_{i}\rangle $ by $|\chi_i\rangle $, bearing in mind $\| \chi_i \| = \sigma_i$:
$$
Y|\phi_{i}\rangle  = \langle \phi_i|Y^{\dagger}|\phi_i\rangle  |\phi_{i}\rangle  + |\chi_i\rangle 
= y_i |\phi_i\rangle  + |\chi_i\rangle. 
$$
We use this in the following:
\begin{eqnarray*}
(y_{2} - y_{1} ) \langle \phi_1| A |\phi_2\rangle  &=&
\langle \phi_1| A |y_{2} \phi_2\rangle  - \langle y_{1} \phi_1| A |\phi_2\rangle \\
&=&
\langle \phi_1| AY |\phi_2\rangle  - \langle \phi_1| A |\chi_2\rangle  - \langle \phi_1| YA |\phi_2\rangle  
+ \langle \chi_1| A |\phi_2\rangle  \\
&=&
\langle \phi_1| [A,Y] |\phi_2\rangle  + \langle \chi_1| A |\phi_2\rangle  - \langle \phi_1| A |\chi_2\rangle. 
\end{eqnarray*}
Estimating with the Cauchy-Schwarz-inequality and the operator norm in each term we obtain
$$
|(y_{2} - y_{1} )| \cdot |\langle \phi_1| A |\phi_2\rangle | \leq ( \delta + \sigma_1 + \sigma_2 ) \| A \|.
$$
\begin{flushright}
\emph{q.e.d.}
\end{flushright}
\end{quote}
 We use lemma~(\ref{vectorredux}) in the same way as lemma~(\ref{snarwak}):
\begin{prop}[Generalized Collapse]\label{snarklop2}
Let $\mathbf{M}^*$ : $\mathscr{S(A) \to S(A \otimes B)}$ be of the form 
$\mathbf{M}^* (\rho) = \alpha^{*}(\rho \otimes \tau)$ for some 
automorphism $\alpha$ of $\mathscr{A \otimes B}$ and $\tau \in \mathscr{S(B)}$.
Let $\psi_i$, $(i = 1,2)$ be vector states on $\mathscr{A}$, let $|\alpha|^2 + |\beta|^2 = 1$
and let $Y \in \mathscr{A \otimes B}$ be Hermitean such that 
$$
\mathbf{Var}_{\mathbf{M}^*(\psi_i)}(Y) = \sigma_i 
\quad \mathrm{and} \quad
\mathbf{M}^*(\psi_i) (Y) = y_i \quad (i = 1,2)
$$ 
with $y_1 \neq y_2$.
Then for all $A \in \mathscr{A \otimes B}$ such that $[A,Y] \leq \delta \|A\|$:
$$
	|\mathbf{M}^*(|\alpha \psi_1 + \beta \psi_2\rangle \langle \alpha \psi_1 + \beta \psi_2|)(A) -
	\mathbf{M}^*(|\alpha|^2|\psi_1\rangle \langle \psi_1| + |\beta|^2 |\psi_2\rangle \langle \psi_2|)(A) | 	
$$
$$
	\leq \quad \frac{\delta + \sigma_1 + \sigma_2}{|y_1 - y_2|} \|A\|.
$$
\end{prop}
\textbf{Proof:}
\begin{quote}
As the proof of proposition~(\ref{snarklop}), but we now estimate equation~(\ref{snarhond}) 
with lemma~(\ref{vectorredux}) instead of lemma~({\ref{snarwak}}).
\begin{flushright}
\emph{q.e.d.}
\end{flushright}
\end{quote}
The ratio $\frac{\sigma_1 + \sigma_2}{|y_1 - y_2|}$ is an indicator of the quality of measurement:
suppose you know that, prior to measurement, 
$\mathscr{A}$ is either in state $\psi_1$ or $\psi_2$. To find out which, you perform measurement.
Suppose $y_1 < y_2$, then you conclude that the state was $\psi_2$ if the pointer $Y$ takes value
$\geq \frac{y_1 + y_2}{2}$.
The probability of deciding $\psi_2$ while the state was really $\psi_1$ is less than 
$\frac{4 \sigma_{1}^2}{|y_1 - y_2|^2}$ by
Chebyshev's inequality:
$\mathbb{P}_{\mathbf{M}^* (\psi_1) , Y}(|Y - y_1| \geq \frac{|y_1 - y_2|}{2}) 
\leq \frac{4 \sigma_{1}^2}{|y_1 - y_2|^2}$. Of course the same goes for $1 \leftrightarrow 2$,
so that $\frac{\sigma_1}{|y_1 - y_2|} + \frac{\sigma_2}{|y_1 - y_2|}$ indicates the quality 
of measurement indeed.

\subsubsection{Conclusions} 

For the case of a pointer outside $\mathscr{A}$, I will summarize some consequences of measurement which
will come in handy in the next chapter:

\begin{itemize}
\item[-] Collapse takes place on the commutant $Y'$ of the pointer. This includes the original algebra 
$\mathscr{A} \otimes \mathbb{I}$.
\item[-] An imperfect collapse will also occur on observables $A$ commuting well with the pointer $Y$ in the sense that  	
	$\|[A,Y]\| \leq \delta \| A \| $ for some small $\delta$. 
\item[-] If the measurement is imperfect ( a $|\psi_i\rangle $-measurement yields outcome $a_i$ only with high probability ),
	then also an imperfect collapse will occur on the commutant of the pointer.
\end{itemize}

\chapter {Macroscopic Observables} \label{macro}
Suppose that an outside observer $\mathscr{C \subset B}$ performs measurement on a \mbox{system $\mathscr{A}$.}
Then we have seen that a collapse must always take place \emph{on the original system} 
$\mathscr{A} \otimes \mathbb{I}$.
This simple and rigorous law of nature is, I believe, the collapse of the wave 
function usually alluded to in elementary textbooks on quantum mechanics 
(e.g. \cite{bj}, \cite{dir}, \cite{boh}, \cite{neu} and even \cite[p.~184]{jau}).
  
But on the combined system $\mathscr{A} \otimes \mathscr{B}$ there always remain observables with respect 
to which no collapse occurs. Indeed, Hepp and later Bell (see \cite{hep} and \cite{bel}) 
have pointed out that all one has to do to track these down is to run time evolution backwards.

On $M_2$, the observable $\sigma_x$ may serve to distinguish $\psi_+ / \psi_-$ mixtures 
from superpositions (p.~\pageref{oegsnok}). But if an outside observer $\mathscr{C \subset B}$
performs a $\psi_+ / \psi_-$ measurement $\mathbf{M}^* :
\mathscr{S}(M_2) \to \mathscr{S}(M_2 \otimes B)$ of the form 
$\mathbf{M}^* (\rho) = \alpha^* ( \rho \otimes \tau)$, then
$$
\mathbf{M}^* (\rho) ( \alpha^{-1}(\sigma_x \otimes \mathbb{I})) 
= \rho \otimes \tau (\sigma_x \otimes \mathbb{I})
= \rho (\sigma_x)
$$     
so by performing an $\alpha^{-1}(\sigma_x \otimes \mathbb{I})$-measurement
on $M_2 \otimes \mathscr{B}$, a second observer $\tilde{\mathscr{C}}$, outside $M_2$ \emph{and} $\mathscr{B}$,
can indeed ascertain that a full collapse has not taken place\footnote{This in contrast to state 
\emph{reduction}, which involves only one observer. See also p.~\pageref{zuuk}}. 
In practice however, collapse is
observed after measurement, even by the second observer.  

In my view (interpretation~\ref{mijnes}), collapse on closed systems
simply does not occur. Ever. Which leaves me to answer:
\begin{ques} \label{q2}
Why are the remaining coherences so hard to observe in practice? 
\end{ques}

Suppose one were to take the point of view that a rigorous collapse on closed systems does occur 
after measurement. (Interpretation~\ref{inter}.)
Then one needs to answer the following question:
\begin{ques} \label{quatsh}
Exactly when does collapse replace unitary time evolution on closed systems, and why is it so hard, in practice,
to see the difference between collapse at one time rather than another?
\end{ques}

Which point of view to take is merely a matter of taste, not of importance.
An answer to question~(\ref{q2}) entails an answer to the last part of question~(\ref{quatsh}): 
If it is hard to see the difference between unitary time evolution and collapse, it is certainly hard to see 
when the former goes into the latter.  

To question~(\ref{q2}), I see three answers. Two of them testify to the weirdness of the 
observables on which no collapse occurs. Or rather, to the occurrence of collapse on 
classes of ordinary observables:  
 
\section{Collapse on the Measurement Apparatus}   
First of all,   
we have seen how collapse comes about on $ \mathscr{A} \otimes \mathbb{I} $:
one simply applies lemma~(\ref{vectorredux}) to 
$
\phi_1 = \mathbf{M}^* (\psi_1)$ and $\phi_2 = \mathbf{M}^* (\psi_2)$. 

\subsubsection{Collapse on the Original System}

The measurement is perfect if, starting with with $\psi_i$, $(i = 1,2)$, the pointer position 
after measurement is always $y_i = \mathbf{M}^* (\psi_{i})(Y)$: then the corresponding variances $\sigma_{i}^2$
of the pointer observable $Y = \mathbb{I} \otimes \tilde{Y}$ equal 0. 
This results in a perfect collapse on $\mathscr{A} \otimes \mathbb{I} \subset (\mathbb{I} \otimes \tilde{Y})'$: 
proposition~(\ref{snarklop}).

Suppose the measurement is flawed, i.e. the input $\psi_i$, $(i = 1,2)$ does not absolutely 
guarantee the pointer output $y_i$. 
Then it may still be possible to draw reliable conclusions 
from the pointer about the examined system, provided that $\sigma_i \ll |y_1 - y_2|$ for $i = 1,2$,
or briefly 
$
\frac{\sigma_1 + \sigma_2} { |y_1 - y_2| } \ll 1
$.
 
We no longer have any reason to expect the `clean' collapse discussed above,
but still an imperfect measurement must surely induce some imperfect collapse on 
$\mathscr{A} \otimes \mathbb{I} \subset Y'$. 
This is proposition~(\ref{snarklop}).

\subsubsection{Collapse on the Measurement Apparatus}

But a wider range of collapse can be obtained with the same ease:
assume for example that a measurement $\mathbf{M}^*$ distinguishes two eigenstates $\psi_{x_1}$ and $\psi_{x_2}$ of 
some $X \in \mathscr{A}$ in a \emph{repeatable} fashion.
This means that another measurement of $X \otimes \mathbb{I}$ (perhaps by another observer) in state 
$\mathbf{M}^* (\psi_{x_i})$, $(i = 1,2)$ will once again yield $x_i$ with certainty: 
$$
\mathbf{M}^* (\psi_{x_i}) (X \otimes \mathbb{I}) = x_i 
\quad \mathrm{and} \quad
\mathbf{Var}_{\mathbf{M}^* (\psi_{x_i})} (X \otimes \mathbb{I}) = 0 \quad (i = 1,2).
$$
Then we can apply proposition~(\ref{snarklop}) to 
the measured observable $X \otimes \mathbb{I}$ instead of the pointer $\mathbb{I} \otimes \tilde{Y}$.
Collapse then occurs not only on $(\mathbb{I} \otimes \tilde{Y})'$, but also on $(X \otimes \mathbb{I})'$, which includes
the algebra of the measurement apparatus, $\mathbb{I} \otimes \mathscr{B}$.

In exactly the same manner as above, an approximate state collapse on 
$(X \otimes \mathbb{I})' \supset \mathbb{I} \otimes \mathscr{B}$ 
follows from proposition~(\ref{snarklop2}), provided that 
$$
\mathbf{Var}_{\mathbf{M}^* (\psi_{x_1})} (X \otimes \mathbb{I}) +
\mathbf{Var}_{\mathbf{M}^* (\psi_{x_2})} (X \otimes \mathbb{I}) \ll
|x_1 - x_2|^2.
$$
So the thoroughness of collapse on $(X \otimes \mathbb{I})'$ is not regulated by the quality of measurement, 
but by how well $\mathbf{M}^*(\psi_i)$, $(i = 1,2)$ 
remain eigenstates of $X$.
Already, we have a first answer to question~(\ref{q2}):
\begin{quote}
\begin{ans}
In a repeatable measurement, the remaining coherence can neither 
be detected on the original system, nor on the measurement apparatus alone. 
\end{ans}
\end{quote}

\newpage \section{Collapse on Local and Global Observables}
We have seen that there \emph{are} observables on which no collapse occurs, but if the 
measurement is repeatable they lie neither in
the original system $\mathscr{A} \otimes \mathbb{I}$ nor in the measurement apparatus
 $\mathbb{I} \otimes \mathscr{B}$. Moreover, the vector space (Not the algebra!)
spanned by the commutant of the pointer $Y$ and that of the measured observable $X$,
denoted $X' + Y'$, allows no coherences to be detected.   
This already testifies to the weirdness of the observables we are looking for: 
I for one would be very interested
to learn about actual measurements performed (with the help of a second measurement apparatus) on
observables in $\mathscr{A} \otimes \mathscr{B}$, but outside $X' + Y'$.
If they do exist, they are certainly quite exotic.
 
But there are two more classes of ordinary observables on which collapse takes place:
the local ones and the macroscopic ones.
This, I believe to be the main point of Klaus Hepp's 1972 article\footnote
{
The author himself did not, for as far as I can tell, seek to make this particular point.
I take the liberty of interpreting his results in a different fashion,
utilizing Hepp's considerable mathematical achievements in a context slightly different from the one originally intended.
The following digression should not be seen as a summary of \cite{hep}, but as a highly personal interpretation. 
}
 `Quantum Theory of Measurement and Macroscopic Observables' \cite{hep}.
\subsection{K.~Hepp: Quasilocal Algebras}
Hepp investigates the possibility of modelling 
time evolution by a weak limit of automorphisms and, as the title suggests, 
pointers by so called `macroscopic observables' in a 
quasilocal algebra. 

Let me try to suppress my sense of guilt about not explaining these notions
properly by giving an example:
imagine a countably infinite chain of quantum spins $M_2$ indexed by $n \in \mathbb{N}$, 
their position on the real line.
Local observables are supposed to affect only a finite amount of spins. For example, 
the spin in the $z$-direction of atom number $i$,
$\sigma_{z}^{i} \isperdef 
\mathbb{I} \otimes \ldots \otimes \mathbb{I} \otimes \sigma_z \otimes
\mathbb{I} \otimes \mathbb{I} \otimes \ldots $~,
is a local observable.
So is $\frac{1}{N} \sum_{i = 1}^{N} \sigma_{z}^{i}$\label{konijn}, 
the average $z$-spin over the first $N$ atoms. 
Now a quasilocal observable is almost local in the sense that, outside a finite amount of sites, 
it is arbitrarily close to $\mathbb{I}$ in norm. 

Macroscopic observables however are \emph{not} supposed to lie in the quasilocal algebra.
We would like them to be something like `averages':
let $Y_n$ be a uniformly bounded sequence of local observables `converging to infinity' in the 
sense that $Y_n$ has to do with spins arbitrarily far away from the origin for $n$ sufficiently large.
Then it would be pleasant to call $\lim_{N \to \infty} \frac{1}{N} \sum_{i =
0}^{N} Y_i$ a macroscopic observable.
For example, take $Y_n = \sigma_{z}^{n}$. Then 
$S_z = \lim_{N \to \infty} \frac{1}{N} \sum_{i = 0}^{N} \sigma_{z}^{i}$ is the average spin in the 
$z$-direction.

Unfortunately, this limit does not exist. At least not in norm. 
But if we choose one particular state on the algebra, we may form its
GNS-representation. (See \cite[p.~278]{ka1}.)
Then we have at our disposal a weak topology, coarser than the norm topology, in which the limit may well exist.        

In short, macroscopic observables lie in the weak closure of some represented 
quasilocal algebra, but not in the algebra itself. 
The crux of Hepp's article is macroscopic difference:
\newpage
\begin{defin}[Macroscopic Difference]
Let $\omega_1$ and $\omega_2$ be states on a quasilocal algebra $\mathscr{D}$.
Then $\omega_1$ and $\omega_2$ are called macroscopically different if there exists
a uniformly bounded sequence of observables $Y_n$ converging to infinity, 
and real numbers $y_1 \neq y_2$ such that
$$
\lim_{N \to \infty} \frac{1}{N} \sum_{n = 1}^{n = N} \omega_{i}(Y_n) = y_{i} \quad (i = 1,2).  
$$
\end{defin}
and the upshot is formed by the following two lemmas:
\begin{lemma}[Lemma 6 of \cite{hep}] \label{hepp6}
Let $\omega_1$ and $\omega_2$ be macroscopically different states on a quasilocal 
algebra $\mathscr{D}$ having short range correlations.
Then $\omega_1$ and $\omega_2$ are disjoint.
\end{lemma}

\begin{lemma}[Lemma 3 of \cite{hep}] \label{hepp3}
Consider two disjoint states $\omega_{1}$ and $\omega_{2}$ on a quasilocal algebra $\mathscr{D}$, and two 
(not necessarily disjoint) sequences $\omega_{1,t}$ and 
$\omega_{2,t}$ such that \\* $\lim_{t \to \infty} \omega_{i,t} =
\omega_{i} \quad (i = 1,2)$.  
Let $(\pi_t , \mathscr{H}_{ \pi_t })$ be representations of $\mathscr{D}$ and 
$\psi_{1,t}, \psi_{2,t} \in \mathscr{H}_{ \pi_t }$ such that 
$\omega_{i,t}(A) = \langle \psi_{i,t} |  \pi_t (A)  | \psi_{i,t} \rangle  \quad (i = 1,2)$ for all 
$A \in \mathscr{D}$. Then for all quasilocal $D \in \mathscr{D}$:
$$
\lim_{t \to \infty} \langle \psi_{1,t} |  \pi_t (D)  | \psi_{2,t} \rangle  = 0.
$$
\end{lemma}
These lemmas may be used as follows:
we are attempting to measure, say, the observable $\sigma_z$ in the algebra $M_2$.
We do this by coupling $M_2$ to a quasilocal algebra $\mathscr{B}$ in a state $\tau$.
We now seek automorphisms $\alpha_t$ of $\mathscr{D} = M_2 \otimes \mathscr{B}$ such that 
$\lim_{t \to \infty} \alpha_{t}^{*} (\psi_{i} \otimes \tau ) = \omega_{i}
$ $(i = 1,2)$, where $\omega_{1}$ 
and $\omega_{2}$ are 
short-range correlated, macroscopically different states on the quasilocal algebra 
$\mathscr{D}$.
(Hepp gives several explicit examples of such constructions.)

We may now use lemmas (\ref{hepp6}) and (\ref{hepp3}) consecutively to see that for each 
fixed quasilocal $ A \in M_{2} \otimes \mathscr{B} $,
all the `cross-terms' go to zero: 
$$
\lim_{t \to \infty} \langle \psi_{1,t} |  \pi_t (D)  | \psi_{2,t} \rangle  = 0
$$
in the sense of lemma (\ref{hepp3}), with $\omega_{i,t} = \alpha_{t}^{*} ( \psi_{i} \otimes \tau )$ for $i = 1,2$.

The relevance of this all to question~(\ref{q2}) is clear: 
\begin{quote}
	\emph{Pointers used in real life are often macroscopic.} 
\end{quote}
Furthermore, macroscopic information is more easily detected than microscopic information. 
This is the very reason for using macroscopic pointers.
A ray of light shining onto a measurement apparatus is almost certain to
record the (macroscopic) position of the pointer, i.e. the average position of some $10^{23}$ atoms.
It may even accidentally record the position of one single atom. 
But it is very unlikely to record the detailed excitation of \emph{each} of these atoms 
from their respective equilibrium positions.  

In other words: Hepp points out the classes of quasilocal and macroscopic observables as ordinary ones.
In his spirit, an answer to question (\ref{q2}) could be:
\begin{quote}
\emph{`In the 
course of measurement with a macroscopic pointer, reduction occurs increasingly well on all quasilocal 
observables'} 
\end{quote}
However, it is not easy to pinpoint the exact physical relevance of Hepp's weak-operator limit 
procedure:
\begin{itemize}
\item[-]First of all, the notion of a 'Macroscopic' observable is only mathematically defined on a quasilocal algebra.
	In reality, the algebra describing an actual measurement apparatus is usually extremely 
	large, but not quasilocal. In what way, if at all, are Hepp's result approximately valid?
\item[-]Secondly, we only come 'close' to macroscopically disjoint states
	in the weak topology.  
	At each fixed time $t$, $\alpha_t$ is still automorphic. 
	So there remain quasilocal observables $A$ for which the cross-terms are large. 
	On the other hand, for each fixed quasilocal observable $A$, the cross-terms do become 
	small in the course of time.
	Putting it more precisely and less clearly:
	$$
	\forall \epsilon > 0 \quad \forall A \in \mathscr{A} \otimes \mathscr{B} 
	\quad
	\exists t \in \mathbb{R}:  
	\quad
	t' \geq t \Rightarrow |\langle \psi_{+,t} |  \pi_t (A)  |
	\psi_{-,t} \rangle | \leq \epsilon\;, 
	$$
	yet
	$$
	\exists \epsilon > 0 \quad \forall t \in \mathbb{R} 
	\quad
	 \exists A \in \mathscr{A} \otimes \mathscr{B} \quad \exists t' \geq t : 
	\quad
	|\langle \psi_{+,t} |  \pi_t (A)  | \psi_{-,t} \rangle | >
	\epsilon\;. 
	$$
\end{itemize}

\subsection{Local Algebras}

We'll give up quasilocal algebras all together, and with it the sharp distinction between 
local and macroscopic observables.
Then we will give estimates on the amount of state collapse on `local'
observables, yet to be defined, based on: 
\begin{itemize}
\item[-]Exactly how local the observable is. 
\item[-]Exactly how macroscopic the pointer is.
\item[-]Exactly how much macroscopic difference there is.
\end{itemize}
Instead of utilizing Hepp's machinery, we shall resort to lemma~(\ref{vectorredux}).
Let us describe our combined system $\mathscr{D}$ by a large but finite number $N$ 
of possibly different atoms, each described by an algebra $\mathscr{D}_i$: 
$\mathscr{D} = \bigotimes_{i=1}^{N} \mathscr{D}_i$. 
Such an algebra, plus its (non-unique) subdivision into atoms\footnote
{By `atom', I just mean some small part of the algebra. It may represent an
electron, atom or molecule, or any other structure small compared to
$\mathscr{D}$. 
The same algebra 
$\mathscr{D} = \bigotimes_{i=1}^{6 \times 10^{23}}(\mathscr{D}_{\mathrm{H}} \otimes
\mathscr{D}_{\mathrm{O}} \otimes \mathscr{D}_{\mathrm{H}})$, describing a mole of water, must be 
considered a different local algebra according to whether one chooses the
hydrogen and oxygen atoms $\mathscr{D}_{\mathrm{H}}$
and $\mathscr{D}_{\mathrm{O}}$ as local atoms, or the water molecules $\mathscr{D}_{\mathrm{H}} \otimes
\mathscr{D}_{\mathrm{O}} \otimes \mathscr{D}_{\mathrm{H}}$.  
} 
may be called a local algebra. 

Let $X^i \in \mathscr{D}_i$. We will denote by $X_i$ the observable 
$\mathbb{I}_1  \otimes \ldots \otimes \mathbb{I}_{i - 1} \otimes X^i \otimes \mathbb{I}_{i + 1} 
\otimes \ldots \otimes \mathbb{I}_N $ in $\mathscr{D}$.   
Although no sharp distinction can be made between macroscopic and microscopic observables, it is intuitively clear 
that for each set $\{\, X^i \,| \, i \in \{1, \ldots , N\}, X^i \in \mathscr{D}_i \,\}$: 
\begin{itemize}
\item[-]
$\frac{1}{N}\sum_{i=1}^{N} X_{i}$ is an average, or very global observable. 
It might well be observed by accident.
 \item[-]
$X_{37}$ is a very local observable, representing detailed information about one particular atom. (Number 37.) A measurement of $X_{37}$ would 
probably cost a lot of effort, and is unlikely to be performed by accident.
\item[-]
$X^1 \otimes X^2 \otimes \ldots \otimes X^{N-1} \otimes X^{N}$ represents an observable giving detailed information about \emph{all} 
atoms in the measurement apparatus. It is unlikely that such a measurement can
ever be performed at all, let alone accidentally. 
\end{itemize}

So we would like to show that, if the pointer is $Y = \frac{1}{N}\sum_{i=1}^{N}
X_{i}$, some very global observable, then observables with respect to
which no collapse takes place are certainly not very global, nor very local, and typically of the third variety mentioned above. 
We will start by quantifying these rather vague notions:
\begin{itemize}
\item[-]
\begin{defin}[$n$-local]
A Hermitean $X$ is called $n$-local iff there exist integers
\mbox{$
1 \leq i_1 < \ldots i_n \leq N  
$} 
such that 
$
X \in \mathscr{D}_{i_{1}} \otimes \ldots \otimes \mathscr{D}_{i_{n}} \subset \mathscr{D}.
$
\end{defin}
An observable $X \in \mathscr{D}$ is `$n$-local' if it only affects $n$ atoms.
$X_{37}$ is $1$-local. Of course, each $n$-local observable is also $\tilde{n}$-local if $\tilde{n} \geq n$. 
\item[-]\begin{defin}[$\kappa$-global]
A Hermitean $Y \in \mathscr{D}$ is called $\kappa$-global iff there exist
\mbox{$M \in \mathbb{N}$},
$1 \leq i_1 < \ldots < i_M \leq N$ and Hermitean $Y^{i_k} \in \mathscr{D}_{i_{k}}$ 
such that: 
	\begin{itemize}
	\item[$ \bullet $] $Y = \frac{1}{M} \sum_{k = 1}^{M} Y_{i_k}$.
	\item[$ \bullet $] $ \kappa \geq \frac{\|Y^{i_k}\|}{M}$ for all $k$.
	\end{itemize}
\end{defin}
For example, $\frac{1}{N}\sum_{1}^{N} \sigma_{z}^{i}$ as on page~\pageref{konijn} is $N^{-1}$-global.
$X_{37}$ is $\|X_{37}\|$-global, for example because it equals $X_{37} = \frac{1}{1}(X_{37})$, or, if you happen 
to be in a troublesome mood, because $X_{37} = \frac{1}{N}(0 + \ldots + 0 + N X_{37} + 0 \ldots + 0)$.
An observable will be called global if it is $\kappa$-global for some $\kappa \in \mathbb{R}$.
Not all observables are global.
\item[-]The amount of difference between two states $\phi_1$ and $\phi_2$ on
some global Hermitean $Y$
can easily be quantified by the ratio $\frac{\sigma_1 + \sigma_2}{|y_1 - y_2|}$, 
where $y_i = \phi_i (Y)$ and $\sigma_{i}^2 =$ \mbox{$ \phi_i
(Y^{2}) - \phi_i (Y)^2$}, $(i = 1,2)$.
\end{itemize}
These definitions allow us to apply lemma~(\ref{vectorredux}) in the situation of a measurement using
a global pointer to distinguish $\psi_i$ from $\psi_j$: 
$\mathbf{M}^*(\psi_i)$ and $\mathbf{M}^*(\psi_j)$ are globally
different. 

\begin{corzel}$\!\!\!$\footnote
{
	Consider the example of a cloud of $N$ particles, with positions $x^i$ and momenta $p^i$.
	These observables are not bounded, but since $[x^i,p^j]$ is, this is merely a technical problem 
	which may be averted by, for example, a cut-off.
	The position of the cloud, $X = \frac{1}{N} \sum_{i = 1}^{N} x_i$, is increasingly macroscopic
	for \mbox{increasing $N$}.
	However, the \emph{total} momentum of the cloud, $P = \sum_{i = 1}^{N} p_i$, is 
	$\|p^i\|$-global, irrespective \mbox{of $N$}. So since $[X,P] = i\hbar$, it would seem that
	reduction on $X$, using $P$ as pointer, is good because $\hbar$ is small, and not because 
	$N$ is large. This is not the case however: of importance is the ratio 
	${\hbar}/{|p_1 - p_2|}$, and typical momentum differences \emph{do} grow as $N$
	increases.   
}
Let $\phi_1, \phi_2 \in \mathscr{S}(\mathscr{D})$ be $\kappa$-globally different
vector states, i.e. there is a $\kappa$-global Hermitean $Y \in \mathscr{D}$ 
such that $\phi_{1}(Y) \neq \phi_{2}(Y)$.
Let $y_{1,2} = \phi_{1,2}(Y)$ be the expectation of $Y$ in $\phi_{1,2}$, and 
$\sigma_{1,2}^2 = \mathbf{Var}_{\phi_{1,2}}(Y)$ the variance.
Let $\alpha, \beta \in \mathbb{C}$ be such that $ |\alpha|^2 + |\beta|^2 = 1$. 
Then, for every $n$-local $A$:
$$
|\langle \alpha \phi_1 + \beta \phi_2| A |\alpha \phi_1 + \beta \phi_2 \rangle - 
(|\alpha|^2 \! \langle \phi_1 | A |\phi_1\rangle + |\beta|^2 \! \langle \phi_2| A |\phi_2\rangle)|
\leq \frac{2 n \kappa + \sigma_1 + \sigma_2}{|y_1 - y_2|}\|A\|.
$$
And for every global observable $Y' = \frac{1}{M'} \sum_{l = 1}^{M'} Y_{j_{l}}$
with $\|Y_{j_{l}}\| \leq y'$:
$$
|\langle \alpha \phi_1 + \beta \phi_2| Y' |\alpha \phi_1 + \beta \phi_2 \rangle - 
(|\alpha|^2 \! \langle \phi_1 | Y' |\phi_1\rangle + |\beta|^2 \! \langle \phi_2| Y' |\phi_2\rangle)|
\leq \frac{2 \kappa  + \sigma_1 + \sigma_2}{|y_1 - y_2|}y'.
$$
\end{corzel}
\textbf{Proof}:
\begin{quote}
Since $2 |\alpha| |\beta| \leq 1$, all we have to do is apply lemma~(\ref{vectorredux}) 
to $\langle \phi_1 | A |\phi_2\rangle$.
For the first inequality, we write $Y = \frac{1}{M} \sum_{k = 1}^{M} Y_{i_{k}} $.
Since at most $n$ of the $Y_{i_k}$ do not commute with $A$, we have
$\|[Y,A]\| = \| \frac{1}{M} \sum_{k = 1}^{M} [Y_{i_{k}} , A] \| \leq 
2 n \|A\| \frac{\max_{k} \|Y^{i_{k}}\|}{M} \leq 2 n \kappa \|A\| $.
For the second inequality, we use that $\|[Y_{i_{k}} , Y_{j_{l}}]\| \leq
2\kappa M y'$, and that it equals zero if $i_k \neq j_l$. We obtain
$\|[Y,Y']\| = \| \frac{1}{M'M} \sum_{k = 1, l=1}^{k=M, l=M'} [Y_{i_{k}} ,
Y_{j_{l}}] \| \leq 
 \frac{1}{M'M} \sum_{k = 1, l=1}^{k=M, l=M'} \delta(i_k, j_l) 2\kappa M y' \leq 2 \kappa y'$. Of course
 $\|Y'\| \leq y'$.
\begin{flushright}
\emph{q.e.d.}
\end{flushright}
\end{quote}
Bear in mind that $\kappa$, for a typical pointer, will have values in the order of
$10^{-23}$.
For a perfect measurement ($\sigma_{1,2} = 0$) this means one may probe detailed information 
about billions and billions of atoms simultaneously 
without running the slightest risk of encountering any lack of collapse.

Furthermore, in the case of non-perfect measurement using a $\kappa$-global pointer, 
as long as $n \kappa \ll \sigma_{1} + \sigma_{2}$, 
all experimentally observed coherence may be attributed to the poor quality of measurement.
All of this constitutes a second answer to question~(\ref{q2}):
\begin{quote}
\begin{ans}
When using a very global pointer, the coherence remaining after measurement can neither be 
detected on very global observables, nor on very local ones. 
\end{ans}
\end{quote}
\newpage \section {Global Information Leakage}  

So suppose we are in the circumstance of a repeatable measurement with the help of a
very global `pointer', say an actual pointer.
We have seen that observables on which no collapse occurs are present 
on the combined system, but they can neither be  
entirely inside the measurement-apparatus, nor entirely in the atom. They can also not be very global, nor very local.
All in all, they are pretty weird indeed. Yet in principle, they \emph{do} exist and they \emph{can} be measured 
by a second observer. 
\subsection{Information Leakage After Measurement}
One should be extremely careful with this kind of reasoning, however.
Imagine, for example, that $M_2$ represents a two-level atom, and $\mathscr{B}$ describes some large measuring apparatus,
measuring eigenstates of $\sigma_z \otimes \mathbb{I}$ with pointer 
$\mathbb{I} \otimes Y$ which represents, literally, the position of a pointer. 
(The average position of all atoms in the pointer is of course rather global.)
Then as soon as $\mathbb{I} \otimes Y$ is measured (with the help of an
ancillary system, e.g. light reflecting on
the pointer and reaching our eyes), collapse on the combined atom-apparatus system takes place. 
It is of course immaterial whether or not someone is actually \emph{looking} at the photons. 
If even the smallest speck of light were to fall on the pointer, the information
about the pointer position would 
already be encoded in the light, causing full collapse on the atom-apparatus system.

So as soon as the information about the pointer position has reached our eyes, we can be mathematically certain that collapse
on the atom-apparatus-system has taken place, even on those extraordinary observables that 
commute poorly with the pointer of the measurement apparatus.
However, if no such information has reached our eyes, we may still be practically sure that
all measurement on the combined atom-apparatus will reveal that collapse has taken place,
unless extreme measures (e.g. shielding, extreme cooling) have been taken to prevent 
pointer-information from leaking out of the system.
A similar line of reasoning may provide a third answer to question~(\ref{q2}), again due to 
lemma~(\ref{vectorredux}): 
\begin{quote}
\begin{ans}
If information leaks from the pointer into the outside world, collapse unavoidably takes place on the 
combination of system and measurement apparatus. In practice, global pointers constantly leak information.
\end{ans}
\end{quote}

\subsection{Information Leakage in General}
A Macroscopic system may be modelled by some local algebra $\mathscr{D} = \bigotimes_{i = 1}^{N} \mathscr{D}_i$.
If this system is interacting normally with the outside world, (the occasional photon happens to 
scatter on it, for instance)
then one may imagine a number of very global observables being measured continually, with a certain 
measurement uncertainty\footnote
{According to proposition (\ref{jm}), simultaneous perfect measurement is not possible,
but global observables normally commute well enough for the limits of accuracy imposed by proposition 
(\ref{jmdelta}) to remain well below the macroscopic scale.} $\sigma$. 
For a decent definition of this `measurement uncertainty' or `quality' $\sigma$, I will have to refer to chapter~\ref{heis},
definitions (\ref{measdef}) and (\ref{qual}). 
But it entails the $\sigma_{1}$ and $\sigma_{2}$ of lemma~(\ref{vectorredux}) being $\leq \sigma$ for eigenstates of $X$. 

\newpage
This enables us to apply lemma~(\ref{vectorredux}).
It tells us that all coherences between eigenstates $\psi_{x_1}$ and $\psi_{x_2}$ of global observables 
$X$ are continually vanishing on all of $\mathscr{D}$, (the pointer, e.g. a
beam of light, is outside the system), 
provided their eigenvalues $x_1$ and $x_2$ satisfy $| x_1 - x_2 | \gg 2\sigma$. 

Take for example a collection of $N$ spins, $\mathscr{D} = \bigotimes_{i = 1}^{N} M_2$.
Suppose that for $\alpha = x,y,z$, the observables $S_{\alpha} = \frac{1}{N}\sum_{i = 1}^{N} \sigma_{\alpha}^{i}$ 
are continually being measured
with an accuracy $N^{-1} \ll \sigma \ll 1$. For $N$ in the order of Avogadro's number, $N \sim 6 \times 10^{23}$,
this allows for extremely accurate measurement. Then between 
globally different eigenstates of $S_{\alpha}$, i.e. states for which the eigenvalues 
$|s_{\alpha} - \tilde{s}_{\alpha}| \gg \sigma$, coherences are constantly disappearing.
However, the measurement need not have any effect\footnote
{Of course it is \emph{possible} for a measurement to destroy coherence
between $\rho \otimes |+ \rangle $ and $\rho \otimes |- \rangle $.
Just take any old measurement, and add a `decoherence operation' by hand.
The net result is a measurement destroying coherence.} 
on states which only differ on a small scale.
Take for instance $\rho \otimes |+ \rangle $ 
and $\rho \otimes |- \rangle $, with $\rho$ some state on $N-1$ spins.
Indeed, $|s_{\alpha} - \tilde{s}_{\alpha}| \leq 2/N \ll \sigma$, so lemma~(\ref{vectorredux}) is 
vacuous in this case. 

We see how $\sigma$ produces a smooth boundary between the macroscopic 
and the microscopic world: global processes (involving $S_{\alpha}$-differences $\gg \sigma$) continually 
suffer from loss of coherence, while local processes (involving $S_{\alpha}$-differences $\ll \sigma$) 
are unaffected. 

\newpage \section{Conclusion}

In order for a second observer to notice the lack of collapse after a first observer
has measured $X \in \mathscr{A}$ with a global pointer, 
he or she must do the following:
\begin{itemize}
\item[-]Keep the original
system $\mathscr{A}$ and the system $\mathscr{B}$, containing the first observer, from interacting with the 
outside world, for instance by shielding or extreme cooling. 
\item[-]Then perform a measurement on an observable $\tilde{X} \in \mathscr{A \otimes B}$
which does not commute with the global pointer: it cannot be in the original
\mbox{system $\mathscr{A}$,} it cannot be very global and it cannot be very local.  If the first measurement is
repeatable, it cannot lie entirely in $\mathscr{B}$ either. 
\item[-]The outcome of the first measurement must remain unknown to the second
observer. Suppose that an observable in the second observer serves as
a pointer to the first measurement. 
The net situation would then be a measurement of $X \in \mathscr{A}$ 
by an observer outside of $\mathscr{A \otimes B}$.
A second measurement will then always show collapse on $\mathscr{A \otimes B}$.
\end{itemize}
Under these circumstances, it cannot be excluded that coherences are
experimentally detected by the second observer. 

\subsubsection{Paradox}

All of this leaves us with one glaring paradox: surely the first observer, remembering 
the outcome of $X$-measurement, can perform $\tilde{X}$-measurement and observe that there is no 
collapse, let alone reduction, contradicting proposition~(\ref{redrum})?

The answer is no. The first observer cannot simultaneously measure $\tilde{X}$ 
\emph{and} remember a value of $X$.
We will elaborate this in proposition~(\ref{crux}), where we will have some more tools at our disposal.

\chapter{Measurement Inequalities}\label{heis}

There are many different concepts of measurement.
Up until now, we have investigated mappings 
$\mathscr{S(A) \to S(A \otimes B)}$ of the form $\mathbf{M}^* (\rho) = \alpha^{*}(\rho \otimes \tau)$
for some automorphism $\alpha$ of $\mathscr{A \otimes B}$ and $\tau \in \mathscr{S(B)}$.
But other points of view are possible. For instance, von Neumann (see \cite{neu}) and Holevo (see \cite{hol}) define
measurement as an affine mapping from $\mathscr{S(A)}$ to $\mathscr{S(C(}\Omega))$, the space of probability 
measures on the Borel $\sigma$-algebra in $\Omega \subset \mathbb{R}$. 

In order to cover all concepts of measurement at the same time, we will investigate completely 
positive operations.   
With their help, we will define perfect measurement and we will define
unbiased measurement. 
We will then rigorously define the quality of unbiased measurement\footnote{   
The attentive reader may have noticed the grotesque ugliness of proposition~(\ref{redrumdelta}).
This results from the ad hoc use of $| \mathbf{M}^* (\rho)(\mathbf{Q}) - \rho(\mathbf{P})|$ 
as a measure of quality.}.

After this, the way will be cleared for quite general statements on the trade-off between measurement quality 
and the amount of disturbance\footnote{
Although propositions and proofs will be different from the ones encountered before, their 
interpretation will be similar if not the same.  
In order not to disturb the flow of reasoning, I've chosen to once again go over details exhaustively 
 mentioned before. My apologies to the reader.}. 
 
\section{Completely Positive Operations}

What do we expect from \emph{any} physical operation $\mathbf{T}^* : \mathscr{S(A) \to S(B)}$ from 
one (quantum) probability space into another?
We formulate three natural requirements (copied from \cite{maa}):
\begin{itemize}
	\item[-] 
	The \emph{stochastic equivalence principle} is common to all interpretations of postulates (\ref{post1})
	and (\ref{post2}). It states that a system that is in state $\rho_1$ with probability
	$\lambda_1$ and in state $\rho_2$ with probability $\lambda_2$ cannot be distinguished from a system in state 	
	$\lambda_1 \rho_1 + \lambda_2 \rho_2$.  Therefore, $\mathbf{T}^*$ must be \emph{affine}:
	for all $0 \leq \lambda_1, \lambda_2  \leq 1$ such that $\lambda_1 + \lambda_2 = 1$,
	$$
	\lambda_1 \mathbf{T}^*(\rho_1) + \lambda_2 \mathbf{T}^*(\rho_2) = 
	\mathbf{T}^*(\lambda_1 \rho_1 + \lambda_2 \rho_2)
	$$
	$\mathbf{T}^*$ can thus be extended to a linear mapping between the full duals of $\mathscr{A}$ 
	and $\mathscr{B}$, so that $\mathbf{T}^*$ is the dual of a linear map 
	$\mathbf{T} : \mathscr{B} \to \mathscr{A} $. This justifies our
	notation: $\mathbf{T}^* (\rho) = \rho \circ \mathbf{T}$. 
	\item[-]
	In order for $\mathbf{T}^*$ to map states to states, it must respect normalization and positivity:
	$
	\mathbf{T}^*(\rho)(\mathbb{I}) = 1 
	$
	and
	$
	\mathbf{T}^*(\rho)(B^{\dagger}B) \geq 0 \quad \forall \quad B \in \mathscr{B}, \quad \rho \in \mathscr{S(A)}
	$.
	Equivalently, \mbox{$\mathbf{T}(\mathbb{I}) = \mathbb{I}$} and $B \geq 0 \Rightarrow \mathbf{T}(B) \geq 0$.
	\item[-] 
	So $\mathbf{T}^*$ is linear, normalized and positive. But it was realized by K.~Krauss in the
	1970's that it must be possible to couple $\mathscr{A}$ and $\mathscr{B}$ to
	another system $\mathscr{C}$ and 
	perform the operation $\mathbf{T}^*$ on $\mathscr{A}$, leaving $\mathscr{C}$ untouched.
	This leads us to the last requirement (see \cite{kra}). 	

	An operation $\mathbf{T}^*$ is called \emph{$n$-positive} if 
	the map 
	$\mathbf{id}^{*}_n \!\otimes \mathbf{T}^* : \mathscr{S}(M_n \otimes \mathscr{B}) 
	\to \mathscr{S}(M_n \otimes \mathscr{A})$
	defined by
	$\tau \otimes \rho \mapsto \tau \otimes (\mathbf{T}^* (\rho))$ is 
	linear, normalized and positive. An operation is called \emph{completely positive}
	if it is $n$-positive for all $n \in \mathbb{N}$.
\end{itemize}
There exist positive operations which are not completely positive. Formulating
the above in the Heisenberg picture, we define a linear, continuous map
$\mathbf{id}_n\! \otimes \mathbf{T} : M_n \otimes\mathscr{B} \to M_n
\otimes\mathscr{A}$  by $\mathbf{id}_{n}\!\otimes \mathbf{T} (A \otimes B) = A \otimes \mathbf{T}(B)$
for all $A \in M_n$ and $B \in \mathscr{B}$. 
\begin{defin}[Complete Positivity]
Let $\mathscr{A}$ and $\mathscr{B}$ be unital C\/$^*\!$-algebras. A linear map $\mathscr{B} \to \mathscr{A}$
is called completely positive if\/\footnote
{We assume all completely positive operations to be automatically unital: $\mathbf{T}(\mathbb{I}) = \mathbb{I}$.
This is not always so in the literature.} 
$\mathbf{T}(\mathbb{I}) = \mathbb{I}$ and if for all 
$D \in M_n \otimes \mathscr{B}$, $n \in \mathbb{N}$:
$$
	D \geq 0 
	\quad \Longrightarrow \quad 
	\mathbf{id}_{n}\!\otimes \mathbf{T}(D) \geq 0. 
$$
\end{defin}

The class of completely positive operations was invented to encompass every
physical operation you could ever want.
For example, it contains all automorphisms, *-ho\-mo\-morphisms and states, as well as dilations to
automorphisms. A positive operation from or to an abelian algebra is automatically 
completely positive, and it hardly needs mentioning that both concepts of measurement mentioned 
above are completely positive too.

We will proceed to investigate completely positive operations. It is surprising how much can be 
said about so general an object. 

\newpage \section{A Cauchy-Schwarz Inequality}
If $\mathscr{A}$, $\mathscr{B}$ are C$^*\!$-algebras, one can  
define a sesquilinear map $\mathbf{F}_{\mathbf{T}} : \mathscr{B} \times
\mathscr{B} \to \mathscr{A}$. 
\begin{defin}
Let $\mathbf{T} : \mathscr{B} \rightarrow \mathscr{A}$ be a 4-positive unital operation.
Let $A, B \in \mathscr{B}$. Then
\begin{displaymath}
\mathbf{F}_{\mathbf{T}}(A,B) \isperdef \mathbf{T}(A^{\dagger}B) -\mathbf{T}(A^{\dagger})\mathbf{T}(B).
\end{displaymath}
\end{defin}
If no confusion is possible, we will often omit the subscript.
$\mathbf{F}$ is a sesquilinear positive semidefinite $\mathscr{A}$-valued form
on $\mathscr{B}$, i.e. 
\begin{itemize}

\item[-]$\mathbf{F}$ is linear in the second argument, anti-linear in the first. 
\item[-]$\mathbf{F}(A,B)^{\dagger} = \mathbf{F}(B,A)$ for all $A, B \in \mathscr{B}$.
\item[-]$\mathbf{F}(B,B) \geq 0$ as an operator inequality for all $B \in \mathscr{B}$.

\end{itemize}
The first and second point follow immediately from
$\mathbf{T}(A)^{\dagger} = \mathbf{T}(A^{\dagger})$. We will derive this and the third
point shortly, along with an $\mathscr{A}$-valued Cauchy-Schwarz-inequality. 
There is also a fourth point of interest, clear from the definition:
\begin{itemize}
\item[-]$\mathbf{F}(\mathbb{I},B)
= \mathbf{F}(B , \mathbb{I}) = 0 \quad \forall B \in \mathscr{B}$
\end{itemize}
The likeness of $\mathbf{F}$ to an inner product incites us to introduce a 
semi-norm on $\mathscr{B}$: 
\begin{defin}[$\mathbf{T}$-norm]
Let $\mathbf{T} : \mathscr{B} \rightarrow \mathscr{A}$ be a 4-positive unital operation.
Let $B \in \mathscr{B}$. Then
\begin{displaymath}
\| B \|_{\mathbf{T}} \isperdef \| \sqrt{\mathbf{F}_{\mathbf{T}}( B , B )} \|.
\end{displaymath}
\end{defin}
$\|B\|_{\mathbf{T}}$ is called the $\mathbf{T}$-norm of $B$. Since $\mathbf{T}$ is always
a contraction, we have $\|B\|_{\mathbf{T}} \leq \|B\|$.
Of course it is possible that $\| B \|_{\mathbf{T}} = 0 $ for $ B \neq 0$.
If, for example, $\mathbf{T}$ happens to be a C$^*\!$-homomorphism, then $\mathbf{F}_{\mathbf{T}}$ is
identically zero and $\| B \|_{\mathbf{T}} = 0$ for all $B \in \mathscr{B}$.
$\mathbf{F}_{\mathbf{T}}$ is in many ways a measure of how well
$\mathbf{T}$ respects multiplication. 

\subsubsection{Real and Imaginary Part}

Like any element of $\mathscr{A}$ we can
split $\mathbf{F}_{\mathbf{T}}(A,B)$ into a Hermitean and an anti-Hermitean part:

$$
\mathbf{F}_{\mathbf{T}}(A,B) = \Re\mathbf{F}_{\mathbf{T}}(A,B) +
\mathrm{i}\Im\mathbf{F}_{\mathbf{T}}(A,B)  
$$
with
\begin{eqnarray*}
2\Re\mathbf{F}_{\mathbf{T}}(A,B) \!\!\! & = & \!\!\! \mathbf{T}(A^{\dagger}B +
B^{\dagger}A) - \big(\mathbf{T}(A)^{\dagger} \mathbf{T}(B) +
\mathbf{T}(B)^{\dagger} \mathbf{T}(A) \big) \\
2i\Im\mathbf{F}_{\mathbf{T}}(A,B) \!\!\! & = & \!\!\! \mathbf{T}(A^{\dagger}B -
B^{\dagger}A) - \big(\mathbf{T}(A)^{\dagger} \mathbf{T}(B) -
\mathbf{T}(B)^{\dagger} \mathbf{T}(A) \big). 
\end{eqnarray*}
The commutator of $A$ and $B$ is defined by $ [A,B] \isperdef AB - BA $.
The anti-commutator by $\{ A , B \}_{+} \isperdef AB + BA$.
In case of Hermitean $A$ and $B$ the above boils down to:
\begin{eqnarray*}
2\Re\mathbf{F}_{\mathbf{T}}(A,B) \!\!\! & = & \!\!\! 
 \mathbf{T}(\{ A, B \}_{+}) -  \{ \mathbf{T}(A) , \mathbf{T}(B) \}_{+}\\
2i\Im\mathbf{F}_{\mathbf{T}}(A,B) \!\!\! & = & \!\!\!  \mathbf{T}( [ A, B ] ) - 
[ \mathbf{T}(A) ,\mathbf{T}(B) ]. 
\end{eqnarray*}
On Hermitean $A$ and $B$, 
$\Re\mathbf{F}_{\mathbf{T}}(A,B)$ indicates how well $\mathbf{T}$ respects the
anti-commutator, while $\Im\mathbf{F}_{\mathbf{T}}(A,B)$ indicates how well
it respects the commutator. ($\Im\mathbf{F}_{\mathbf{T}}$~is identically zero on the Hermiteans if and only if 
$\mathbf{T}$ is a Lie-algebra homomorphism.) 

\subsubsection{A Cauchy-Schwarz Inequality}

This sesquilinear, positive semidefinite $\mathbf{F}_{\mathbf{T}}$  
allows for an $\mathscr{A}$-valued Cauchy-Schwarz\footnote
{Reminiscent of the Cauchy-Schwarz inequality for Hilbert C$^*\!$-modules (see \cite{lan}), but not quite the same:
the result is identical, but the conditions differ.} inequality:

\begin{lemma}[C$^*\!$-Cauchy-Schwarz inequality] \label{cs}
Let $\mathbf{T}$ be a four-positive unital operation $\mathscr{B} \to \mathscr{A}$.
Then for all $ A,B \in \mathscr{B}$, we have in operator ordering:
\begin{displaymath}
\mathbf{F}_{\mathbf{T}}(A,B) \mathbf{F}_{\mathbf{T}}(B,A) \leq \| \mathbf{F}_{\mathbf{T}}(B,B) \| \mathbf{F}_{\mathbf{T}}(A,A). 
\end{displaymath}
\end{lemma}
\textbf{Proof}:
\begin{quote}
First we prove that $\mathbf{T}(A)^{\dagger} = \mathbf{T}(A^{\dagger})$ and that 
$\mathbf{F}(A,A) \geq 0 \quad \forall A \in \mathscr{B}$.
Since $\mathbf{T}$ is two-positive on $\mathscr{B}$, $\mathbf{id}_{2} \otimes \mathbf{T}$ is positive on $M_{2} \otimes \mathscr{B}$, 
and we see that
$$
   \mathbf{id}_{2} \otimes \mathbf{T}    \left(         \left( 	\begin{array}{cc}
 								A^{\dagger}	& 0	\\
								\mathbb{I} 	& 0	\\
								\end{array}
							\right)
							\left( 	\begin{array}{cc}
 								A & \mathbb{I}	\\
								0 & 0	\\
								\end{array}
							\right)					\right)
							\geq 0
$$
or 
$$
\left(
	\begin{array}{cc}
	\mathbf{T}(A^{\dagger}A)	&	\mathbf{T}(A^{\dagger})\\
	\mathbf{T}(A)			&	\mathbb{I}\\	
	\end{array}
\right)
\geq 0
$$
in the operator ordering. In particular it must be Hermitean so that $\mathbf{T}(A)^{\dagger} = \mathbf{T}(A^{\dagger})$.
For each $X \geq 0$, also $Y^{\dagger} X Y \geq 0$ for any Y. Making a convenient choice for $Y$ :
$$Y = \left(
	\begin{array}{cc}
	\mathbb{I}	&	0	\\
	-\mathbf{T}(A)	&	0	\\	
	\end{array}
\right)
$$
we obtain 
$$\left(
	\begin{array}{cc}
	\mathbf{T}(A^{\dagger}A) - \mathbf{T}(A)^{\dagger} \mathbf{T}(A)	&	0	\\
	0									&	0	\\	
	\end{array}
\right)
\geq 
0
$$
so that $\mathbf{F}(A,A) \geq 0 $ for all two-positive $\mathbf{T}$.
Now since $\mathbf{T}$ is four-positive on $\mathscr{B}$, $\mathbf{id}_{2} \otimes \mathbf{T}$ is again two-positive on 
$M_{2} \otimes \mathscr{B}$. Consequently, making a convenient choice of `$A$' in $M_{2} \otimes \mathscr{B}$: 
$$
\mathbf{F}_{\mathbf{id}_{2} \otimes \mathbf{T}} 
\left(
 \left(
	\begin{array}{cc}
	A	&	B	\\
	0	&	0	\\	
	\end{array}
\right)
,
 \left(
	\begin{array}{cc}
	A	&	B	\\
	0	&	0	\\	
	\end{array}
\right)
\right)
\geq
0 \quad \forall \quad A,B \in \mathscr{B}. 
$$
Working out this expression explicitly:
\begin{eqnarray*}
	\lefteqn{
		\mathbf{F}_{\mathbf{id}_{2} \otimes \mathbf{T}} 
		\left(
 			\left(
			\begin{array}{cc}
			A	&	B	\\
			0	&	0	\\	
			\end{array}
			\right)
			,
 			\left(
			\begin{array}{cc}
			A	&	B	\\
			0	&	0	\\	
			\end{array}
			\right)
		\right) =
	}\\
&=&
\mathbf{id}_{2} \otimes \mathbf{T} 
\left(
	\left(
		\begin{array}{cc}
		A^{\dagger}	&	0	\\
		B^{\dagger}	&	0	\\	
		\end{array}
	\right)
	\left(
	\begin{array}{cc}
	A	&	B	\\
	0	&	0	\\	
	\end{array}
	\right)
\right) \, -\\
& &
\mathbf{id}_{2} \otimes \mathbf{T} 
\left(
	\left(
		\begin{array}{cc}
		A^{\dagger}	&	0	\\
		B^{\dagger}	&	0	\\	
		\end{array}
	\right)
\right)	
\mathbf{id}_{2} \otimes \mathbf{T} 
\left(
\left(
	\begin{array}{cc}
	A	&	B	\\
	0	&	0	\\	
	\end{array}
	\right)
\right)\\
&=&
\left(
	\begin{array}{cc}
	\mathbf{T}(A^{\dagger}A)	&	\mathbf{T}(A^{\dagger}B)	\\
	\mathbf{T}(B^{\dagger}A)	&	\mathbf{T}(B^{\dagger}B)	\\	
	\end{array}
\right)
-
\left(
	\begin{array}{cc}
	\mathbf{T}(A^{\dagger}) \mathbf{T}(A)	&	\mathbf{T}(A^{\dagger}) \mathbf{T}(B)	\\
	\mathbf{T}(B^{\dagger})\mathbf{T}(A)	&	\mathbf{T}(B^{\dagger}) \mathbf{T}(B)	\\	
	\end{array}
\right)\\
&=&
\left(
	\begin{array}{cc}
	\mathbf{F}(A,A)	&	\mathbf{F}(A,B)	\\
	\mathbf{F}(B,A)	&	\mathbf{F}(B,B)	\\	
	\end{array}
\right)\\
&\geq&
0.
\end{eqnarray*}
Once again using $X \geq 0 \Rightarrow Y^{\dagger} X Y \geq 0$, this time with 
$$
Y 
=
\left(
	\begin{array}{cc}
	\mathbb{I}		&	0	\\
	-\mathbf{F}(B,A)	&	0	\\	
	\end{array}
\right) 
$$
we obtain, proceeding as above: 
\bq\label{mastbank}
\mathbf{F}(A,A) - 2\mathbf{F}(A,B) \mathbf{F}(B,A) + \mathbf{F}(A,B) \mathbf{F}(B,B) \mathbf{F}(B,A) \geq 0.
\eq
In the case that $\mathbf{F}(B,B) = 0$, we need to prove 
that $\mathbf{F}(A,B)\mathbf{F}(B,A) = 0$.
Now $\mathbf{F}(NB,NB) = 0$ for $N \in \mathbb{N}$. Applying inequality~\ref{mastbank} to 
$A$ and $NB$, we see that  
$$
	N^2 \mathbf{F}(A,B)\mathbf{F}(B,A) \leq \mathbf{F}(A,A)
$$
for all $N$, so that $\mathbf{F}(A,B)\mathbf{F}(B,A) = 0$.
In the case that $\mathbf{F}(B,B) \neq 0$, we rephrase inequality~\ref{mastbank} as
$$
\mathbf{F}(A,B) \mathbf{F}(B,A) \leq \mathbf{F}(A,A) + \mathbf{F}(A,B)(\mathbf{F}(B,B) - \mathbb{I}) \mathbf{F}(B,A).
$$
So, putting $ B' = B / \| \sqrt{\mathbf{F}(B,B)}  \| $ and noting $\mathbf{F}(B',B') - \mathbb{I} \leq 0$:
$$
\frac{\mathbf{F}(A,B)\mathbf{F}(B,A)}{\| \mathbf{F}(B,B) \|} \leq \mathbf{F}(A,A) + 
\mathbf{F}(A,B')(\mathbf{F}(B',B') - \mathbb{I}) \mathbf{F}(B',A) \leq \mathbf{F}(A,A)
$$
yielding the required expression.
\begin{flushright}
$q.e.d.$\\
\end{flushright}
\end{quote}

\subsubsection{Covariance and Uncertainty}

We shall give a few easy corollaries to clarify the nature of $\mathbf{F}$.
First of all, it resembles the covariance of a state. Classically, a
probability distribution $\mathbb{P}$ on a probability space $(\Omega,
\mathscr{B},\mathbb{P})$ induces a covariance on pairs of random variables
$\mathfrak{a, b}$:
\bq \label{wortel1}
\mathbf{Cov}_{\mathbb{P}}(\mathfrak{a},\mathfrak{b}) = \mathbb{E}_{\mathbb{P}}( \mathfrak{a}\mathfrak{b}) -  
\mathbb{E}_{\mathbb{P}}(\mathfrak{a})  \mathbb{E}_{\mathbb{P}}(\mathfrak{b}).   
\eq
Where $\mathbb{E}_{\mathbb{P}}$ is the expectation with respect to $\mathbb{P}$.
In a quantum probability space, observables are not represented by random
variables, but by Hermitean elements of a C$^*\!$-algebra. If one chooses to
represent the product observable of $A$ and $B$ by $(AB + BA)/2$, one can
generalize (\ref{wortel1}) to arbitrary C$^*\!$-algebras.
\begin{defin}[Covariance]\label{defcov}
Let $\rho \in \mathscr{S}(\mathscr{B})$. Let $A,B \in \mathscr{B}$ Hermitean. Then the
covariance of $A$ and $B$ in $\rho$ is defined by:
\begin{displaymath} 
\mathbf{cov}_\rho (A,B) = \rho\big({ \textstyle \frac{1}{2} }(AB + BA)\big)
- \rho(A) \rho(B)\;. 
\end{displaymath}
\end{defin}
Of course there is no conflict with proposition (\ref{simple}) for commuting $A$ and $B$.  
From the C$^*\!$-Cauchy-Schwarz inequality, we now have two easy corollaries. 
The first is a standard result, known as the `covariance inequality':
\begin{cor}[Covariance Inequality]\label{ziekkonijn} 
Let $\rho \in \mathscr{S(B)}$. Then for all Hermitean $ A,B \in \mathscr{B}$:
\begin{displaymath}
|\mathbf{cov}_{\rho}(A,B)|^2 \leq \mathbf{var}_{\rho}(A) \mathbf{var}_{\rho}(B). 
\end{displaymath}
\end{cor}
The second standard result is known as the `Heisenberg uncertainty relation':
\begin{cor}[Heisenberg Inequality]\label{hur}
Let $\rho \in \mathscr{S(B)}$. Then for all Hermitean $ A,B \in \mathscr{B}$:
\begin{displaymath}
\left| \rho \left( { \textstyle \frac{[A,B]}{2i} } \right) \right|^2 \leq \mathbf{var}_{\rho}(A) \mathbf{var}_{\rho}(B).
\end{displaymath}
\end{cor}
In particular, if $A$ = $i\hbar \partial_{x}$ and $B = x$, It follows\footnote
{Since $ \partial_{x}$ and $x$ are not bounded, we are not
allowed to apply the C$^*\!$-Cauchy-Schwarz inequality directly. The statement is true
nonetheless.} that $\sigma_{A}\sigma_{B} \geq \hbar/2$.
We prove both corollaries at the same time:\\*
\textbf{Proof}:
\begin{quote}
A state $\rho$ on $\mathscr{B}$ is
just a completely positive map $\mathscr{B} \to \mathbb{C}$.  So we can form
$\mathbf{F}_{\rho}$ and note that 
$$
\Re \mathbf{F}_{\rho}(A,B) =
{ \textstyle \frac{1}{2}}\big(\mathbf{F}_{\rho}(A,B) + \mathbf{F}_{\rho}(B,A) \big) =
\mathbf{cov}_{\rho}(A,B),
$$
$$
\Im \mathbf{F}_{\rho}(A,B) = 
{ \textstyle \frac{1}{2i}}\big(\mathbf{F}_{\rho}(A,B) - \mathbf{F}_{\rho}(B,A) \big) =
\rho\left(\frac{[A,B]}{2i}\right).
$$
Therefore, both $|\mathbf{cov}_{\rho}(A,B)|^2 \leq |\mathbf{F}_{\rho}(A,B)|^2$ and 
$|\rho(\frac{[A,B]}{2i})|^2 \leq |\mathbf{F}_{\rho}(A,B)|^2$.

The two corollaries above now follow from the C$^*\!$-Cauchy-Schwarz inequality:
$$
|\mathbf{F}_{\rho}(A,B)|^2 \leq |\mathbf{F}_{\rho}(A,A)||\mathbf{F}_{\rho}(B,B)| =
\mathbf{var}_{\rho}(A) \mathbf{var}_{\rho}(B).
$$ 
\begin{flushright}
\emph{q.e.d.}
\end{flushright}
\end{quote}
In words, corollary~(\ref{ziekkonijn}) is the real part of the C$^*\!$-Cauchy-Schwarz inequality,
corollary~(\ref{hur}) its imaginary part.  

\subsubsection{Multiplication Theorems}

Up to this point, we've used a state $\rho$ to construct $\mathbf{F}_{\rho}$, but we will encounter
$\mathbf{F}$ descendant from 
more general positive operations $\mathscr{B} \rightarrow \mathscr{A}$ later on. 
The C$^*\!$-Cauchy-Schwarz inequality was inspired by a `multiplication theorem' due to R.~Werner
(see \cite{wer}):
\begin{cor}[Multiplication Theorem]\label{multi}
Let $\mathbf{T}$ be a four-positive unital operation $\mathscr{B} \to \mathscr{A}$.
Let $B \in \mathscr{B}$ such that $\|B\|_{\mathbf{T}} = 0$.
Then for all $ A \in \mathscr{B}$:
\begin{displaymath}
\mathbf{F}_{\mathbf{T}}(A,B) = \mathbf{F}_{\mathbf{T}}(B,A) = 0  \nonumber
\end{displaymath}
\begin{displaymath}
\mathrm{i.e. } \quad
\mathbf{T}(A^{\dagger}B) = \mathbf{T}(A)^{\dagger} \mathbf{T}(B) \quad
\mathrm{and} \quad \mathbf{T}(B^{\dagger}A) = \mathbf{T}(B)^{\dagger} \mathbf{T}(A).
\end{displaymath}
\end{cor}
The proof is immediate from the following generalization, the
`almost multiplication theorem'.  
\begin{cor}
Let $\mathbf{T}$ be a four-positive unital operation $\mathscr{B} \to \mathscr{A}$.
Let $B \in \mathscr{B}$.
Then for all $ A \in \mathscr{B}$:
\begin{displaymath}
\| \mathbf{F}_{\mathbf{T}}(A,B)  \| \leq  \|A\| \|B\|_{\mathbf{T}}.  
\end{displaymath}
\end{cor}
\textbf{Proof}:
\begin{quote}
For any $A,B \in \mathscr{B}$ we have by the C$^*\!$-Cauchy-Schwarz inequality
$
\mathbf{F}_{\mathbf{T}}(A,B) \mathbf{F}_{\mathbf{T}}(B,A) \leq
\mathbf{F}_{\mathbf{T}}(A,A) \|B\|^{2}_{\mathbf{T}}
$, so certainly 
$
\| \mathbf{F}_{\mathbf{T}}(A,B) \| \leq
\|A\|_{\mathbf{T}} \|B\|_{\mathbf{T}}
$.
But since  
$
0 \leq \mathbf{F}(A,A) \leq \mathbf{T}(A^{\dagger}A) \leq \|A\|^2 \mathbb{I} 
$,
we also have $\|A\|_{\mathbf{T}} \leq \|A\|$. 
The corollary follows.
\begin{flushright}
\emph{q.e.d.}
\end{flushright}
\end{quote}
This is the form of the C$^*\!$-Cauchy-Schwarz inequality we will utilize most often.

\newpage \section {Quantum Measurement}
 
With the help of the C$^*\!$-Cauchy-Schwarz inequality we will investigate quantum measurement. 
But we will first define it.

\subsection {Introduction}

We will distinguish perfect and unbiased measurement. The former is a special case 
of the latter.

\subsubsection{Perfect Measurement }
 
In order for an operation to be a measurement, it must transport information from $X$, the 
observable to be measured, to $Y$, the pointer-observable. Observation of $Y$ in state 
$\mathbf{M}^*(\rho)$ must be equivalent to observation of $X$ in state $\rho$:   
\begin{defin}[Perfect Measurement] \label{measdefper} 
Let $X \in \mathscr{A}$, $Y \in \mathscr{B}$ be Hermitean. 
A perfect measurement of $X$ with pointer $Y$ is by definition
a completely positive map  $ \mathbf{M} : \mathscr{B} \to \mathscr{A}$
such that
$$
	\mathbb{P}_{\mathbf{M}^* (\rho), Y} = \mathbb{P}_{\rho, X} \quad \forall \rho \in \mathscr{S}( \mathscr{A} ).
$$
\end{defin}
In the Heisenberg picture, this makes $\mathbf{M}_{\mathscr{C}(Y)}$ an
injective $^*\!$-ho\-mo\-morphism:
\begin{prop} \label{perfert}
Let $ \mathbf{M} : \mathscr{B} \to \mathscr{A}$ be completely positive. Let 
$X \in \mathscr{A}$, $Y \in \mathscr{B}$ be Hermitean. Then $\mathbf{M}$ is a 
perfect measurement of $X$ with pointer $Y$ if and only if 
$$
\mathbf{Spec}(X) = \mathbf{Spec}(Y)
\quad \mathrm{and} \quad
\mathbf{M}\big(f(Y)\big) = f(X) \quad \forall f \in \mathscr{C}\big(\mathbf{Spec}(Y)\big).
$$
\end{prop} 
\textbf{Proof}:
\begin{quote}
If $\mathbb{P}_{\rho, X} = \mathbb{P}_{\mathbf{M}^* (\rho), Y}$, then they certainly live on 
the same measure-space: $\mathbf{Spec}(X) = \mathbf{Spec}(Y)$.
By the proof of proposition~(\ref{kloek}),
$\mathbb{P}_{\rho, X} = \mathbb{P}_{\mathbf{M}^* (\rho), Y}$ iff their expectation values on
$f \in \mathscr{C}(\mathbf{Spec}(Y))$ are the same. This is so for all $\rho \in \mathscr{S(A)}$
iff $\rho(\mathbf{M}(f(Y))) = \rho(f(X)) \quad \forall \rho \in \mathscr{S(A)}$, or
equivalently, iff $\mathbf{M}(f(Y)) = f(X)$. 
\begin{flushright}
\emph{q.e.d.}
\end{flushright}
\end{quote} 
In particular, a pointer $Y$ measures only one $X = \mathbf{M}(Y)$.  

\subsubsection{Unbiased Measurement} 
 
We shall broaden our view to include `measurements' that do not transfer the entire probability 
distribution from $X$ to $Y$, but only the average.   
 
\begin{defin}[Unbiased Measurement] \label{measdef} 
Let $X \in \mathscr{A}$, $Y \in \mathscr{B}$ be Hermitean. 
An unbiased measurement $\mathbf{M}$ of $X$ with pointer $Y$ is by definition a completely positive 
map $ \mathbf{M} : \mathscr{B} \to \mathscr{A} $ such that
$$
	\mathbf{M}^* (\rho)(Y) = \rho(X) \quad \forall \rho \in \mathscr{S}( \mathscr{A} )
\quad \mathrm{or \; equivalently } 
\quad 
\mathbf{M}(Y) = X.
$$
\end{defin}
Equivalence is easily established with \cite[p.~257]{ka1}.
Observing $Y$ in state $\mathbf{M}^*(\rho)$ results in the same \emph{average} 
as observing $X$ in \mbox{state $\rho$}. 
But the probability distributions need not be the same. 

Take \emph{any} operation $\mathbf{T}: \mathscr{B \to A}$. Take \emph{any} $Y \in \mathscr{B}$.
Then $\mathbf{T}$ is automatically a measurement of $\mathbf{T}(Y)$ with
pointer $Y$. Unbiased
measurements are not hard to find.

\subsubsection{Quality of Unbiased Measurement}

Next, we define the quality $\sigma$ of this unbiased measurement:
\begin{defin}[Quality] 
\label{qual}
Let $\mathbf{M}: \mathscr{B \to A}$ be an unbiased\/\footnote
{
In order to claim complete generality, we could abandon the demand that a
measurement be unbiased: we would then introduce a \emph{maximal bias}  
$\epsilon = \sup \{ \, |\mathbf{M}^*(\rho)(Y) - \rho(X)| \hspace{0,15 cm}|\hspace{0,15 cm} \rho \in \mathscr{S(A)} \, \} 
= \| \mathbf{M}(Y) - X\|$. 
This would make \emph{any} completely positive operation $\mathbf{M} : \mathscr{B \to A}$ a measurement of 
\emph{any} $X \in \mathscr{A}$ with \emph{any} pointer $Y \in \mathscr{B}$ and with a 
certain quality $\sigma$ and maximal bias $\epsilon$.
The interested reader may consider it a home exercise to adapt the estimates to come, adding $\epsilon$'s along the way.
Good luck.
} measurement of $X$ with pointer $Y$. 
Then its quality $\sigma$ is defined by
$$
\sigma^2 \isperdef \sup \{ \,\mathbf{var}_{\mathbf{M}^* (\rho)}(Y) - 
\mathbf{var}_{\rho}(X) \, | \, \rho \in \mathscr{S}( \mathscr{A} ) \,\}.  
$$
\end{defin}
So $\sigma$ tells us how much the uncertainty of the measurement result maximally exceeds the unavoidable 
amount of uncertainty inherent in the state $\rho$. In particular, when measuring an eigenstate of $X$, 
the variance in the measurement outcome will be less than or equal to $\sigma^2$. 
It is intuitively clear (and we will prove shortly) that $\sigma^2 \geq 0$: the uncertainty of $\rho$ in $X$ is 
inherent and can never be diminished by some clever choice of measurement.  

\begin{lemma} \label{sigma}
The quality $\sigma$ of an unbiased measurement 
$\mathbf{M} : \mathscr{B} \to \mathscr{A}$ with 
Hermitean pointer $Y \in \mathscr{B}$ is $\|Y\|_{\mathbf{M}}$. 
\end{lemma}
\textbf{Proof}:
\begin{quote}
If $\mathbf{M}$ is a measurement of $X$, then $X$ equals $\mathbf{M}(Y)$. 
This is apparent from $\rho(\mathbf{M}(Y) - X) = 0$ for all $\rho \in \mathscr{S(A)}$. 
Now for all $\rho \in \mathscr{S(A)}$
\begin{eqnarray*}
\lefteqn{\mathbf{var}_{\mathbf{M}^* (\rho)}(Y) - \mathbf{var}_{\rho} \big( \mathbf{M}(Y) \big) =}\\
&=&
\Big( \mathbf{M}^* (\rho) (Y^2) - (\mathbf{M}^* (\rho))(Y)^2 \Big) - 
\Big( \rho \left( \mathbf{M}(Y)^2 \right) - \rho \big( \mathbf{M}(Y) \big)^2 \Big)\\
&=&
\rho \big(\mathbf{M}(Y^2) \big) - \rho \big( \mathbf{M}(Y)^2 \big)\\
&=&
\rho \big( \mathbf{F}(Y,Y) \big)
\end{eqnarray*}
so that
$$
\sigma^2 = \sup\{ \, \rho \big( \mathbf{F}(Y,Y) \big) \, | 
\, \rho \in \mathscr{S}( \mathscr{A} ) \, \} = \|Y\|^{2}_{\mathbf{M}}\;.
$$
This proves the assertion, as well as the positivity of $\sigma^2$.
\begin{flushright}
\emph{q.e.d.}
\end{flushright} 
\end{quote}
From this, it is clear that if a measurement is perfect (in the sense of definition~(\ref{measdefper})),
then it has optimal quality: $\sigma = 0$. One need only apply proposition~(\ref{perfert}) with
$f(x) = x^2$. 

It is time to look at some examples of unbiased measurement.
   
\subsection{Examples of Unbiased Measurement}

\begin{enumerate}
\item \label{dirmesj}
A direct\footnote
{This kind of measurement is often called `von Neumann measurement'. 
We will reserve this designation for example \ref{vone} instead.} 
observation of $X \in \mathscr{A}$, denoted $\mathbf{M} : \mathscr{C}(\mathbf{Spec}(X)) \to \mathscr{A}$ 
is defined by $f \mapsto f(X)$. In the dual (Schr\"odinger) picture, $\mathbf{M}^*$ 
maps $\mathscr{S(A)}$ to $\mathscr{S}(\mathscr{C}(\mathbf{Spec}(X)))$, the 
probability distributions on $\mathbf{Spec}(X)$. It is completely positive.
According to proposition~(\ref{kloek}),
$\mathbf{M}^* (\rho) = \mathbb{P}_{\rho, X}$. 
If $\mathfrak{x}$ in $\mathscr{C}(\mathbf{Spec}(X))$ is the random variable 
$\mathfrak{x}(\lambda) = \lambda$, then $\mathbf{M}$ is 
a perfect measurement of $X$ with pointer $\mathfrak{x}$. Its quality is therefore $\sigma = 0$.
\item
Indirect observation is also measurement: If $\mathbf{M}^* : \mathscr{S(A) \to S(A \otimes B)}$ 
is defined by $\mathbf{M}^* (\rho) = \alpha^{*} (\rho \otimes \tau)$ for some automorphism 
$\alpha$ of $\mathscr{A \otimes B}$ and for some $\tau \in \mathscr{S(B)}$, then it is 
completely positive. If $\mathbf{M}^*(\rho)(Y) = \rho(X)$ for all $\rho \in \mathscr{S(A)}$, 
then it is an unbiased measurement of $X$ with pointer $Y$.
It is perfect in the sense of definition~(\ref{measdefper}) if and only if it is perfect in 
the sense of page~\pageref{klantenklop}. 
\item
Each automorphism is completely positive. If $\alpha$ is an automorphism of $\mathscr{D}$
such that $\alpha (Y) = X$, then it is a perfect measurement of $X$ with pointer~$Y$.
Its quality is automatically $\sigma = 0$.
\item
Let $U\subset \mathbb{R}$ be compact in the Euclidean topology, and let $\mathscr{A}$ be some von Neumann algebra.
A `Positive Operator Valued Measure' (POVM) (see \cite[p.~51]{hol}) is defined\footnote{In \cite{dav} and \cite{hol} 
even for non-compact U.} 
as a mapping $M$ from the Borel-measurable subsets
of $U$ into $\mathscr{A}$ satisfying:
	\begin{itemize}
	\item[-]$M(\emptyset) = 0,\quad M(U) = \mathbb{I}$
	\item[-]$M(V) \geq 0$ as an operator inequality for all measurable $V \subset U$.
	\item[-]Suppose $\{\,V_j \,|\, j \in J \,\}$ is a countable decomposition of $V$, then 
	$M(V) =$ \\*$ \sum_{j \in J} M(V_j)$,
	where the sum converges in the weak sense.
	\end{itemize}
By integrating bounded measurable functions $\mathscr{L}^{\infty}(U)$ on
$U$ over the POVM, $M$ may be extended to a unital, positive operation 
$\mathbf{M}: \mathscr{L}^{\infty}(U) \to \mathscr{A}$. In short:
$\mathbf{M}(f) \isperdef \int f(x)M(dx)$. 
It is completely positive due to the commutativity of $\mathscr{L}^{\infty}(U)$.
Thus, we have an unbiased measurement of $\int x M(dx)$ with pointer $f : x
\mapsto x$ and
quality $\sigma^2 = \| \int x^2 M(dx) - (\int x M(dx))^2) \|$.
The POVM is projection-valued iff $\sigma = 0$. It then reduces to the direct
observation of example~(\ref{dirmesj}).  
\item
Davies (see \cite[ch.~3]{dav}) also adopts the POVM as measurement.
He gives a particularly nice example of an unbiased position measurement:
Let $\mathscr{H} = L^2(\mathbb{R})$, and $\mathscr{A = B(H)}$.
Then the position observable $X$ is defined by $(X\psi)(x) = x \psi(x)$.
\mbox{$X$ has spectral} measure $V \mapsto \mathbf{P}(V)$, defined by 
$(\mathbf{P}(V) \psi)(x) = \mathbb{I}_{V}(x)\psi(x)$.

Let $f$ be a probability density on $\mathbb{R}$ with zero mean.
For Borel-sets $V \subset \mathbb{R}$, $M(V)$ is defined\footnote{ 
$f \ast g$ is the convolution of $f$ and $g$. Explicitly, 
$(f \ast g)(x) = \int_{- \infty}^{\infty} f(y) g(x - y) dy$.} by 
$M(V) = \int_{- \infty}^{\infty} (f \ast \mathbb{I}_{V})(x) \mathbf{P}(dx)$. $f$ has the 
effect of blurring the outcome. In the limiting case that $f$ is the Dirac $\delta$-distribution 
on zero, $M(V) = \mathbf{P}(V)$.
  
$M$ turns out to be a measurement of the position observable $X$
indeed: \\* \mbox{$\int x M(dx) = X$}.
Furthermore, Davies shows that the quality of $M$ is exactly the r.m.s. value of $f$: $\sigma^2 = \mathbf{Var}(f)$.
\item \label{vone}
Suppose that the observable being measured has discrete spectrum, 
$X = \sum_{i} x_i \mathbf{P}_{i}$. 
Then there exists a so-called von Neumann measurement 
\mbox{$\mathbf{N}: \mathscr{C}(\mathbf{Spec}(X)) \otimes \mathscr{A} \to
\mathscr{A}$}.
It is defined by
$f \otimes A \mapsto \sum_{i} f(i) \mathbf{P}_{i} A \mathbf{P}_{i}$.
Constrained to $\mathscr{C}(\mathbf{Spec}(X)) \otimes \mathbb{I}$, it reduces 
again to direct observation.
But $\mathbf{N}$ also gives information 
about how the system is left behind. 

$\mathbf{N}$ is a perfect measurement of $X$ with pointer 
$\mathfrak{x} \otimes \mathbb{I}$ (where $ \mathfrak{x}(i) = x_i $) and quality $\sigma = 0$. 
The state $\mathbf{N}^*(\rho) |_{\mathscr{A}\otimes \mathbb{I}}$
is exactly the \emph{collapsed state} of $\rho$ after $X$-measurement: 
$\mathbf{N}^*(\rho) (A \otimes \mathbb{I}) = \rho (\sum_{i} \mathbf{P}_{i} A \mathbf{P}_{i}) = 
\mathbf{C}^*(\rho) (A)$. 

Note that if $[A,X]=0$ for some $A \in \mathscr{A}$, then  
$A = \sum_{i} \mathbf{P}_{i} A \mathbf{P}_{i}$ since it also commutes
with the spectral projections.
Consequently, $\mathbf{N}(A \otimes \mathbb{I}) = A$ and 
$\mathbf{N}^*(\rho) (A \otimes \mathbb{I}) = \rho (A)$.
\item \label{voorbeeld}
Let $\mathbf{M}: M_2 \otimes C_2 \to M_2$ ($C_2$ are the $2 \times 2$-diagonal matrices) be defined by
$$
\mathbf{M}(A \otimes D) = \sum_{i = 0,1} d_{ii} X_{i}^{\dagger} A X_i
$$
with the matrices ($0 \leq\epsilon \leq 1/2$)
$$
X_0 = 
\left( 	\begin{array}{cc}
 		\sqrt{1 - \epsilon} 	& 0			\\
		0 			&\sqrt{\epsilon} 	\\
		\end{array}
		\right)
\quad
X_1 =
\left( 	\begin{array}{cc}
 		\sqrt{\epsilon} 	& 0			\\
		0 			&\sqrt{1 - \epsilon} 	\\
		\end{array}
		\right)
$$
$M_2 \otimes C_2$ is isomorphic to $M_2 \oplus M_2$. There $\mathbf{M}$ reads
$$
\mathbf{M}(A \oplus B) = X_{0}^{\dagger} A X_0 + X_{1}^{\dagger} B X_1
$$
From which one can verify complete positivity of $\mathbf{M}$.
It is an unbiased measurement of $\sigma_z$ with pointer $(1-2\epsilon)^{-1} \mathbb{I} \otimes \mathit{diag}( \, 1 , -1 \, )$ 
and quality $\sigma = \frac{ 2 \sqrt{\epsilon(1 - \epsilon)} }{1 - 2\epsilon}$.
For $\epsilon = 0$, $\mathbf{M}$ reduces to the (perfect) von Neumann measurement of example~\ref{vone}.
For $\epsilon = 1/2$, it has become completely useless: It produces a random outcome on $C_2$, unrelated to 
the measured object.
We'll come back to this example after proposition~(\ref{hpdelta}). 
\item
There are also silly examples of measurement: The identity is a perfect ($\sigma = 0$) measurement of
any observable $X$ with pointer $X$.
\item
Any completely positive operation is a perfect ($\sigma = 0$) measurement of $\mathbb{I}$ with pointer~$\mathbb{I}$.
\end{enumerate}
Definition~(\ref{measdef}) even seems to admit operations one would not call measurement.
In spite of its generality though, much can be said about unbiased measurement and its quality.
We shall give some examples of this, 
leaning heavily on lemmas (\ref{sigma}) and (\ref{cs}).

\subsection{Structure of a Perfect Measurement}

But before that, we will exploit the fact that any completely positive operation $\mathbf{M}$ acts as a 
homomorphism on the elements of $\mathbf{M}$-norm 0. For any Hermitean $H$ we will denote by 
$\mathscr{C}(H)$ the C$^*\!$-algebra generated by $\mathbb{I}$ and $H$. 

\newpage
\begin{lemma} \label{structure}
Let $\mathbf{M} : \mathscr{B} \to \mathscr{A}$ be a 4-positive operation, 
let $B \in \mathscr{B}$ be Hermitean. Among

\begin{enumerate}
\item$\|B\|_{\mathbf{M}} = 0$.	\label{gnar1}
\item$\mathbf{M}$ is an isomorphism $\mathscr{C}(B) \to
\mathscr{C}\big(\mathbf{M}(B)\big)$.	\label{gnar2}
\item$\mathbf{Spec}(B) = \mathbf{Spec}\big(\mathbf{M}(B)\big)$ and 
$
\mathbf{M}\big(f(B)\big) = f\big( \mathbf{M}(B) \big) \label{gnar3}
$
for all  
$f \in \mathscr{C}\big(\mathbf{Spec}(B)\big)$.
\item \label{gnarx}
$
\|f(B)\|_{\mathbf{M}} = 0
$
for all $f \in \mathscr{C}\big(\mathbf{Spec}(B)\big)$.
\item$\mathbf{M}$ maps the relative commutant $B'$ into $\mathbf{M}(B)'$.
\label{gnar4}
\end{enumerate}
The following relations hold:
$$ \mathbf{ (\ref{gnar1}) \iff (\ref{gnar2})  \iff (\ref{gnar3}) \, \iff (\ref{gnarx})  
\Longrightarrow \, (\ref{gnar4}) }$$
\end{lemma}
\textbf{Proof}:
\begin{description} 
\item[$\mathbf{(\ref{gnar1}) \Rightarrow (\ref{gnar2}):} \quad$]
$\mathscr{C}(B)$ is the norm-closure in $\mathscr{B}$ of the collection of polynomials in $B$. 
Similar\-ly, $\mathscr{C}(\mathbf{M}(B))$ is the norm-closure in $\mathscr{A}$ of 
the polynomials in $\mathbf{M}(B)$. By the multiplication theorem~(\ref{multi}),
we see that $\mathbf{M}(B^n) = \mathbf{M}(B)^n$.   From this and linearity,  one
verifies that $\mathbf{M}(p(B)) = p(\mathbf{M}(B))$ and $\| p(B)\|_{\mathbf{M}} = 0$
for all polynomials $p$.

Now let $Q \in \mathscr{C}(B)$. By the Weierstrass theorem, there exist polynomials $p_n$ such that 
$ p_n (B) \to Q $ in norm. Since 
${\mathbf{M}}$ is automatically norm-continuous and since it maps $\mathscr{C}(B)$ densely into 
$\mathscr{C}(\mathbf{M}(B))$, we can verify that 
$ \mathbf{M}(Q) \in \mathscr{C}(\mathbf{M}(B)) $:
$$ \mathbf{M}(Q) = \lim_{n \to \infty} \mathbf{M}\big(p_n(B)\big) = 
\lim_{n \to \infty} p_n \big(\mathbf{M}(B)\big) \in \mathscr{C}\big(\mathbf{M}(B)\big). $$
Similarly, \mbox{$\|Q\|_{\mathbf{M}} = 0$} since  
\begin{eqnarray*}
\mathbf{M}(Q^2) &=& 
\mathbf{M}\big(\lim_{n \to \infty} p_{n}^2(B)\big) =
\lim_{n \to \infty} \mathbf{M}\big( p_{n}^2(B)\big) = 
\lim_{n \to \infty} p_{n}^2 \big( \mathbf{M}(B)\big)\\
 &=& 
\Big(\lim_{n \to \infty} p_n \big( \mathbf{M}(B)\big) \Big)^2 
= \big( \mathbf{M}(Q) \big)^2.
\end{eqnarray*}

This means that the restriction of $\mathbf{M}$ to $\mathscr{C}(B)$ is a C$^*\!$-isomorphism: 
if $Q \in \mathscr{C}(B)$,
then even $\mathbf{M}(Q  A) = \mathbf{M}(Q)\mathbf{M}(A)$ for any $A \in \mathscr{B}$ by the
multiplication theorem~(\ref{multi}). 
Its image is therefore automatically norm-closed (see \cite[p.~242]{ka1}), and thus equal to
$\mathscr{C}(\mathbf{M}(B))$.
\item[$\mathbf{ (\ref{gnar2}) \Rightarrow (\ref{gnar3}):} \quad$]
By the canonical isomorphism $f \mapsto f(B)$ known as the Gel'fand
transform, $\mathscr{C}(B)$ is isomorphic to 
$\mathscr{C}(\mathbf{Spec}(B))$, the C$^*\!$-algebra of 
continuous functions on $\mathbf{Spec}(B)$ equipped with the supremum norm (see \cite[p.~271]{ka1}).
Similarly, $\mathscr{C}(\mathbf{M}(B)) \sim \mathscr{C}(\mathbf{Spec}(\mathbf{M}(B))) $. 
$\mathbf{M}$ acts as a continuous isomorphism mapping polynomials $p$ on $\mathbf{Spec}(B)$ to 
the same $p$ on $\mathbf{Spec}(\mathbf{M}(B))$ since $\mathbf{M}(p(B)) = p(\mathbf{M}(B))$. 
By continuity of $\mathbf{M}|_{\mathscr{C}(B)}$ and $\mathbf{M}^{-1}|_{\mathscr{C}(\mathbf{M}(B))}$, 
polynomials converge on $\mathbf{Spec}(B)$
if and only if they do on $\mathbf{Spec}(\mathbf{M}(B))$.
Since the spectra are closed they must be the same, and by continuity of $\mathbf{M}$
and the theorem of Stone-Weierstrass (the polynomials form a norm-dense set in the space of
continuous functions), we now see that each continuous function is
mapped to itself. 
\item[$\mathbf{(\ref{gnar3}) \Rightarrow (\ref{gnarx})}: \quad$]

Let $f \in \mathscr{C}(\mathbf{Spec}(B))$. Let $g(x) = f(x)^2$. Then by
$\mathbf{(\ref{gnar3})}$ we have
$ \mathbf{M}(f(B)^2) = \mathbf{M}(g(B)) = g(\mathbf{M}(B)) =
f(\mathbf{M}(B))^2  = \mathbf{M}(f(B))^2$.
\item[$\mathbf{(\ref{gnarx}) \Rightarrow (\ref{gnar1})}: \quad$]
Trivial: choose $f(x) = x$. 

\item[$\mathbf{(\ref{gnar1}) \Rightarrow (\ref{gnar4}):} \quad$] 
Suppose that $A \in B'$, i.~e.\ $[A,B] = 0$.
Then by the multiplication theorem~(\ref{multi}),
$$
[\mathbf{M}(B), \mathbf{M}(A)] = \mathbf{M}([A,B]) - [\mathbf{M}(A), \mathbf{M}(B)] = 2i \mathbf{F}_{\mathbf{M}}(A,B) = 0.
$$
\begin{flushright}
\emph{q.e.d.}
\end{flushright}
\end{description}
It is clear that perfect measurement (in the sense of definition~(\ref{measdefper}))
satisfies $\sigma = 0$.
But combining lemma~(\ref{structure}) with proposition~(\ref{perfert}), we also obtain the converse:
\begin{cor}
Unbiased measurement (in the sense of definition~(\ref{measdef})) is perfect 
(in the sense of definition~(\ref{measdefper})) if and only if it has quality $\sigma = 0$.
\end{cor} 
Which is the moral obligation of any definition of quality. In the particular case of von Neumann algebras:
\begin{cor}\label{spectromeas}
Let $\mathscr{A}$ and $\mathscr{B}$ be von Neumann algebras, and $A \in \mathscr{A}$,
$B \in \mathscr{B}$ Hermitean. Suppose $\mathbf{M} : \mathscr{B} \to \mathscr{A}$ measures
$X$ with pointer $Y$ and quality $\sigma = 0$. 
Then $\mathbf{M}$ is also a perfect measurement of all spectral 
projections $\mathbf{P}(V)$ of $X$, with pointer $\mathbf{Q}(V)$, the corresponding
spectral projection of $Y$. 
\end{cor} 
\textbf{Proof}:
\begin{quote}
For any Borel set $V$ and for any $\rho \in \mathscr{S(A)}$, we have seen that 
$\mathbb{P}_{\rho, X}(V) = \mathbb{P}_{\mathbf{M}^*(\rho), Y}(V)$.
In particular, for all normal states $\rho$, this means that 
$\rho(\mathbf{P}(V)) = \mathbf{M}^*(\rho)(\mathbf{Q}(V))$, or 
$\rho \big( \mathbf{P}(V) - \mathbf{M}(\mathbf{Q}(V))\big) = 0 $ for 
all normal states $\rho$. Therefore $ \mathbf{M}(\mathbf{Q}(V)) = \mathbf{P}(V)$.
Automatically, $\| \mathbf{Q}(V) \|_{\mathbf{M}} = 0$, since
$ \mathbf{M}(\mathbf{Q}^2(V)) = \mathbf{M}(\mathbf{Q}(V)) = \mathbf{P}(V) = \mathbf{P}^2(V)$.
\begin{flushright}
\emph{q.e.d.}
\end{flushright}
\end{quote}

I hope that the paragraph above has given some credibility to our definitions of measurement and quality.
They will form the basis of the rest of this thesis.

\subsection{Simultaneous Measurement}
As an appetizer, we'll use the C$^*\!$-Cauchy-Schwarz inequality to generalize a well-known theorem.
Perfect simultaneous measurement (i.e. measurement of
two observables using two commuting pointers) can only be performed 
on commuting observables. The mathematical formulation below is based on Werner (see \cite{wer}),
but the physical statement was known long before, see e.g. \cite{neu}.
  
\begin{prop}[Joint Measurement] \label{jm}
Let $\mathbf{M}: \mathscr{B} \to \mathscr{A}$ be a perfect 
\mbox{($\sigma_{Y} = \|Y\|_{\mathbf{M}} = 0$}, \mbox{$ \sigma_{\tilde{Y}} = \|
\tilde{Y} \|_{\mathbf{M}} = 0$)} 
measurement of both $X \in \mathscr{A}$  
and $\tilde{X} \in \mathscr{A}$ with commuting pointers $Y, \tilde{Y} \in \mathscr{B}$ respectively.
(All Hermitean.) Then
$$
[X, \tilde{X}] = 0.
$$
\end{prop}
For example, let $\mathbf{M}^*$ be an affine map from $\mathscr{S(A)}$ to 
the space of probability distributions on $\mathbf{Spec}(X) \times \mathbf{Spec}(\tilde{X})$
such that $\mathbb{P}_{\rho, X}$ and $\mathbb{P}_{\rho, \tilde{X}}$ are the marginal 
probability distributions of $\mathbf{M}^* (\rho)$.
By the proof of proposition~(\ref{kloek}), the space of probability distributions on a probability 
space $\Omega$ can be identified with $\mathscr{S(C} (\Omega)) $, the state-space of the C$^*\!$-algebra
$\mathscr{C} (\Omega)$. Due to the abelianness of $\mathscr{C} (\Omega)$, $\mathbf{M}^*$ must be the dual of a completely positive map.
It is therefore a joint measurement in the sense of proposition~(\ref{jm}).
We see that the kind of mapping constructed in proposition~(\ref{simple}), vital to the interpretation 
of quantum mechanics, simply does not exist if $[X , \tilde{X}] \neq 0$.

We'll prove proposition~(\ref{jm}) along with a Heisenberg relation-like generalization\footnote
{ \cite[p.~90]{hol} already gives a generalization for POVM's. However, this involves
only $\mathbf{Var}_{\mathbf{M}^* (\rho)}(Y)$ instead of
$\mathbf{Var}_{\mathbf{M}^* (\rho)}(Y) - \mathbf{Var}_{\rho}(X)$, thus staying
much closer to the Heisenberg uncertainty relations.}, relating the product of both 
measurement qualities with the lack of commutativity. 
\begin{prop}[Generalized Joint Measurement] \label{jmdelta}
Let $\mathbf{M}: \mathscr{B} \to \mathscr{A}$ be an unbiased  
measurement of $X \in \mathscr{A}$  
and $\tilde{X} \in \mathscr{A}$, both Hermitean, with commuting 
Hermitean pointers $Y, \tilde{Y} \in \mathscr{B}$ respectively.
Then for the qualities $\sigma_{Y} = \|Y\|_{\mathbf{M}}$ and $\sigma_{\tilde{Y}} = \| \tilde{Y} \|_{\mathbf{M}}$ 
the following relation holds:
$$
2 \sigma_{Y} \sigma_{\tilde{Y}} \geq \|[X, \tilde{X}]\|\;.
$$
\end{prop}
\textbf{Proof}:
\begin{quote}
$
\|[X, \tilde{X}]\| =  \|\mathbf{M}([Y,\tilde{Y}]) - [\mathbf{M}(Y), \mathbf{M}(\tilde{Y})]\|
= \|2 \Im ( \mathbf{F} (Y,\tilde{Y}) )\| \leq 2 \|Y\|_{\mathbf{M}} \| \tilde{Y} \|_{\mathbf{M}}
= 2 \sigma_{Y} \sigma_{\tilde{Y}}
$, proving both propositions (\ref{jm}) and (\ref{jmdelta}). 
\begin{flushright}
\emph{q.e.d.}
\end{flushright}
\end{quote}

\newpage \section {The Heisenberg Principle}

The so-called Heisenberg principle\footnote{Not to be confused with the
Heisenberg uncertainty relations in 
corollary (\ref{hur}).} may be formulated as follows:
\begin{quote}
\emph{When an outside observer extracts quantum-information from a system, 
it is impossible to leave all states unaltered.}
\end{quote}
As for a mathematical formulation and proof, due to R.~Wer\-ner (see \cite{wer}):
\begin{prop}[Heisenberg Principle] \label{hp}  
Let $\mathbf{M} : \mathscr{A} \otimes \mathscr{B} \to \mathscr{A}$ be an 
unbiased measurement of any Hermitean $X \in \mathscr{A}$ with any Hermitean pointer $\mathbb{I} \otimes Y \in
\mathbb{I} \otimes \mathscr{B}$. 
Suppose that $\mathbf{M}$ leaves states on $\mathscr{A}$ undisturbed:  
$\mathbf{M}^* (\rho)( A \otimes \mathbb{I} )
=  \rho(A) \quad \forall A \in \mathscr{A} \quad \forall \rho \in \mathscr{S(A)}$. Then $X$ is in the centre \mbox{of $\mathscr{A}$}.  
\end{prop}
\textbf{Proof}:
\begin{quote}
In circumstances above,  $\mathbf{M}(A \otimes \mathbb{I}) = A$ for all $A \in
\mathscr{A}$.  Therefore also \mbox{$\mathbf{M}(A^{\dagger}A \otimes
\mathbb{I})$} $=
A^{\dagger}A$ for all $A \in \mathscr{A}$.
Thus  $\|A \otimes \mathbb{I}\|^{2}_{\mathbf{M}} = 0$, which
entails  $[\mathbf{M}(A \otimes \mathbb{I}), \mathbf{M}(\mathbb{I} \otimes Y)] =0$,  
i.e. $[A,X] = 0$ for all $A$ in $\mathscr{A}$.
\begin{flushright}
\emph{q.e.d.}
\end{flushright}
\end{quote}
So the only information (of any quality) that can be obtained without
disturbing the original system is information about central elements. 

We have good reason to consider `central information' as `classical information':
In the fully classical case, the algebra $\mathscr{A}$ is abelian. All observables are central, 
so all information is freely accessible.
In the archetypal quantum case however, 
the algebra $\mathscr{A}$ in question is $\mathscr{B}(\mathscr{H})$, the
algebra of bounded linear operators on some Hilbert space $\mathscr{H}$.
In this case, $\mathbb{I}$ is the only central element (modulo $\mathbb{C}$), so that 
no information can be gained without disturbing the system.
\subsection{Global Generalization}

So we have grounds to examine the norm distance of $X$ to the centre,
$d(X,\mathscr{Z}) = $ \mbox{$\inf \{\, \|X - Z\|  \, | \, 
Z \in \mathscr{Z} \, \}$}.
One may think of $d(X,\mathscr{Z})$ as quantifying the amount of `quantumness' in $X$, as it 
determines the maximal amount of non-commutativity with $X$ in the sense below: 

\begin{lemma} \label{qn}
Let $X$ be a Hermitean element of a finite-dimensional von Neumann \mbox{algebra
$\mathscr{A}$}.
Then 
$d(X,\mathscr{Z})$ is the smallest number $c$ such that  
$\|[A,X]\| \leq 2 c \, \|A\|$ for all $A \in \mathscr{A}$. 
\end{lemma}
\textbf{Proof}:
\begin{quote}
Finite or infinite dimensional algebra, it is clear that for any $A \in \mathscr{A}$, 
$\|[A , X ]\| = \|[A , X - Z]\| \leq 2\|A\| \| X - Z \|$ for any  $Z \in \mathscr{Z}(\mathscr{A})$.
Taking the infimum over $Z$, we obtain $\|[A,X]\| \leq 2 d(X,\mathscr{Z}) \|A\| $. 
This means that we are finished if we find $A \in \mathscr{A}$, $A \neq 0$ such that
$\|[A,X]\| = 2 d(X,\mathscr{Z}) \|A\|$. 

Now for any von Neumann algebra $\mathscr{A}$, 
there exists a so-called decomposition over the centre (see \cite[ch.~14]{ka1}).
For finite-dimensional von Neumann algebras, this simply means that
$\mathscr{A}$ is isomorphic to 
$M_{n_1} \oplus \ldots \oplus M_{n_k}$ 
for some $k; n_1, \ldots n_k \in \mathbb{N}$,
where $M_{n_i}$ is the algebra of $n_i \times n_i$ matrices. 
So $X = X_1 \oplus \ldots \oplus X_k$ with $X_k \in M_{n_k}$.
Each $X_k$ can be brought in diagonal form by some unitary transformation
$U_k$. So by the isomorphism $U_1 \oplus \ldots \oplus U_k$, we may think of $X$ 
as 
$$
\mathit{diag}(\lambda_{(1 , 1)}, \ldots ,\lambda_{(1 , n_1)}) \oplus \ldots \oplus 
\mathit{diag}(\lambda_{(k , 1)}, \ldots ,\lambda_{(k , n_k)})
$$ 
with $\lambda_{(i , 1)} \geq \ldots \geq \lambda_{(i , n_i)}$ the eigenvalues
of $X_i$ in decreasing order. Let 
$r_i \isperdef \frac{1}{2} \left( \lambda_{(i , 1)} - \lambda_{(i , n_i)} \right)$ be the spectral
radius of $X_i$. 
Let $t_i \isperdef \frac{1}{2} \left( \lambda_{(i , 1)} + \lambda_{(i , n_i)} \right)$.
Let 
\begin{eqnarray*} 
\tilde{X} &\isperdef& X - \left( t_1 \mathbb{I} \oplus \ldots \oplus t_k \mathbb{I} \right) \\
 &=& 
\mathit{diag} \left( \tilde{\lambda}_{(1 , 1)}, \ldots ,\tilde{\lambda}_{(1 , n_1)} \right) \oplus \ldots \oplus 
\mathit{diag} \left( \tilde{\lambda}_{(k , 1)}, \ldots ,\tilde{\lambda}_{(k , n_k)} \right)
\end{eqnarray*}
with $\tilde{\lambda}_{(i , 1)} = r_i$, $\tilde{\lambda}_{(i , n_i)} = -r_i$.
$ \tilde{X} $ and $ X $ differ only by the central element 
$t_1 \mathbb{I} \oplus \ldots \oplus t_k \mathbb{I} \in \mathscr{Z}$. 
Furthermore, $\|\tilde{X}\| = \max_i r_i$.
Take that maximal $r_i$ and isolate the highest and lowest eigenvectors
$$
0 \oplus \ldots \oplus 0 \oplus \psi_{+,-} \oplus 0 \oplus \ldots \oplus 0
$$  
in $\mathbb{C}^{n_1}\oplus \ldots \oplus \mathbb{C}^{n_k}$. Construct 
$$
A = 0 \oplus \ldots \oplus 0 \oplus (|\psi_+\rangle\langle \psi_-| +
|\psi_-\rangle\langle \psi_+|) \oplus 0 \oplus \ldots \oplus 0$$ 
in $\mathscr{A}$. Now, since $X - \tilde{X} \in \mathscr{Z}$, we see that 
$$
[X,A] = [\tilde{X},A] = 
0 \oplus \ldots \oplus 0 \oplus 2r_i (|\psi_+\rangle\langle \psi_-| -
|\psi_-\rangle\langle \psi_+|) \oplus 0 \oplus \ldots \oplus 0.
$$
So $\|[X,A]\| = \|[\tilde{X},A]\| = 2r_i = 2 \|\tilde{X}\| \|A\|$. 
But since already $d(X,\mathscr{Z}) \leq \|\tilde{X}\|$ and $\|[\tilde{X},A]\| \leq 2 d(X,\mathscr{Z}) \|A\|$,
we see that
$$
2 d(X,\mathscr{Z}) \|A\| \leq 2 \|\tilde{X}\| \|A\| = \| [X , A] \| \leq 2 d(X,\mathscr{Z}) \|A\|.
$$
So $\|\tilde{X}\| = d(X,\mathscr{Z})$ and $\|[X,A]\| = 2d(X,\mathscr{Z}) \|A\|$.
\begin{flushright}
\emph{q.e.d.}
\end{flushright}
\end{quote}
We are now looking for generalizations of proposition (\ref{hp}) of the
following form: suppose you allow some (small) disturbance of the states on the
original algebra. How much information can be gained maximally?
Note that in the proposition below, all that is used about the pointer 
is that it commutes with $\mathscr{A}\otimes \mathbb{I}$.
\newpage
\begin{prop}[Generalized Heisenberg Principle]\label{hpdelta}
Let $\mathscr{A}$ be a finite-dimensional von Neumann algebra with centre $\mathscr{Z}$.
Let $\mathscr{B}$ be an arbitrary von Neumann algebra 
and let $Y \in \mathscr{B}$, $X \in \mathscr{A}$ Hermitean. 
Let $\mathbf{M}$ be completely positive $\mathscr{A} \otimes \mathscr{B} \to
\mathscr{A}$ such that:
\begin{itemize}
\item[-]
$\mathbf{M}(\mathbb{I} \otimes Y) = X$ and $\|\mathbb{I} \otimes Y\|_{\mathbf{M}} = \sigma$, i.~e.\
$\mathbf{M}$ is an unbiased measurement of $X$ with pointer $\mathbb{I} \otimes Y$ 
and quality $\sigma$.  
\item[-]$\| \mathbf{M}^* (\rho)|_{\mathscr{A} \otimes \mathbb{I}} - \rho \| \leq \Delta 
\quad \forall \rho \in \mathscr{S}(\mathscr{A})$ for some $0 < \Delta < 1$.
\end{itemize}
Then
$$	 \sigma \geq d(X, \mathscr{Z}) \frac{1 - \Delta}{\sqrt{3\Delta}}. $$

\end{prop}
\textbf{Proof}:
\begin{quote}
We move to the Heisenberg picture: $\|\mathbf{M} (A \otimes \mathbb{I}) - A\| \leq \Delta \|A\|$
because 
\mbox{$\rho( \mathbf{M} (A \otimes \mathbb{I}) - A)$}$ \leq \Delta \|A\|$ for all $\rho \in \mathscr{S(A)}$.
For notational convenience, we introduce an operation
$\mathbf{T} : \mathscr{A} \to \mathscr{A}$ defined by 
$\mathbf{T}(A) \isperdef\mathbf{M}(A \otimes \mathbb{I})$
and the map $\mathbf{D} : \mathscr{A} \to \mathscr{A}$ defined by 
$\mathbf{D}(A) \isperdef\mathbf{T}(A) - A$. The former is the effect of measurement
on the measured system $\mathscr{A}$, the latter satisfies $\|\mathbf{D}(A)\| \leq \Delta \|A\|$. 
Since $\mathbf{D}(A)^{\dagger} \mathbf{D}(A) \geq 0$, we may estimate
\begin{eqnarray*}
\|  A \otimes \mathbb{I} \|_{\mathbf{M}}^{2} & = & \|\mathbf{F}_{\mathbf{T}}(A,A)\| \\
& \leq & \| \mathbf{F}_{\mathbf{T}}(A,A) + \mathbf{D}(A)^{\dagger} \mathbf{D}(A) \| \\
& = & \|  \mathbf{T}(A^{\dagger}A) - \mathbf{T}(A^{\dagger}) \mathbf{T}(A) 
+ \mathbf{D}(A)^{\dagger} \mathbf{D}(A) \| \\
& = & \|
(\mathbf{D}(A^{\dagger} A) + A^{\dagger}A)  - 
(\mathbf{D}(A) + A)^{\dagger}(\mathbf{D}(A) + A)
+ \mathbf{D}(A)^{\dagger} \mathbf{D}(A)\| \\ 
& = & \| 
(\mathbf{D}(A^{\dagger} A) -
\mathbf{D}(A)^{\dagger}A -
A^{\dagger} \mathbf{D}(A)
\|  \\
& \leq & 3 \Delta \|A\|^2.
\end{eqnarray*}

Since $\Delta < 1$, $\mathbf{T}$
must be injective because it is linear and because
$$
\mathbf{T}(A) = 0 \quad \Longrightarrow \quad 
\|A\| = \| \mathbf{D}(A) \| \leq  \Delta \|A\| 
\quad \Longrightarrow \quad \|A\| = 0. 
$$
Since $\mathscr{A}$ is finite-dimensional, this implies that $\mathbf{T}$ is
onto.
%
% Perhaps an infinite-dimensional version is possible... 
Furthermore, $\|\mathbf{T}(A) - A\| \leq \Delta \|A\|$ implies $| \|\mathbf{T}(A)\| - \|A\|
| \leq \Delta \|A\|$ and hence $\|A\| \leq \|\mathbf{T}(A)\|/(1 - \Delta)$.
By the C$^*\!$-Cauchy-Schwarz inequality, we deduce 
\begin{eqnarray*}
\|\, [X, \mathbf{T}(A)] \,\| & = & \|\,[\mathbf{M}(\mathbb{I}\otimes Y) , \mathbf{M}(A \otimes \mathbb{I}) ]\,\| \\
& \leq & 2 \| (\mathbb{I}\otimes Y) \|_{\mathbf{M}} \| 
(A \otimes \mathbb{I}) \|_{\mathbf{M}} \\
& \leq &
2 \sigma \sqrt{3\Delta} \|A\| \\
& \leq & 
2 \sigma \frac{\sqrt{3\Delta}}{1 - \Delta} \|\mathbf{T}(A)\|.
\end{eqnarray*}
So, since $\mathbf{T}$ is onto, $\sigma \frac{\sqrt{3\Delta}}{1 - \Delta}$ is a number $c$ such that 
$\| [X,A] \| \leq 2 c \|A\| \quad \forall A \in \mathscr{A}$.
By lemma (\ref{qn}), $d(X, \mathscr{Z})$ is the smallest such number. 
Thus $d(X, \mathscr{Z}) \leq \sigma \frac{\sqrt{3\Delta}}{1 - \Delta}$. 
\begin{flushright}
\emph{q.e.d.}
\end{flushright}
\end{quote}

If $\Delta = 0$, the last line of the proof reduces to proposition (\ref{hp}) for finite dimensional 
algebras: unbiased measurement is only possible if $X$ is central. 

If $\Delta \neq 0$, proposition (\ref{hpdelta}) says that $\sigma \geq d(X, \mathscr{Z}) \frac{1-\Delta}{\sqrt{3\Delta}} $: given 
a non-central $X \in \mathscr{A}$ to be measured with maximal disturbance $\Delta$. Then the 
attainable measurement quality $\sigma$ is worse than $d(X, \mathscr{Z}) \frac{1-\Delta}{\sqrt{3\Delta}}$.
We see that $\sigma$ becomes deplorable if $\Delta$
is lowered to zero. 

For example, let's look again at the unbiased measurement $\mathbf{M} : M_2 \otimes C_2 \to M_2$, 
discussed in example~\ref{voorbeeld} on page~\pageref{voorbeeld}.
Explicit calculation shows that $\mathbf{M}$ satisfies the conditions of proposition~(\ref{hpdelta})
for $\Delta = 1 - 2 \sqrt{\epsilon(1 - \epsilon)}$. It yields the estimate 
$\sigma \geq 2 \sqrt{
\frac	{\epsilon(1-\epsilon)}
	{3 - 6\sqrt{\epsilon(1-\epsilon)}} } $
whereas the real quality of $\mathbf{M}$ equals $\frac{ 2 \sqrt{\epsilon(1 - \epsilon)} }{1 - 2\epsilon}$.
In this case (and probably in general) the
estimate is rather crude\footnote
{ Note that the hideous $\sqrt{3\Delta}$ comes from the estimate 
$\|A \otimes \mathbb{I}\|_{\mathbf{M}} \leq \sqrt{3\Delta}\|A\|$.
This is the part of the proof where the crudeness comes in: In this particular example   
$\|A \otimes \mathbb{I}\|_{\mathbf{M}} \leq (1 - 2\epsilon)\|A\| \quad \forall A \in \mathscr{A}$. 
Taking the proof of proposition (\ref{hpdelta})
from there would yield the true $\sigma$ as an estimate.
}. 
But it does contain some general features of the curve $\sigma(\epsilon)$, notably
$\lim_{\epsilon \uparrow \frac{1}{2}} \sigma(\epsilon) = \infty$  and 
$\lim_{\epsilon \downarrow 0} \sigma(\epsilon) = 0$.   

\subsection{Local Generalization}

We have extended the Heisenberg principle by demanding that all states on
$\mathscr{A}$ are perturbed in norm only slightly instead of not at all.
Another way of `extending' it is by demanding that all states
are left exactly in place, but only with respect to some obser\-vables:
\begin{prop}[Generalized Heisenberg Principle]
Let $X \in \mathscr{A}$,
$Y \in \mathscr{B}$ be Hermitean.
Let $\mathbf{M}$ : $\mathscr{A} \otimes \mathscr{B} \to \mathscr{A}$
be an unbiased measurement of $X$ with pointer $\mathbb{I} \otimes Y$ 
and quality $\sigma$.
Let $0 \neq A \in \mathscr{A}$ with $\|[X,A]\| = \delta \|A\|$ be such that
$$
\mathbf{M}^* (\rho)(A \otimes \mathbb{I}) = \rho(A) \quad 
\forall \rho \in \mathscr{S}(\mathscr{A}).
$$ 
Then
$$ 	\sigma \geq \delta / 2.	$$
\end{prop} 
\textbf{Proof}:
\begin{quote}
Of course $\mathbf{M}(A \otimes \mathbb{I}) = A$.
By the C$^*\!$-Cauchy-Schwarz inequality,
\begin{eqnarray*}
\delta \|A\| &=& \|[ X , A ]\| = \|[\mathbf{M}(\mathbb{I} \otimes Y) , \mathbf{M}(A \otimes \mathbb{I})] - 
\mathbf{M}\left( [\mathbb{I} \otimes Y , A \otimes \mathbb{I}] \right) \| \\
&=& \| 2 i \Im \mathbf{F_{M}}(\mathbb{I} \otimes Y , A \otimes \mathbb{I} ) \|
\leq 2 \| \mathbb{I} \otimes Y \|_{\mathbf{M}} \| A\otimes\mathbb{I} \|_{\mathbf{M}}
\leq 2 \sigma \|A\|.
\end{eqnarray*}
\begin{flushright}
\emph{q.e.d.}
\end{flushright}
\end{quote}
For example, let $\mathscr{A} = \bigotimes_{i = 1}^{N} M_2 $.
Let $\sigma_x , \sigma_y , \sigma_z$ be the Pauli spin-matrices in $M_2$,
and denote by $\sigma_{\alpha}^{i}$ the observable 
$\mathbb{I}\otimes \ldots \otimes \mathbb{I}\otimes \sigma_{\alpha} \otimes \mathbb{I}\otimes
\ldots \otimes \mathbb{I}$.
Let $X = \frac{1}{N} \sum_{i = 1}^{N} \sigma_{x}^i$ and $A = \frac{1}{N} \sum_{i = 1}^{N} \sigma_{y}^i$
be the average spin in the $x$- and $y$-directions.
Then $\|[X,A]\| = \frac{2}{N} \|A\|$, so that any measurement of $X$ leaving $A$ untouched
automatically has quality $\sigma \geq \frac{1}{N}$.  
Accurate average spin measurement in all directions simultaneously is only possible in large systems.

\newpage \section{State Reduction and Collapse}

Until now we have only looked upon the Heisenberg principle from one side:
given a certain amount of disturbance, how much information can one gain from a system?
On the flip side, we may consider the following question. 
Given a measurement of a certain quality, how does this perturb the system?
In case of a perfect measurement ($\sigma = \|Y\|_{\mathbf{M}} = 0$), lemma (\ref{structure}) 
and its corollaries give fairly detailed restrictions on the structure of $\mathbf{M}$.
For example, $\mathbf{M}$ has to map $Y'$ 
into $\mathbf{M}(Y)'$.  
In the Schr\"odinger picture, this translates into a particularly nice 
answer to the above question known as `state collapse'.

\subsection{State Reduction}

Let $\rho$ be a state on $\mathscr{A}$, and $X \in \mathscr{A}$.
In definition~(\ref{condit}), 
we have defined the state $\rho_{X}$ on $\mathscr{A}$ by
$$
\rho_{X}(A) = \frac{\rho(X^{\dagger}AX)}{\rho(X^{\dagger}X)}.
$$
Let $\mathscr{A}$ and $\mathscr{B}$ be von Neumann algebras, and let
$\mathbf{M}$ : $\mathscr{B} \to
\mathscr{A}$ be a measurement of $X \in \mathscr{A}$ with pointer 
$Y \in \mathscr{B}$ and quality $\sigma = 0$. Let $\mathbf{P}(V)$ be the spectral projections
of $X$, $\mathbf{Q}(V)$ those of $Y$.
In corollary (\ref{spectromeas}), we have seen that $\mathbf{M}$ measures perfectly the spectral 
projections of $X$ with the corresponding ones \mbox{of $Y$:}
$\mathbf{M}(\mathbf{Q}(V)) = \mathbf{P}(V)$ and $\|\mathbf{Q}(V)\|_{\mathbf{M}} = 0$.
Under these circumstances, and if both $\rho$ and $\mathbf{M}$ are normal (i.e. weakly continuous), 
we have seen 2 examples of the above definition:

\begin{itemize}

\item[-]We may look at the normal state $\mathbf{M}^* (\rho)$ 
resulting from measurement. An observer $\mathscr{C} \ni Y$ may condition all its observations 
on the pointer outcome $V$. On page~\pageref{condex}, the conditioned probability 
is shown to be induced by the state $(\mathbf{M}^* (\rho))_{\mathbf{Q}(V)} $:
$$
\mathbb{P}_{\mathbf{M}^*(\rho) , B ,Y}( [B \, \mathrm{in} \, W] \,| \, [Y \, \mathrm{in} \, V])  =  
\mathbb{P}_{(\mathbf{M}^* (\rho))_{\mathbf{Q}(V)} , B}([B \, \mathrm{in} \, W]). 
$$

\item[-]
If \emph{any} perfect measurement of $Y$ yields outcome in $V$, the system
is experimentally known to be in state $\rho_{\mathbf{P}(V)}$ prior to measurement.
This is called the \emph{reduced state}. 
\end{itemize}
We prove proposition~(\ref{redrum}) in the completely positive setting, showing that the commuting diagram on 
page~\pageref{diagramke} remains valid for all completely positive perfect measurements $\mathbf{M}$:
\begin{prop}[Reduction] \label{reduction}
Let $\mathbf{M}$ : $\mathscr{B} \to \mathscr{A}$ be a measurement of $X \in \mathscr{A}$ with pointer 
$Y \in \mathscr{B}$ and quality $\sigma = 0$.
Then for all states $\rho \in \mathscr{A}$: 
$$
(\mathbf{M}^{*}\rho)_{Y} = \mathbf{M}^{*}(\rho_{X}). 
$$
\end{prop}
\newpage
\textbf{Proof}:
\begin{quote}
Writing out the definitions, we need to prove that for all $A \in \mathscr{A}$:
$$
\rho \big( \mathbf{M}(Y^{\dagger} A Y) \big) \rho(Y^{\dagger} Y)  = 
\rho \big( X^{\dagger} \mathbf{M}(A) X \big) \rho \big( \mathbf{M}(Y^{\dagger} Y) \big).
$$
Since $\|Y\|_{\mathbf{M}}$ = 0,
$
\mathbf{M}(Y^{\dagger} Y) = 
X^{\dagger} X 
$.
By the multiplication theorem~(\ref{multi}), 
$
\mathbf{M}(Y^{\dagger} A Y) = 
X^{\dagger} \mathbf{M}(A) X 
$.
Letting $\rho$ act on the above proves the assertion.
\begin{flushright}
\emph{q.e.d.}
\end{flushright}
\end{quote}
Taking $\mathbf{Q}(V)$ for $Y$ and $\mathbf{P}(V)$ for $X$ in the
 above proposition, 
$(\mathbf{M}^* (\rho))_{\mathbf{Q}(V)} = 
\mathbf{M}^* (\rho_{\mathbf{P}(V)})$: 
reducing a measured state according to pointer outcome 
and measuring a reduced state
amounts to the same thing. 
In particular, for $[B , Y] = 0$,
$$
\mathbb{P}_{\mathbf{M}^*(\rho) , B ,Y}( [B \, \mathrm{in} \, W] \,| \, [Y \, \mathrm{in} \, V])  =  
\mathbb{P}_{(\mathbf{M}^* (\rho_{\mathbf{P}(V)}) , B}([B \, \mathrm{in} \, W]). 
$$
If you measure $X$ with pointer $Y$ and register
an outcome in $V$, then direct observation of all $B \in \mathscr{C}$ will be as if, prior to measurement, 
the system had been in the reduced \mbox{state $\rho_{\mathbf{P}(V)}$}.

Of course this is also true for all indirect observations, as long as the outcome 
is not erased from the original pointer $Y$. If, after $\mathbf{M}^*$, an operation $\mathbf{N}^*$
takes place, leaving the pointer $Y$ and its projections untouched, then we may simply apply 
proposition~(\ref{reduction}) to $\mathbf{N}^* \circ \mathbf{M}^*$ instead of $\mathbf{M}^*$.  
This explains why state reduction $\rho \mapsto \rho_{\mathbf{P(V)}}$ is observed by $\mathscr{C}$
as long as $\mathscr{C}$ observes the outcome $X$ in $V$ indirectly. 

\subsection{State Collapse}

Note the essential difference between conditioning on the outcome [$X$ in $V$] and forming the 
$X$-reduced state: the former can only be done on $X'$, the latter on all of $\mathscr{A}$.  
Let $X$ have spectral measure $V \mapsto \mathbf{P}(V)$.
On $X'$, $\rho$ is a harmless classical superposition of $X$-reduced states: 
$$
\rho(A) = \sum_{I}\rho \big( \mathbf{P}(V_i) A \big) = \sum_I \rho \big( \mathbf{P}(V_i) A \mathbf{P}(V_i) \big) = 
\sum_I \rho \big( \mathbf{P}(V_i) \big) \rho_{{\mathbf{P}(V_i)}}(A) \quad 
\forall A \in X'.
$$
As far as $A \in X'$ is concerned, a system in state $\rho $ is simply in state $\rho_{\mathbf{P}(V_i)}$
with probabi\-lity $\rho(\mathbf{P}(V_i))$.
On page~\pageref{koliek}, we have introduce the \emph{collapse operation} $\mathbf{C} : \mathscr{A} \to \mathscr{A}$, defined by
$\mathbf{C}(A) \isperdef \sum_I \mathbf{P}(V_i) A \mathbf{P}(V_i)$. Then
$\mathbf{C}^* (\rho)$ is known as the 
\emph{collapsed} state: $\mathbf{C}^* (\rho) = \sum_I \rho(\mathbf{P}(V_i)) \rho_{{\mathbf{P}(V_i)}}$.

When you remove the restriction $A \in X'$, $\rho$ is 
\emph{not} just the classical superposition of its reduced states. The difference between
$\rho$ and $\mathbf{C}^* (\rho)$ on $\mathscr{A}$ is 
experimentally\footnote{Of course, reduction on classical (central) observables is the exception to the rule. 
	  For a central observable $Z$, the relative commutant $Z' $ equals
	  $ \mathscr{A}$, and any state may at any 
	  time be safely considered a classical superposition of $Z$-reduced states.} observable, 
e.g.  by two-slit-experiments or along the lines of page~\pageref{oegsnok}.   

In summary, a state $\rho$ is only a classical superposition of the $X$-reduced states $\rho_{\mathbf{P}(V_i)}$ 
on $X'$, 
the commutant of $X$. But proposition~(\ref{reduction}) shows that after measurement the tables are turned:
$\mathbf{M}^* (\rho)$ is a classical superposition of the $\mathbf{M}^* (\rho_{\mathbf{P}(V_i)})$ on  
$Y'$, not $X'$. In other words,  
$
\mathbf{M}^{*}(\rho) = 
\mathbf{M}^{*} \!\circ \mathbf{C}^{*} (\rho)$ on $Y'$. This can also be seen directly from the 
structure of $\mathbf{M}$.
\begin{prop}[Collapse]
Let $\mathscr{A}$, $\mathscr{B}$ be von Neumann algebras.
Let $\mathbf{M}$ : $\mathscr{B} \to \mathscr{A}$ be a perfect ($\sigma = 0$) measurement of a Hermitean $X \in \mathscr{A}$ with pointer 
$Y \in \mathscr{B}$. 
Let 
$V \mapsto \mathbf{P}(V)$ be the spectral measure of $X$. Let $\{ \, V_i \, | \, i \in I  \, \}$ 
be a countable decomposition of $\mathbf{Spec}(X)$. Then a collapse operation
$\mathbf{C}$ : $\mathscr{A} \to \mathscr{A}$ is defined by 
$\mathbf{C}(A) \isperdef \sum_{I} \mathbf{P}(V_i) A \mathbf{P}(V_i)$.
In this situation, we have for any $\rho \in \mathscr{S}(\mathscr{A})$:
$$
\mathbf{M}^{*}(\rho) = 
\mathbf{M}^{*} \! \circ \mathbf{C}^{*} (\rho) \quad \mathrm{on} \quad Y'.
$$
\end{prop}
\textbf{Proof}:
\begin{quote}
$\mathbf{M}$ maps $Y'$ into $X'$, and $\mathbf{C}$ leaves
$X'$ pointwise fixed. Therefore, if $B \in Y'$, we have $\mathbf{M}^{*} \! \circ \mathbf{C}^{*} (\rho)(B) = 
\rho \big( \mathbf{C} \circ \mathbf{M} (B) \big) =  \rho \big( \mathbf{M} (B) \big) = 
\mathbf{M}^{*}(\rho) (B)$. 
\begin{flushright}
\emph{q.e.d.}
\end{flushright}  
\end{quote}
The above proposition states that when you measure $X$ in state $\rho$, and 
then restrict attention to the commutant of the pointer, 
then the system will behave as if it had been in the collapsed state $\mathbf{C}^{*} (\rho)$ prior to measurement.
It generalizes the diagram on page~\pageref{diagramketje} to all completely positive perfect measurements. 

A measurement of $X$ with pointer $Y$ is called \emph{repeatable} if immediate repetition of the 
measurement would yield the same result,
i.e. if $\mathbf{M}^* (\rho) = \rho $ on $\mathscr{C}(X)$. 
Then $\|X\|_{\mathbf{M}}$ must be\footnote
{Since $\mathbf{M}(f(X)) = f(X)$ for all continuous $f$ on $\mathbf{Spec}(X)$, this is certainly true for
$f(x) = x^2$, whence $\mathbf{F}(X,X) = 0$.}~0, so according to the definitions $\mathbf{M}$ 
is also a perfect measurement of $X$ with \mbox{pointer $X$}.
According to the above proposition then, 
$\mathbf{M}^{*}(\rho)$
equals 
$\mathbf{M}^{*} \! \circ \mathbf{C}^{*} (\rho)$  
not only on $Y'$, but also on $X'$.

For example, if $\mathbf{M}$ : $\mathscr{A} \otimes \mathscr{B} \to \mathscr{A}$
measures $X$ with pointer $\mathbb{I} \otimes Y$ in a repeatable way,
then the distinction between states that are collapsed or intact prior to measurement 
can, after measurement, neither be made by observables of the form $A \otimes \mathbb{I}$ nor 
of the \mbox{form $\mathbb{I} \otimes B$}.

\subsection{Generalized State Collapse}

Very well. For perfect ($\sigma = 0$) measurements of $X$ with pointer $Y$, the essence of state-collapse 
is that $\mathbf{M}$ maps $Y'$ into $X'$. The way to think of $X'$ is the following:
in a finite-dimensional algebra, $X$ can be decomposed into projections as $X = \sum_{i} \lambda_i \mathbf{P}_i$.
And of course $\mathbf{M}(B) = \sum_{i,j} \mathbf{P}_i \mathbf{M}(B)
\mathbf{P}_j$ for any $B \in Y'$. 
Now $\mathbf{M}(B) \in X'$ 
means $\mathbf{M}(B) = \sum_{i} \mathbf{P}_i \mathbf{M}(B) \mathbf{P}_i$; then $\mathbf{M}(B)$ contains only 
diagonal blocks, like
$$
X = 	\left( 
		\begin{array}{ccccc}
		\lambda_1& 0&0&0&0 \\
		0&\lambda_1 &0&0&0 \\
		0& 0&\lambda_2&0&0 \\
		0& 0&0&\lambda_2&0 \\
		0& 0&0&0&\lambda_2 \\
		\end{array}
	\right)
	 \quad \Longrightarrow \quad
\mathbf{M}(B) = \left( 
		\begin{array}{ccccc}
		{}*&*&0&0&0 \\
		{}*&*&0&0&0 \\
		0&0&*&*&* \\
		0&0&*&*&* \\
		0&0&*&*&* \\
		\end{array}
	\right).	
$$
And of course this is just the disappearance of coherences between eigenstates of $X$ with different 
eigenvalues. In this light, an approximate collapse proposition would have to be something that 
says how small the `off-diagonal blocks' of $\mathbf{M}(B)$ get, provided that $B$ commutes rather well 
with the pointer $Y$, and that the quality of measurement $\sigma$ is not too bad.  
In other words: a Heisenberg-equivalent of lemma~(\ref{vectorredux}).
\begin{prop}[Generalized Collapse in the Heisenberg-Picture] \label{collapsedelta}
Let $\mathscr{A}$ be a \\*\mbox{von Neumann algebra}.
Let $\mathbf{M} : \mathscr{B} \to \mathscr{A}$ be an unbiased measurement of  $X \in \mathscr{A}$ with pointer 
$Y \in \mathscr{B}$ (both Hermitean)
and quality $\sigma$.  
Let $\mathbf{Spec}(X) \supseteq V \mapsto \mathbf{P}(V)$ denote the 
projection valued measure belonging to $X$. Suppose $B \in \mathscr{B}$ is a Hermitean 
element such that $\| [Y,B] \| = \delta \|B\|$.
Then
$$
\| \mathbf{P}( [x , x + \epsilon] ) \mathbf{M}(B) \mathbf{P}( [y , y + \epsilon] ) \| \leq
\frac{\delta + 2\sigma + \epsilon}{|x - y|} \|B\|.
$$  
\end{prop}
\textbf{Proof}:
\begin{quote}
By the C$^*\!$-Cauchy-Schwarz inequality,
\begin{eqnarray} 
\| [ X , \mathbf{M}(B)] \| &=& \| \mathbf{M}([Y,B]) + 2i\Im \mathbf{F}(B,Y) \| \nonumber \\
	& \leq & \| [ Y , B ] \| + 2 \|Y\|_{\mathbf{M}}\|B\|_{\mathbf{M}} \nonumber \\
	& \leq & (\delta + 2\sigma)\|B\|. \label{paaidans}
\end{eqnarray}
In order to dehorrify our formulas, we introduce some notation.
First of all, we define $u_n \isperdef u + n \epsilon$ for $u,\in \mathbb{R}$, $n \in \mathbb{Z}/2$.
Secondly, $\mathbf{P}_{u,n} \isperdef \mathbf{P} \big( [u_n , u_{n+1}] \big)$.
And finally,
we define  
$X_u = \sum_{n \in \mathbb{Z}} u_{n + \frac{1}{2}} \mathbf{P}_{u,n}$.
This is an approximation of $X$ by a step function operator, so that
\bq \label{tafel}
\| X - X_{u} \| \leq \epsilon/2
\eq
This leads us to 
\begin{eqnarray}
\lefteqn{ \| X_x \mathbf{M}(B) - \mathbf{M}(B) X_y \|  = }   \nonumber\\ 
&=& \| [X , \mathbf{M}(B)] + \mathbf{M}(B)(X - X_y)  - (X - X_x)\mathbf{M}(B)\| \nonumber \\
& \leq & (\delta + 2\sigma)\|B\| + \epsilon \| \mathbf{M}(B) \| \nonumber \\
& \leq & (\delta + \epsilon + 2\sigma ) \|B\|. \label{kolnoot}
\end{eqnarray}
Now for any 2 projections $\mathbf{P}$ and $\mathbf{Q}$ and for any $A$ in $\mathscr{A}$, 
we have $\|\mathbf{P} A \mathbf{Q}\| \leq \|\mathbf{P} \|\| A \|\|\mathbf{Q}\| \leq \|A\|$.
We use this in the second step below. In the first step, we use that  
 $\mathbf{P}_{u,n} \mathbf{P}_{u,0} = \delta_{n,0}\mathbf{P}_{u,0}$ for $n \in \mathbb{N}$.
And in the third step, we make use of $\sum_{n \in \mathbb{Z}} \mathbf{P}_{u,n}
= \mathbb{I}$.
We then obtain
\begin{eqnarray}
\lefteqn{ |x - y| \, \| \mathbf{P} \big( [x , x + \epsilon] \big) \mathbf{M}(B) 
\mathbf{P}\big( [y , y+\epsilon] \big) \| = } \nonumber \\
& = & \bigg\| \mathbf{P}_{x,0}  
\bigg( \sum_{m,n \in \mathbb{Z}}  
\big( x_{n + \frac{1}{2}} - y_{m + \frac{1}{2}} \big) 
\mathbf{P}_{x,n} 
\mathbf{M}(B)  
\mathbf{P}_{y,m}  \bigg) 
\mathbf{P}_{y,0} \bigg\|  \nonumber\\
& \leq & \bigg\|  
\sum_{m,n \in \mathbb{Z}}  
\big( x_{n + \frac{1}{2}} - y_{m + \frac{1}{2}} \big) 
\mathbf{P}_{x,n}
\mathbf{M}(B)  
\mathbf{P}_{y,m}
\bigg\|  \nonumber \\
&  = & \bigg\| \sum_{n \in \mathbb{Z}}  
 x_{n + \frac{1}{2}}  
\mathbf{P}_{x,n}
\mathbf{M}(B)    
- \sum_{m \in \mathbb{Z}}  
 y_{m + \frac{1}{2}} 
\mathbf{M}(B)  
\mathbf{P}_{y,m}  \bigg\| \nonumber \\
& = & \big\| X_x \mathbf{M}(B) -  X_y \mathbf{M}(B) \big\|. \label{oorknie}
\end{eqnarray} 
With inequality (\ref{kolnoot}), we now have what we wanted.
\begin{flushright}
\emph{q.e.d.}
\end{flushright}
\end{quote}
Even if $\delta = 0$, the norm distance between $\mathbf{M}(B)$ and $X'$ does not go to zero 
as $\sigma \downarrow 0$.
Suppose that $X$ has a continuous spectrum. If $\sigma \neq 0$, however small, 
we can always choose $x$ and $y$ in $\mathbf{Spec}(X)$ so that $|x - y| \leq 2 \sigma$.
The above proposition then becomes trivial. It allows for large off-diagonal elements as 
long as they are close to the diagonal.

This is physically relevant: suppose that the internal energy $X$ 
of a block of iron is measured with an accuracy $\sigma$ of few microjoules. 
Clearly this measurement does not produce decoherence between energy-states inside the atoms, 
i.e. eigenstates with energies $x$ and $y$ differing several $eV$.
Indeed, the estimates `kick in' only if the energy difference approaches the quality of measurement:
$|x - y| \sim \sigma$.

\subsubsection{Almost Classical Observables}

Collapse with respect to a central observable $X$ is meaningless. 
This has nothing to do with measurement whatsoever:   
since the spectral projections $\mathbf{P}(V)$ of a central observable $X$ are central,
we see that
$\mathbf{C}^* (\rho) (A) = \rho \left( \sum_{i} \mathbf{P}(V_i) A \mathbf{P}(V_i) \right) = 
\rho \left( \sum_{i} \mathbf{P}(V_i) A \right) = \rho(A)$.
Thus $\mathbf{C}^* (\rho) = \rho$ for all states $\rho$.

This can be generalized for almost classical observables $X$, i.~e.\ observables for which
$d(X , \mathscr{Z})$ is small.
\begin{prop}
Let $\mathscr{A}$ be a von Neumann algebra with centre $\mathscr{Z}$.
Let $A,X \in \mathscr{A}$, $X$ Hermitean.
Let $\mathbf{P}(V)$ be the spectral projections of $X$. Then
$$
\left\| \mathbf{P}\big( [x , x + \epsilon] \big) A \mathbf{P}\big( [y , y+\epsilon] \big) \right\| \leq
\frac{  \epsilon + 2 d(X , \mathscr{Z})  }{|x - y|} \| A \|.
$$  
\end{prop}
\textbf{Proof}: 
\begin{quote}
Pretty much the same as that of proposition (\ref{collapsedelta}).
Under the assumptions above, inequality~(\ref{tafel}) remains 
valid, as does (\ref{oorknie}) with $\mathbf{M}(B)$ replaced by $A$.
Inequality (\ref{paaidans}) is replaced by 
\bq 
\| \, [ X , A] \, \| \leq 2 d(X , \mathscr{Z}) \| A \|
\eq
and (\ref{kolnoot}) by
\begin{eqnarray}
\| X_x A - A X_y \|  
&=& \| [X , A] + A(X - X_y)  - (X - X_x)A\| \nonumber \\
 & \leq & (2 d(X , \mathscr{Z}) + \epsilon) \| A \|. 
\end{eqnarray}
\begin{flushright}
\emph{q.e.d.}
\end{flushright}
\end{quote}
The same caveat as before applies: if $d(X , \mathscr{Z})$ is non-zero, then off-diagonal
blocks close to the diagonal can remain large, so that no bound for $\|\rho
- \mathbf{C}^* (\rho)\|$ 
is obtained. 

\subsection{Generalized State Reduction}

In view of proposition (\ref{reduction}), it is tempting to speculate 
that for measurements with good quality ($\sigma \ll 1$),
perhaps also $\| (\mathbf{M}^* (\rho))_{Y} - \mathbf{M}^* (\rho_{X} ) \| \ll 1$ for all $\rho$.
Alas, nature is cruel and hard:
consider again the \ref{voorbeeld}$^{\mathrm{th}}$ example on page~\pageref{voorbeeld}.
\mbox{$\mathbf{M}$ (also)} measures
$
X =
\left( 	\begin{array}{cc}
 		1 - \epsilon 	& 0		\\
		0 		&\epsilon 	\\
	\end{array}
\right)
$
with pointer
$
Y =
\mathbb{I}\otimes \left( 	\begin{array}{cc}
				1  	& 0	\\
				0 	& 0 	\\
				\end{array}
			\right)$. 
One easily calculates $\sigma = \sqrt{\epsilon(1 - \epsilon)}$.
But, taking for $\rho$ the spin-down state with density matrix
$
\left( 	\begin{array}{cc}
				0  	& 0	\\
				0 	& 1 	\\
				\end{array}
			\right)
$
, one may figure out (identifying $M_2 \otimes C_2$ with $M_2 \oplus M_2$) that
$(\mathbf{M}^* ( \rho))_Y$ is represented by the density matrix  
$
\left( 	\begin{array}{cc}
				0  	& 0	\\
				0 	& 1 	\\
				\end{array}
			\right)
\oplus 
\left( 	\begin{array}{cc}
				0  	& 0	\\
				0 	& 0 	\\
				\end{array}
			\right)			
$
and $\mathbf{M}^* (\rho_X)$ by 
$
\left( 	\begin{array}{cc}
				0  	& 0		\\
				0 	& \epsilon 	\\
				\end{array}
			\right)
\oplus 
\left( 	\begin{array}{cc}
				0  	& 0		\\
				0 	& 1 - \epsilon 	\\
				\end{array}
			\right)			
$.
Consequently $\| (\mathbf{M}^*( \rho))_{Y} - \mathbf{M}^* (\rho_{X} ) \| = 1 - \epsilon$.
Thus, by choosing $\epsilon$ small, it is possible to have measurements with $\sigma \ll 1$ 
yet $\| (\mathbf{M}^* (\rho))_{Y} - \mathbf{M}^* (\rho_{X} ) \| \approx 1$.
In the example above however, the ratio 
$\frac
{\mathbf{var}_{\mathbf{M}^* (\rho)}(Y) - \mathbf{var}_{\rho}(X)}
{\mathbf{var}_{\mathbf{M}^* (\rho)}(Y)}
$ 
equals 1 for all $ \epsilon $. And smallness of this ratio \emph{does} force an 
approximate reduction, as we will see below. Once again it is not the quality
$\sigma$ `an sich' that 
regulates reduction, but the quality divided by the typical variations in pointer outcome, 
cf. proposition (\ref{collapsedelta}).  
\begin{prop}[Generalized reduction] \label{appred}
Let $X \in \mathscr{A}$, $Y \in \mathscr{B}$ be Hermitean.
Let $\mathbf{M}: \mathscr{B} \to \mathscr{A}$ be an unbiased measurement of $X$
with pointer $Y$. Then
$$
\| (\mathbf{M}^* (\rho))_{Y} - \mathbf{M}^* (\rho_{X} ) \| \leq 
2
\sqrt{
\frac
{\mathbf{var}_{\mathbf{M}^* (\rho)}(Y) - \mathbf{var}_{\rho}(X)}
{\mathbf{M}^* (\rho) (Y^{\dagger} Y)}
}
\left( 1 +
\sqrt
{
	\frac
	{\mathbf{var}_{\mathbf{M}^* (\rho)}(Y) - \mathbf{var}_{\rho}(X)}
	{\mathbf{M}^* (\rho) (Y^{\dagger} Y)}
}
\right). 
$$
\end{prop}
\textbf{Proof}:
\begin{quote}
Brutally applying the C$^*\!$-Cauchy-Schwarz inequality would get the job done. That is to say it yields 
$
\| (\mathbf{M}^* (\rho))_{Y} - \mathbf{M}^* (\rho_{X} ) \| \leq 3 \|B\|
\frac{\sigma}{\sqrt{\mathbf{M}^{*} (\rho) (Y^{\dagger} Y)}}
$.
Partly because I don't like the numerator being independent of $\rho$ 
(allowing  
$\frac{\sigma^2}{\mathbf{M}^* (\rho) (Y^{\dagger} Y)}$ 
to blow up whereas
$\frac
	{\mathbf{var}_{\mathbf{M}^* (\rho)}(Y) - \mathbf{var}_{\rho}(X)}
	{\mathbf{M}^* (\rho) (Y^{\dagger} Y)}$
is nicely bounded by 1)  
and partly to keep you from dozing off, we'll go about it another way\footnote
{I was put on this track by M$\breve{\mathrm{a}}$d$\breve{\mathrm{a}}$lin
Gu\k{t}$\breve{\mathrm{a}}$, who suggested a simple proof of
the C$^*\!$-Cauchy-Schwarz inequality in the case of completely positive maps.} for a
change.
By the GNS-representation (see \cite{ka1}[p.~278]), we may assume $\mathscr{A}$ to be an algebra of operators on some Hilbert
space $\mathscr{H}_{\rho}$, with $\rho$ a vector state $\psi_{\rho}$.
By the Stinespring theorem (see \cite[p.~194]{tak}), we may assume $\mathscr{B}$ to be 
an algebra of operators on some Hilbert space
$\mathscr{R}$, and the existence of a contraction $V: \mathscr{H}_{\rho} \to \mathscr{R}$ such 
that $\mathbf{M}$ is of the form $\mathbf{M}(B) = V^{\dagger} B V$.
Then $\mathbf{M}^{*}\rho$ is a vector state with vector $V \psi_{\rho}$, since 
$\mathbf{M}^* (\rho)(B) = \langle \psi_{\rho}|V^{\dagger}BV | \psi_{\rho} \rangle$.
In the proof of lemma (\ref{sigma}), we have seen that 
$$
\mathbf{var}_{\mathbf{M}^* (\rho)}(Y) - \mathbf{var}_{\rho} \big( \mathbf{M}(Y)
\big) = \rho \big( \mathbf{F}(Y,Y) \big).
$$
If we introduce the notation $W \isperdef \sqrt{(\mathbb{I} - V^{\dagger}V)} $,
we have 
$$
\rho(\mathbf{F}(Y,Y)) = 
\langle V \psi_{\rho}| Y^{\dagger} W^{2} Y |V \psi_{\rho} \rangle = 
\|  W Y V \psi_{\rho}  \|^2.
$$
The rest is hardly exhilarating: for any $B \in \mathscr{B}$,
\begin{eqnarray*}
\lefteqn{
\mathbf{M}^{*}\rho(Y^{\dagger} B Y) - \rho\big(\mathbf{M}(Y^{\dagger})\mathbf{M}(B)\mathbf{M}(Y)\big) = }\\
&=& \langle V \psi_{\rho}| Y^{\dagger} B Y - Y^{\dagger} V^{\dagger}V B V^{\dagger}V Y | V \psi_{\rho} \rangle  \\
& = &
\langle V \psi_{\rho}| 
Y^{\dagger} W^{2} B Y  + 
Y^{\dagger} B W^{2} Y  -
Y^{\dagger} W^{2} B W^{2} Y
| V \psi_{\rho} \rangle \\
 & \leq & 
 2 \| B \| \| YV\psi_{\rho} \| \|  W Y V \psi_{\rho}  \| +
\| B \| \|  W Y V \psi_{\rho}  \|^2. 
\end{eqnarray*}
So, since 
$$
\mathbf{M}^* (\rho)(Y^{\dagger}Y) = \rho \big(\mathbf{M}(Y)^{\dagger} \mathbf{M}(Y) \big) +
\|  W Y V \psi_{\rho}  \|^2 
$$
we see that
\begin{eqnarray*}
\lefteqn{ \| (\mathbf{M}^* (\rho))_{Y} (B) - \mathbf{M}^* (\rho_{\mathbf{M}(Y)} )(B)  \| =} \\
&=&	\left\|
\frac
{\mathbf{M}^{*}\rho(Y^{\dagger} B Y) \rho\big(\mathbf{M}(Y)^{\dagger} \mathbf{M}(Y) \big) -
 \rho \big( \mathbf{M}(Y)^{\dagger}\mathbf{M}(B)\mathbf{M}(Y) \big) \mathbf{M}^* (\rho)(Y^{\dagger}Y)}
{\rho\big(\mathbf{M}(Y)^{\dagger} \mathbf{M}(Y) \big) \mathbf{M}^* (\rho)(Y^{\dagger}Y) }
	\right\|   \\
&=& \bigg\|
\frac
{
 \rho \big(\mathbf{M}(Y)^{\dagger} \mathbf{M}(Y) \big)
\Big(\mathbf{M}^{*}\rho(Y^{\dagger} B Y)  -
 \rho\big(\mathbf{M}(Y)^{\dagger}\mathbf{M}(B)\mathbf{M}(Y) \big) \Big)
}
{\rho\big(\mathbf{M}(Y)^{\dagger} \mathbf{M}(Y) \big) \mathbf{M}^* (\rho)(Y^{\dagger}Y) }
  \\
& & - \quad \frac
{ 
 \rho\big(\mathbf{M}(Y)^{\dagger}\mathbf{M}(B)\mathbf{M}(Y) \big) 
 \big\|  W Y V \psi_{\rho}  \big\|^2
 }
{\rho\big(\mathbf{M}(Y)^{\dagger} \mathbf{M}(Y) \big) \mathbf{M}^* (\rho)(Y^{\dagger}Y) }
\bigg\|  \\
& \leq &2 \| B \|
\frac
{ \|  W Y V \psi_{\rho}  \| }
{ \| YV\psi_{\rho} \| } \quad + \quad
2\| B \|
\frac
{ \|  W Y V \psi_{\rho}  \|^2 }
{ \| YV\psi_{\rho} \|^2 }   \\
& = &2 \|B\|
\Bigg( 
\sqrt
{
	\frac
	{\mathbf{var}_{\mathbf{M}^* (\rho)}(Y) - \mathbf{var}_{\rho}\big(\mathbf{M}(Y)\big)}
	{\mathbf{M}^* (\rho) (Y^{\dagger} Y)}
}
 +
\frac
	{\mathbf{var}_{\mathbf{M}^* (\rho)}(Y) - \mathbf{var}_{\rho}\big(\mathbf{M}(Y)\big)}
	{\mathbf{M}^* (\rho) (Y^{\dagger} Y)}
\Bigg).  
\end{eqnarray*}
\begin{flushright}
\emph{q.e.d.}
\end{flushright}
\end{quote}
Since $0 \leq \mathbf{var}_{\mathbf{M}^* (\rho)}(Y) \leq \mathbf{M}^* (\rho) (Y^{\dagger} Y)$, we may
also write down a weaker version, starring the ratio 
$\frac
{\mathbf{var}_{\mathbf{M}^* (\rho)}(Y) - \mathbf{var}_{\rho}(X)}
{\mathbf{var}_{\mathbf{M}^* (\rho)}(Y)}
$ discussed above:
\begin{cor}
Let $X \in \mathscr{A}$, $Y \in \mathscr{B}$ be Hermitean.
Let $\mathbf{M}$: $\mathscr{B} \to \mathscr{A}$ be an unbiased measurement of $X$
with pointer $Y \in \mathscr{B}$ such that $ \mathbf{var}_{\mathbf{M}^* (\rho)}(Y) \neq 0$.
Then
$$
\| (\mathbf{M}^* (\rho))_{Y} - \mathbf{M}^* (\rho_{X} ) \| \leq 
2
\sqrt{
\frac
{\mathbf{var}_{\mathbf{M}^* (\rho)}(Y) - \mathbf{var}_{\rho}(X)}
{\mathbf{var}_{\mathbf{M}^* (\rho)}(Y)}
}
\left( 1 +
\sqrt
{
	\frac
	{\mathbf{var}_{\mathbf{M}^* (\rho)}(Y) - \mathbf{var}_{\rho}(X)}
	{\mathbf{var}_{\mathbf{M}^* (\rho)}(Y)}
}
\right). 
$$
\end{cor}

If $\mathbf{M}$ is a measurement with outcome 0 or 1, i.e. $Y$ is a projection, 
then automatically $\mathbf{var}_{\mathbf{M}^* (\rho)}(Y) = p(1 - p)$ if $p$
is the probability of measuring outcome 1.
From this and   
$\mathbf{var}_{\mathbf{M}^* (\rho)}(Y) - \mathbf{var}_{\rho}(X) \leq \sigma^2$,
we obtain another corollary.
\begin{cor}
Let $X \in \mathscr{A}$ be Hermitean.
Let $\mathbf{M}$: $\mathscr{B} \to \mathscr{A} $ be an unbiased measurement of $X$ of quality $\sigma$ 
which only allows outcomes 0 and 1, i.e. the pointer $Y$ is a projection.
Then for all states $\rho$ with probability $p$ of measuring outcome 1:
$$
\| (\mathbf{M}^* (\rho))_{Y} - \mathbf{M}^* (\rho_{X} ) \| \leq 
2 \frac{\sigma}{\sqrt {p(1 - p)}} \left( 1 + \frac{\sigma}{\sqrt{p(1 - p)}}
\right).
$$ 
\end{cor}

\newpage \section{A Paradox Resolved} \label{zuuk}

Imagine the following thought experiment.
The universe is described by the algebra $\mathscr{D = A \otimes B}$.
An observer $\mathscr{C \subset B}$ contains two separate pointers 
$Y_1$ and $Y_2$.
(One may think of a computer memory consisting of 2 classical bits, for example.)
A perfect measurement $\mathbf{M} : \mathscr{C \to A \otimes B}$ is
performed on $X \in \mathscr{A}$ using $Y_1$ as a pointer. 
(Information is stored in the first bit.) 
Since all time-evolution is automorphic, $\mathbf{M}$ must have as dilation 
some automorphism $\alpha$ of $\mathscr{D}$. Then there must still be
observables  $D \in \mathscr{A} \otimes \mathscr{B}$ on which no collapse
occurs. 

Having learnt the outcome of the first measurement, $\mathscr{C}$  
performs a second perfect measurement $\mathbf{N} : \mathscr{C \to A \otimes B}$ 
but now on $D$, using $Y_2$ as pointer.
(This information is stored in the second bit.)
Comparing information stored on $Y_1$ with that on $Y_2$, 
$\mathscr{C}$ has solved the riddle of reduction once and for
all:
\begin{itemize}
\item[-]Either there is a full and objective \emph{reduction} after the first measurement,
	and the state of $\mathscr{D}$ jumps into an eigenstate of $X$.
\item[-]Or all time evolution is automorphic, and purity on $\mathscr{D}$ is conserved
\end{itemize}
The difference cannot be seen on observables commuting with $Y_1$, but it can be seen on $D$. 
(Un?)fortunately, such a crucial experiment is not possible: apply the 
next proposition \mbox{to $\alpha \circ \mathbf{N}$}.
\begin{prop} \label{crux}
Suppose $Y_1, Y_2 \in \mathscr{B}$ are Hermitean elements such that \mbox{$[Y_1
, Y_2] = 0$}. 
Suppose $\mathbf{M}: \quad \mathscr{C} \to \mathscr{A}\otimes\mathscr{B}$ 
is a measurement of $D \in  \mathscr{A}\otimes\mathscr{B}$ with pointer
$Y_2$ such that $\mathbf{M}(Y_1) = \mathbb{I} \otimes Y_1$. Then 
$$ [D, \mathbb{I} \otimes Y_1 ] = 0. $$
\end{prop}
\textbf{Proof}:
\begin{quote}
$\sigma_2 = \| Y_2 \|_{\mathbf{M}} = 0$, so 
$0 = \mathbf{M}([Y_1 , Y_2 ]) = 
[ \mathbf{M}(Y_1) , \mathbf{M}(Y_2)] 
= [\mathbb{I} \otimes Y_1 , D]$ 
\begin{flushright}
\emph{q.e.d.}
\end{flushright}
\end{quote}
It is true that a measurement can be performed on $D \not\in (\mathbb{I} \otimes
Y_{1})'$. But this necessarily erases
the information that was gained on $X$ from the pointer $Y_1$. \label{dolsnok}

\backmatter

\chapter{Epilogue}

The subject of quantum measurement is particularly susceptible to misunderstanding.
\mbox{I would} therefore like to clarify (perhaps superfluously) my view on
the so-called `measurement problem' in relation to the interpretation of quantum mechanics 
ventilated on page~\pageref{mijnes}. 
The problem of measurement is commonly defined as follows: 
\begin{quote}
\emph{How and when do observables take one particular value out of all the possibilities allowed by quantum mechanics?}
\end{quote}
On page~\pageref{niretsch} as well as on page~\pageref{jm}, I have briefly sketched some of the problems one 
would have to overcome when answering this question. I do not make any attempt
to do so.
In my mind, a more relevant question seems to be: 
\begin{quote}
\emph{How and when is one particular value of an observable
observed by one particular observer, out of all the possibilities allowed by quantum mechanics?}
\end{quote}  

In order to consider this question, one needs a theory with mathematical representatives of both the primitive notion of 
`observable' and of `observer'.
Observables are commonly modelled by Hermitean elements. This seems rather
sensible to me.
But how to model an observer? 

In my mind, the key property of any `observer' is that it is able to 
directly observe a number of observables. I therefore represent an abstract `observer' by the set $\mathscr{C} \subset \mathscr{D}$ of 
all observables which it can detect directly. 
Since an observer can construct sums, products and limits from the 
observed values of observables in $\mathscr{C}$, it seems plausible that $\mathscr{C}$ is a C$^*\!$-algebra.
And since simultaneous observation of observables in $\mathscr{C}$
necessarily induces a map of the form discussed in the example following proposition~(\ref{jm}), $\mathscr{C}$
may only contain commuting observables. $\mathscr{C}$ is an Abelian C$^*\!$-algebra.   
   
One would like to assign a value to each $D \in \mathscr{D}$.
This cannot be done in a consistent\footnote{At least if $\mathscr{D} = \mathscr{B(H)}$ with $\mathit{dim}(\mathscr{H} > 2)$. See \cite{koc}.} 
manner.  But with the interpretation on page~\pageref{mijnes}, consistency is only necessary within Abelian algebras.
It suffices to have a random generator do the following:   
\begin{itemize}
\item[-] At time 0, choose an Abelian $\mathscr{C \subset D}$.
\item[-] Assign values to all Hermitean $C \in \mathscr{C}$ in a consistent manner, according to the joint probability
	 measures induced by $\rho$. These are the values observed by $\mathscr{C}$
	 at time 0.
\item[-] Repeat this for all possible abelian $\mathscr{C \subset D}$.
Quantum mechanics does not prescribe joint probability distributions for
non-commuting observables, so we have some freedom in our choice of random
generator. The procedure need not be independent for different observers:
there may well be some consistency in their observations. 
But according to \cite{koc}, full consistency is impossible in general.
\end{itemize}
Now the question `What value of $C \in \mathscr{C}$ is observed by $\mathscr{C}$ at time $t$?' is ans\-wered as follows.
Look at the Abelian algebra $\alpha_t (\mathscr{C})$ at time 0. The Hermitean observable $\alpha_t(C)$ gets 
assigned a value in $\mathbf{Spec}(\alpha_t(C)) = \mathbf{Spec}(C)$ with probability distribution
$\mathbb{P}_{\rho, \alpha_t(C)} = \mathbb{P}_{\alpha_{t}^{*}(\rho), C}$. This is the value of $C$ 
observed by $\mathscr{C}$ at time $t$. 

This is a deterministic procedure: at time 0, the random generator determines what each observer gets to observe at time $t$.
But observations are not objective.
If $A \in \mathscr{C}$ and $A \in \tilde{\mathscr{C}}$, it may well be that $A$ gets different values
with $\mathscr{C}$ and $\tilde{\mathscr{C}}$: different observers observe different values of the same 
observable at the same time. From this, you see that the same observer $\mathscr{C}$ may also observe 
different values of the same $C$ at different times. The trick is to show that these observations are always 
consistent with all other observations of the same observer at the same time, and that they allow an observer 
to store information about the world around it.
This can be done entirely within 
the framework of (quantum) probability theory, without any reference to the nature of the random generator, 
c.f. \ page~\pageref{polpokje}, proposition~\ref{redrum}, proposition~\ref{reduction} and page~\pageref{zuuk}.      

This is my way to interpret these propositions. But I would like to emphasize for one last time that the interpretation 
above is merely a tool. Any consistent interpretation is just as good as any other.
The structure of nature is engraved in mathematics, and interpretations only serve to tie abstract structure to daily experience.
This is why I have chosen to be brief in explanation and tedious in calculation.     
In particular, it explains why the above exposition is muffled away in this epilogue.

\cleardoublepage

\bigskip


\begin{thebibliography}{K\&R}

\bibitem[DAr]{ari}G.~M.~D'Ariano, On the Heisenberg Principle, Namely on the Information-disturbance 
			Trade-off in a Quantum Measurement, \textsl{Fortschr.\ Phys.\ }\textbf{51}, 
			No.~4--5 (2003), 318--330.  

\bibitem[Dav]{dav}E.~B.~Davies, Quantum Theory of Open Systems, Academic Press, London, 1976.


\bibitem[Bel]{bel}J.~S.~Bell, On Wave Packet Reduction in the Coleman-Hepp Model, 
			\textsl{Helv.\ Phys.\ Acta} \textbf{48} (1975), 93--98.

\bibitem[B\"oh]{boh}A.~B\"ohm, Quantum Mechanics, Springer Verlag, New York, 1979.

\bibitem[B\&J]{bj}B.~H.~Bransden and C.~J.~Joachain, Introduction to Quantum Mechanics, 
	Langman Scientific\&technical and Wiley\&Sons inc, New York, 1989 

\bibitem[Coh]{coh}Donald Cohn, Measure Theory, Birkh\"auser, Boston, 1980.

\bibitem[Dir]{dir}P.~A.~M.~Dirac, The Principles of Quantum Mechanics, Oxford University Press, London, 1958.

\bibitem[Hep]{hep}K.~Hepp, Quantum Theory of Measurement and Macroscopic Observables, 
			\textsl{Helv.\ Phys.\ Acta} \textbf{45} (1972), 237--248.  
			
\bibitem[Hol]{hol}A.~S.~Holevo, Probabilistic and Statistical Aspects of Quantum Theory, North-Holland 			
		Publishing Company, Amsterdam, 1982.	

\bibitem[Jau]{jau}J.~M.~Jauch, Foundations of Quantum Mechanics, Addison-Wesley, Reading Massachusetts, 1968.			
		
\bibitem[K\&S]{koc}S.~Kochen and E.~P.~Specker, The Problem of Hidden Variables in Quantum Mechanics,		
		\textsl{ J.~Math.~Mech.\ }\textbf{17} (1976), 59--87.
\bibitem[K\&R]{ka1}R.~V.~Kadison and J.~R.~Ringrose, Fundamentals of the theory of Operator Algebras I and II,
			Academic Press, London, 1983/1986.		
\bibitem[Kra]{kra}K.~Kraus, General State Changes in Quantum Theory, \textsl{ Ann.\ Phys.\ } \textbf{64} (1971), 311--335.  

\bibitem[Lan]{lan}E.~C.~Lance, Hilbert C$^*\!$-modules: a toolkit for operator algebraists, Cambridge University Press, 1995.
			
\bibitem[Tak]{tak}M.~Takesaki, Theory of Operator Algebras I, Springer-Verlag, New York, 1979. 

\bibitem[Maa]{maa}
J.~D.~M.~Maassen, Quantum Probability, Quantum Information and Quantum Computing, 
\texttt{www.math.kun.nl/medewerkers/maassen}, (2004). 

\bibitem[Neu]{neu}J.~von Neumann, Mathematische Grundlagen der Quantenmechanik, Springer-Verlag, Berlin, 1932. 

\bibitem[Wer]{wer}R.~F.~Werner, Quantum Information Theory -- an Invitation, 
		\textsl{Springer Tracts in Modern Physics} \textbf{173} (2001), 14--57. Or alternatively
		\texttt{xxx.lanl.gov/abs/quant-ph/0101061>quant-ph/0101061}.
		 

\end{thebibliography}
\end{document}